\documentclass[longauth]{aa}  

\usepackage{graphicx}
\usepackage{txfonts}
\usepackage{hyperref}
\usepackage{multirow}

\newcommand{\jm}{$j_\mathrm{M}$}
\newcommand{\jb}{$j_B$}

\bibpunct{(}{)}{;}{a}{}{,}
\graphicspath{{Figures/}}

\begin{document} 


\title{The miniJPAS survey. Evolution of the luminosity and stellar mass functions of galaxies up to $z\sim0.7$}

\authorrunning{L.~A.~D\'iaz-Garc\'ia et al.}
\titlerunning{Evolution of the luminosity and stellar mass functions up to $z\sim0.7$}

\author{
L.~A.~D\'iaz-Garc\'ia\inst{\ref{IAA}}
\and
R.~M.~Gonz\'alez Delgado\inst{\ref{IAA}}
\and
R.~Garc\'ia-Benito\inst{\ref{IAA}}
\and
G.~Mart\'inez-Solaeche\inst{\ref{IAA}}
\and
J.~E.~Rodr\'iguez-Mart\'in\inst{\ref{IAA}}
\and
C.~L\'opez-Sanjuan\inst{\ref{CEFCA2}}
\and
A.~Hern\'an-Caballero\inst{\ref{CEFCA}}
\and
I.~M\'arquez\inst{\ref{IAA}}
\and
J.~M.~V\'ilchez\inst{\ref{IAA}}
\and
R.~Abramo\inst{\ref{IFUSP}}
\and
J.~Alcaniz\inst{\ref{ON}}
\and
N.~Benítez\inst{\ref{IAA}}
\and
S.~Bonoli\inst{\ref{CEFCA2},\ref{DIPC},\ref{IKER}}
\and
S.~Carneiro \inst{\ref{IFUFB}}
\and
A.~J.~Cenarro\inst{\ref{CEFCA2}}
\and
D.~Crist\'obal-Hornillos\inst{\ref{CEFCA2}}
\and
R.~A.~Dupke\inst{\ref{ON}}
\and
A.~Ederoclite\inst{\ref{CEFCA}}
\and
A.~Mar\'in-Franch\inst{\ref{CEFCA}}
\and
C.~Mendes de Oliveira\inst{\ref{IAGUSP}}
\and
M.~Moles\inst{\ref{CEFCA2},\ref{IAA}}
\and
L.~Sodr\'e\inst{\ref{IAGUSP}}
\and
K.~Taylor\inst{\ref{Instruments4}}
\and
J.~Varela\inst{\ref{CEFCA2}}
\and
H.~V\'azquez Rami\'o\inst{\ref{CEFCA}}
}

\institute{
Instituto de Astrof\'{\i}sica de Andaluc\'{\i}a (CSIC), P.O.~Box 3004, 18080 Granada, Spain\label{IAA}
\newline\email{luis@iaa.es}
\and
Centro de Estudios de F\'isica del Cosmos de Arag\'on (CEFCA), Unidad Asociada al CSIC, Plaza San Juan 1, E-44001, Teruel, Spain\label{CEFCA2}
\and
Centro de Estudios de F\'isica del Cosmos de Arag\'on (CEFCA), Plaza San Juan 1, E-44001, Teruel, Spain\label{CEFCA}
\and
Instituto de F\'isica, Universidade de S\~{a}o Paulo, Rua do Mat\~{a}o 1371, 05508-090, S\~{a}o Paulo, Brazil\label{IFUSP}
\and
Observat\'orio Nacional -- MCTI (ON), Rua General Jos\'e Cristino, 77, S\~ao Crist\'ov\~ao, 20921-400, Rio de Janeiro, Brazil\label{ON}
\and
Donostia International Physics Center (DIPC), Manuel Lardizabal Ibilbidea 4, San Sebasti\'an, Spain\label{DIPC}
\and
IKERBASQUE, Basque Foundation for Science, 48013 Bilbao, Spain\label{IKER}
\and
Instituto de F\'isica, Universidade Federal da Bahia, 40210-340, Salvador, BA, Brazil\label{IFUFB}
\and
Universidade de S\~{a}o Paulo, Instituto de Astronomia, Geof\'isica e Ci\^encias Atmosf\'ericas, Rua do Mat\~{a}o 1226, 05508-090, S\~{a}o Paulo, Brazil\label{IAGUSP}
\and
Instruments4, 4121 Pembury Place, La Canada Flintridge, CA 91011, USA\label{Instruments4}
}

\date{Received ? / Accepted ?}

 
\abstract
{}
{We aim at developing a robust methodology for constraining the luminosity and stellar mass functions (LMFs) of galaxies by solely using data from multi-filter surveys and testing the potential of these techniques for determining the evolution of the Javalambre Physics of the Accelerating Universe Astrophysical Survey (J-PAS) LMFs up to $z \sim 0.7$.}
{As J-PAS is still an ongoing survey, we use the miniJPAS dataset (a stripe of $1$~deg$^2$ dictated according to the J-PAS strategy) for determining the LMFs of galaxies up to $z \sim 0.7$. Stellar mass and $B$-band luminosity for each of the miniJPAS galaxies are constrained using an updated version of our fitting code for spectral energy distribution, MUlti-Filter FITting (MUFFIT), whose values are based on non-parametric composite stellar population models and the probability distribution functions of the miniJPAS photometric redshifts. Galaxies are classified according to their star formation activity through the stellar mass versus rest-frame colour diagram corrected for extinction (MCDE) and we assign a probability to each source of being a quiescent or star-forming galaxy. Different stellar mass and luminosity completeness limits are set and parametrised as a function of redshift, for setting the limitations of our flux-limited sample ($r_\mathrm{SDSS} \le 22$) for the determination of the miniJPAS LMFs. The miniJPAS LMFs are parametrised according to Schechter-like functions via a novel maximum likelihood method accounting for uncertainties, degeneracies, probabilities, completeness, and priors.}
{Overall, our results point to a smooth evolution with redshift ($0.05 \le z \le 0.7$) of the miniJPAS LMFs in agreement with previous work. The LMF evolution of star-forming galaxies mainly involve the bright and massive ends of these functions, whereas the LMFs of quiescent galaxies also exhibit a non-negligible evolution on their faint and less massive ends. The cosmic evolution of the global $B$-band luminosity density decreases $\sim 0.1$~dex from $z=0.7$ to $0$, whereas for quiescent galaxies this quantity roughly remains constant. In contrast, the stellar mass density increases $\sim0.3$~dex at the same redshift range, where such evolution is mainly driven by quiescent galaxies owing to an overall increasing number of this kind of galaxies, which in turn includes the majority and most massive galaxies ($60$--$100$~\% fraction of galaxies at $\log_{10}(M_\star/M_\odot) \gtrsim 10.7$).}
{}


\maketitle
%

\section{Introduction}\label{sec:intro}

The luminosity and stellar mass functions of galaxies (LMFs in the following) establish the comoving number density of galaxies per luminosity and stellar mass interval, respectively. Luminosity functions are open to be defined according to any photometric band or spectral range, mostly rest-frame luminosities and absolute magnitudes, whereas stellar mass functions are only attached to this stellar population parameter, which is subject to data analysis and scaled according to a stellar population synthesis model set. Primarily, these functions are largely used for a preliminary estimation of the expected number of galaxies included in an imaging survey, as well as for defining their strategy plans. Owing to the large amount of information contained in the LMFs, these are largely employed for a wide variety of purposes in multiple researches. 

Amongst the many applications, galaxy luminosity functions can be used as priors to improve the precision of photometric redshifts (hereafter photo-$z$) and reduce the number of photo-$z$ outliers at fainter magnitudes \citep[see e.g.][]{Benitez2000,Ilbert2006,Brammer2008,Arnouts2011,Molino2014,HernanCaballero2021}. This is thanks to luminosity functions are able to account for how frequent are the templates embedded in photo-$z$ codes (which are related to the different spectral-types of galaxies) and that the volume sampled at lower redshifts is smaller \citep[see e.g.][]{Benitez2000,Brammer2008}. Regarding other cosmological applications, the LMFs can be also used for supporting the development and creation of mock catalogues with realistic galaxy clustering \citep[see e.~g.][]{Somerville2015}. In brief, these functions help to test, fine-tune, and calibrate schemes for populating halos from dark-matter-only simulations with galaxies, such as semi-analytic models, the subhaloe abundance matching, the halo occupation distribution schemes, and others \citep[][]{Yang2003,Vale2004,Zheng2005,Baugh2006,Conroy2006,Benson2010,Guo2011,Smith2017}. In the same sense, results from hydrodynamical simulations are often cross-checked with the local stellar mass function of galaxies \citep[][]{Genel2014,Schaye2015,Cui2018}.

As the LMFs and their evolution are tightly related to the star formation processes taking place in galaxies \citep[see e.g.][]{Wilkins2008,Behroozi2013,Ilbert2013,Bernardi2016}, these functions have been mainly discussed and included in galaxy evolution studies. In particular, the cosmic evolution of the LMFs or the evolution of the comoving number densities of galaxies with redshift have been a powerful tool to provide hints on the evolutive paths of certain type of galaxies \citep[see e.g.][]{Ferreras2009a,Ferreras2009b,Ilbert2010,Ilbert2013,Brammer2011,Moustakas2013,DiazGarcia2019b}. In this regard, processes such as quenching \citep[i.e.~cessation of star formation][]{Faber2007,Peng2015} must be also imprinted in the LMFs. Therefore, these functions have been employed to understand the physical processes leading to the bimodal distribution of galaxies in colour--magnitude, stellar mass--colour, and $UVJ$-like diagrams and their evolution with redshift \citep[see e.g.][]{Faber2007,Ilbert2010,Ilbert2013,Moustakas2013,Schawinski2014,DiazGarcia2019a}. In this regard and thanks to the LMFs, some authors figured out that massive quiescent galaxies ($\log_{10}M_\star \gtrsim 10.7$) show a rapid and active increase in number since $z\sim 3$ and efficient up to $z\sim 1$, that is referred as `mass quenching' and independent of environment \citep[see e.g.][]{Peng2010,Ilbert2013}. On the other hand, multiple works found that the variation in number density of the LMFs of this kind of galaxies at $z < 1$ is mainly driven by less massive quiescent galaxies \citep[][]{Pozzetti2010,Ilbert2010,Ilbert2013,Maraston2013,Moresco2013,Moustakas2013,DiazGarcia2019b}. Also based on these diagrams, the number density of the so-called `green valley' galaxies can be also used for setting contraints on the typical time scales of quenching processes, where some works apparently point to a fast transition \citep[see e.g.][]{Bell2004,Muzzin2013,Schawinski2014,Angthopo2019,DiazGarcia2019b}. Complementarily, many authors have investigated whether LMFs depend on the environment or the density of galaxies within a projected radius. Mostly, it is well accepted that the LMFs of galaxies vary according to the local density or environment in which they reside, in the sense that more luminous and massive galaxies, especially red luminous galaxies, typically reside in more dense environments  \citep[][]{Eke2004,Baldry2006,Peng2010,Zandivarez2011,Guo2014,Nantais2016,VazquezMata2020}. Interestingly, there is evidence of non-negligible excess of low mass quiescent galaxies and spheroids in high density environments \citep[][]{Peng2010,Scoville2013,Moutard2018} with respect to the field. In fact, this result led to propose a second quenching mechanism dubbed `environment quenching' acting in low mass systems at $z<1$, which would result from ram pressure stripping \citep[][]{Gunn1972,Abadi1999,Quilis2000} or strangulation \citep[][]{Larson1980,Balogh2000} and may explain the low-mass upturn in the stellar mass function of quiescent galaxies \citep[][]{Drory2009,Taylor2015}. These facts are also reflected in the fraction of quiescent galaxies in galaxy groups and clusters, where this fraction is higher for more dense environments \citep[][]{Woo2013,Nantais2016,Liu2021,GonzalezDelgado2022,RodriguezMartin2022,Sobral2022}.

For tackling the determination of LMFs, a large variety of methodologies have been proposed and developed during the last decades, which can be used for constraining both functions due to their similarities. Some of them do not assume any functional form for the LMFs and are referred as non-parametric, hence the non-parametric LMF are discretised in stellar mass and/or absolute magnitude bins. One of the most extended non-parametric methods for the determination of the LMFs is the so-called $1/V_\mathrm{max}$ method \citep[][]{Schmidt1968,Marshall1985}. Alternatively, more non-parametric methods have been proposed such as the $C^-$ \citep[][]{LyndenBell1971}, $C^+$ \citep[][]{Zucca1997}, and SWML \citep[][]{Efstathiou1988} methods for the determination of discretised LMFs. On the other hand, when observations are deep enough and the sample is complete in a wide range of stellar mass or luminosity, many authors prefers to develop methods that assume a parametric LMF based on the empirical Schechter function \citep[][]{Schechter1976}. Briefly, a Schechter function adopts a power-law function for describing the number density of faint galaxies, whereas the number density of bright galaxies decays exponentially. Thus, a Schechter function is defined according to three parameters: the power or slope of the faint end, the characteristic stellar mass or luminosity, and the normalisation in number density units; parameters that, in general, are correlated or degenerated \citep[][]{Beare2015,LopezSanjuan2017}. During the last decades, many works showed that the LMFs of galaxies are properly described by parametric Schechter-like functions since very high redshift \citep[see e.g.][]{PerezGonzalez2008,Ilbert2013,Muzzin2013,Moustakas2013,Wright2018,Ishigaki2018,Bhatawdekar2019,Bowler2020,Donnan2023,PerezGonzalez2023}. Furthermore, recent works also show that for a general case, sometimes the use of a double-Schechter function (i.e.~a combination of two Schechter functions) is more adequate than the single case, which is commonly interpreted as some galaxy populations are assembled according to different galaxy formation scenarios \citep[see e.g.][]{Drory2009,Peng2010,Ilbert2013}. In this sense, the LMFs of quiescent galaxies are often described or fitted by a double Schechter function, while the star-forming ones are parametrised through making use of a single Schechter function \citep[e.g.][]{Li2009,Ilbert2010,Baldry2012,Muzzin2013,Weigel2016,LopezSanjuan2017,Ishigaki2018}.

Recent large scale and/or deep multi-filter surveys including or combining broad, intermediate, and narrow-band filters \citep[e.g.~COMBO-17\footnote{Classifying Objects by Medium-Band Observations in 17 filters}, COSMOS\footnote{Cosmological Evolution Survey}, ALHAMBRA\footnote{Advanced Large Homogenous Area Medium Band Redshift Astronomical Survey}, PAU\footnote{Physics of the Accelerating Universe}, SHARDS\footnote{Survey for High-\textit{z} Absorption Red and Dead Sources}, J-PAS\footnote{Javalambre Physics of the Accelerating Universe Astrophysical Survey}, J-PLUS\footnote{Javalambre Photometric Local Universe Survey}, and S-PLUS\footnote{Southern Photometric Local Universe Survey}][]{Wolf2003,Scoville2007,Moles2008,Benitez2009a,PerezGonzalez2013,Benitez2014,Cenarro2019,MendesOliveira2019} have proven to be very adequate for the determination of precise galaxy photo-$z$ \citep[$\sigma_z \sim 0.01$;][]{Wolf2004,Ilbert2009,Marti2014,Molino2014,Molino2020,HernanCaballero2021}. This fact, together with the advantages that this kind of surveys offer, makes them very attractive for the study of the LMFs of galaxies since intermediate redshift. Some of these advantages comprise: (i) there is no pre-selection of the sample since all the galaxies imaged down to the magnitude limit of the detection band are included in the analysis; (ii) images allow the use of non-fixed aperture photometry for estimating the total flux of each surveyed galaxy, meaning that there is overall no aperture bias in the determination of the total luminosity or stellar mass; and (iii) these surveys can easily map large areas of the sky in a wide redshift range that yield unbiased and statically robust samples of galaxies, meaning that systematic effects as those coming from Poisson errors and cosmic variance may be largely reduced or even neglected in the best cases. 

In this paper, we aim at developing a robust methodology for determining the so-called luminosity and stellar mass functions of galaxies, which must be properly adapted to the particularities of large scale multi-filter surveys like J-PAS, putting emphasis on the proper determination of these functions up to $z \sim 0.7$. Whenever possible, our method must include the outputs or value-added catalogues available from the J-PAS collaboration. In particular, we specially require for the analysis the inclusion of the typical outputs from our trusted fitting codes for spectral energy distribution (SED) in the J-PAS collaboration, such as the ones described in \citet[][]{GonzalezDelgado2021}, in order to perform a solid and statistical analysis accounting for the uncertainties, degeneracies, and correlations amongst all the involved parameters. The ultimate goal consists in checking that all our techniques, J-PAS inputs, and results are self-consistent as to provide proper LMF results that are in agreement with previous studies for a flux-limited sample up to $r = 22.0$~AB-magnitudes and $z=0.7$. As the development of our methods for determining the LMFs is based on a preliminary J-PAS-like data set, which is dubbed miniJPAS \citep[][further details in Sect.\ref{sec:minijpas}]{Bonoli2021}, our conclusions and results are focused on introducing our techniques, as well as showing the J-PAS potential and LMF studies that can be achieved once the J-PAS footprint is largely imaged rather than shedding light on new LMF results. As a consequence of our analysis, we also provide a statistical spectral-type classification of galaxies (quiescent versus star-forming), as well as the stellar mass and luminosity constraints for each of the galaxies that will be made publicly available in a miniJPAS value-added catalogue.

This paper is organised as follows. In Sect.~\ref{sec:minijpas}, we introduce the miniJPAS survey and the value-added catalogues employed throughout this research. The SED-fitting techniques and methodologies developed to constrain the stellar population properties that are needed to determine the miniJPAS LMFs are summarised in Sect.~\ref{sec:sedfit}. The process to perform the classification of galaxies according to their spectral type, as well as the luminosity and stellar mass completeness of our sample are detailed in Sects.~\ref{sec:qprob} and \ref{sec:completeness}, respectively. The novel methodology to statistically determine the characteristic parameters defining the LMFs is detailed in Sect.~\ref{sec:zsty}. In Sect.~\ref{sec:results}, we present our main results, which include the miniJPAS LMFs of galaxies up to $z=0.7$ and the cosmic evolution of the stellar mass and luminosity densities in the same redshift range. Finally, we discuss our results and conclusions in Sect.~\ref{sec:discussion}, while a brief summary of this work is included in Sect.~\ref{sec:summary}.

Throughout this paper we adopt a spatially flat lambda cold dark matter ($\Lambda$CDM) cosmology with parameters based on the recent \textit{Planck} results \citep[][]{Planck2020}, namely: Hubble constant of $H_0=67.4$~km~s$^{-1}$~Mpc$^{-1}$ (i.e.~$h=0.674$), $\Omega_\mathrm{M}=0.315$ (total matter density), and $\Omega_\Lambda=0.685$ (cosmological constant density). Stellar masses are quoted in solar mass units [$M_\odot$] and are scaled according to a universal \citet{Chabrier2003} initial stellar mass function (IMF). Magnitudes are expressed in the AB system \citep{Oke1983}. 


\section{The miniJPAS survey}\label{sec:minijpas}


The miniJPAS survey \citep[][]{Bonoli2021} was conceived to primarily test the performance of the Javalambre Survey Telescope (JST250) optical system and determine the potential and applications of the J-PAS spectro-photometric data. Until the main camera of the J-PAS survey (JPCam) is up and running, we are carrying out a first scientific exploitation of the miniJPAS dataset. Therefore, all the data used throughout this research comes from the miniJPAS survey and its value-added catalogues. Even though miniJPAS slightly differs in some aspects with respect to J-PAS, the miniJPAS strategy and configuration (Sect.~\ref{sec:minijpas:observations}) was dictated according to the baseline survey strategy of J-PAS \citep[further details in][]{Benitez2014, Bonoli2021} in order to comply with the J-PAS requirements. Owing to the large number of parameters involved in the determination of the LMFs, we take advantage of the miniJPAS value-added catalogues containing contraints on the galaxy photo-$z$ (Sect.~\ref{sec:minijpas:photoz}) and the star/galaxy classification (Sect.~\ref{sec:minijpas:starclass}) of the sources that belong to our sample of miniJPAS sources (Sect.~\ref{sec:minijpas:sample}).


\subsection{Observations and instrumentation}\label{sec:minijpas:observations}

The miniJPAS survey was conducted at the Javalambre Astronomical Observatory\footnote{\url{https://oajweb.cefca.es}} (OAJ), a site with an excellent median seeing and darkness that is dedicated to large scale multi-filter surveys to image the northern hemisphere \citep[][]{Moles2010}. In particular, all the miniJPAS observations were done with the JPAS-Pathfinder camera, that is, the first scientific instrument installed at the JST250 telescope. This camera includes a single CCD of $9.2k \times 9.2k$ pixels, with an effective field-of-view (FoV) of $0.27$~deg$^2$, and a pixel scale of $0.23\arcsec$~pix$^{-1}$. In fact, the greatest difference between miniJPAS and J-PAS resides in the camera employed for the observations (JPCam, a $1.2$~Gpixel detector composed of 14 CCDs with a FoV of $4.2$~deg$^{2}$ and the same pixel scale than JPAS-Pathfinder). The JST250 telescope is a wide-field telescope of $2.55$~m based at the OAJ with an effective collecting area of $3.75$~m$^2$ and an optimised éttendue of $26.5$~m$^2$~deg$^2$ as to perform large FoV surveys \citep[][]{Cenarro2018}.

One of the main characteristics of the miniJPAS survey is that it was performed making use of the J-PAS filter system. This unprecedented photometric system comprises $54$ narrow-band filters with a full-width-at-half-maximum of $FWHM \sim 145$~\AA\ (equally spaced every $100$~\AA), one broad band ($u_\mathrm{JAVA}$, $FWHM \sim 495$~\AA) and one high-pass filter ($J1007$) extending to the UV and near-infrared ends of the optical range, which yields an effective wavelength range of $3500$--$9300$~\AA\ \citep[further details in][]{Bonoli2021}. The narrow band filters are properly characterised by transmission curves with very steep side slopes and flat tops, which are usually referred as top-hat filters. As a result, the photometric system delivers spectral energy distributions (SEDs) or J-spectra with an equivalent wavelength resolution of $R\sim 60$ for each of the imaged sources. For the miniJPAS case, the filter set was complemented with the `standard' broad-band filters $u_\mathrm{JPAS}$, $g_\mathrm{SDSS}$, $r_\mathrm{SDSS}$, and $i_\mathrm{SDSS}$. Only $u_\mathrm{JPAS}$ differs a little from the SDSS-like $u$ band, since it has a redder cut-off.

The miniJPAS field lies on the well-known AEGIS\footnote{All-wavelength Extended Groth Strip International Survey} field, centred at (RA, Dec) = ($215\degr$, $+53\degr$), and is composed of a stripe of four tiles or pointings in such a way that covers the EGS\footnote{Extended Groth Strip} field. There is a little overlapping by $3.6\arcmin$ between contiguous pointings ($0.09$~deg$^2$ of total overlapping area) and the total area observed amounts to $\sim1$~deg$^{2}$ of the sky. However, after masking bright stars, the window frame, and artefacts (typically from internal light reflections in the system), the effective area of the miniJPAS surveys amounts to $0.895$~deg$^2$. The average image depths, defined for a $5\sigma$ level and a circular aperture of $3\arcsec$, of the narrow-band filter set range from $23.5$ for the bluer bands to $22$~AB-magnitudes for the redder parts of the photometric system. Regarding the broad-band filters, the image depths range from $23$ to $24$~AB-magnitudes. Owing to the observational campaign, the FWHM of the point spread function (PSF) ranges from $0.6\arcsec$ to $2\arcsec$, with most of the bands showing FWHMs below $1.5\arcsec$.


\subsection{Photometric redshifts}\label{sec:minijpas:photoz}

The J-PAS photometric system was specifically designed to measure Baryon Acoustic Oscillations \citep[][]{Benitez2009b,Benitez2014} along the line of sight, which theoretically requires a redshift precision of $\sigma_z \sim 0.003 \times (1+z)$ \citep[][]{Benitez2009a}. To achieve the photo-$z$ precision for this aim, the photo-z group within the J-PAS collaboration worked on different photo-$z$ codes. For this research, we preferentially focus on the JPHOTOZ package \citep[][]{HernanCaballero2021}, which was developed at Centro de Estudios de Física del Cosmos de Aragón\footnote{\url{https://www.cefca.es/}} (CEFCA) as part of the data reduction pipeline for J-PAS, also known as JYPE. It is worth mentioning that there exist alternative photo-$z$ codes that were tested in the collaboration such as TOPz \citep[Tartu Observatory photo-$z$, details in][]{Laur2022}, which also satisfies the photo-$z$ precision requirements via SED-fitting techniques. Throughout this work, we only make use of the photo-$z$ predictions obtained by the JPHOTOZ package, since checking the impact over LMFs due to the use of different photo-$z$ constraints is beyond the scope of this work.

In brief, JPHOTOZ is a set of \texttt{Python} scripts that interface between the database and a customised version of the SED-fitting code \textsc{LePhare} \citep[][]{Arnouts2011}, which is preferentially adapted to work with J-PAS-like data. The set of templates used for estimating the J-PAS photo-$z$ is composed of $50$ synthetic templates, which were generated by using \textsc{CIGALE}\footnote{Code Investigating GALaxy Emission} \citep[][]{Boquien2019}. This set of templates was carefully defined through an iterative method based on scores evaluating the combination of templates that optimised the photo-$z$ of miniJPAS galaxies after comparison with spectroscopic redshifts \citep[for further details see][]{HernanCaballero2021}. It is also of note that the miniJPAS photo-$z$ were determined using the \textsc{LePhare} redshift priors from spectroscopic redshifts from the VIMOS VLT survey \citep[VVDS,][]{LeFevre2005}, meaning that according to the $i$-magnitude of the galaxy and the rest-frame $(g-i)$ colour of each template a prior is applied. For a better photo-$z$ determination, a recalibration offset was applied to the input photometry, which is recomputed pointing per pointing by an iterative process.

To perform a robust statistical analysis about the determination of the miniJPAS LMFs, we require of the photo-$z$ probability distribution functions (PDZ in the following), and its associated \textit{odds} value, for each of the miniJPAS sources. This is neccesary for including all the uncertainties and peculiarities of the miniJPAS PDZs. Therefore, for this work, we make use of the PDZ of miniJPAS galaxies rather than the photo-$z$ with the highest probability (maximum of the PDZ or $z_\mathrm{best}$) for each miniJPAS source (see also Sect.~\ref{sec:sedfit}). The $odds$ \citep[parameter first defined in][]{Benitez2000,Molino2014}, which is also provided by the JPHOTOZ package, is the probability enclosed by the PDZ in an interval around its maximum value, which is defined according to $z_\mathrm{best} \pm 0.03\times(1+z_\mathrm{best})$ for the miniJPAS case. Certainly, this parameter has proven to correlate with the quality of each photo-$z$ estimation \citep[][]{HernanCaballero2021,HernanCaballero2023}, as well as to be sensitive to catastrophic outliers. Consequently, $odds$ is included in our analysis for a better determination of the LMFs (see Sect.~\ref{sec:zsty}). As a reference, the typical error for the photo-$z$ of miniJPAS galaxies is set at $\sigma_\mathrm{NMAD}=0.013$ with an outlier rate of $\eta=0.39$ for $r_\mathrm{SDSS} < 23$ \citep[][]{HernanCaballero2021}. For $\textit{odds} > 0.82$, the typical error decreases up to $\sigma_\mathrm{NMAD}=0.003$ and $\eta=0.05$, which results in $\sim 5\,200$ galaxies per deg$^{-2}$ with this photo-$z$ accuracy. We note that the miniJPAS PDZs range from $z=0$ to $z=1.5$, that is, a much larger redshift range than the redshift upper-limit ($z=0.7$) of the LMFs in this work. All the data related to the photo-$z$ constraints can be downloaded from the CEFCA web portal\footnote{\url{http://archive.cefca.es/catalogues/minijpas-pdr201912}} via Astronomical Data Query Language (ADQL) queries, which are available in the table \texttt{minijpas.PhotoZLephare\_updated} with keywords \texttt{SPARSE\_PDF} and \texttt{ODDS}. 


\subsection{Star and galaxy classification}\label{sec:minijpas:starclass}

For a correct determination of the LMFs, specially at brighter magnitudes, it is required a star and galaxy classification. For this purpose, we take advantage of the Bayesian classification by probability distribution function (PDF) analysis initially developed for the J-PLUS survey by \citet[][]{LopezSanjuan2019}. This Bayesian analysis accounts for morphological information from the miniJPAS $g_\mathrm{SDSS}$, $r_\mathrm{SDSS}$, and $i_\mathrm{SDSS}$ broad-band filters that combines with prior probabilities on the $r_\mathrm{SDSS}$ magnitude, which is related to the larger number of galaxies at fainter magnitudes in this kind of surveys. This method is performed for each of the pointings to avoid issues related to variations in the observing conditions and sky positions. To complement this classification, the miniJPAS catalogues includes parallax information from \textit{Gaia} catalogues. The star and galaxy classification method assigns a Bayesian probability between unity and zero to determine that a source is classified as a star or galaxy ($P_\mathrm{\star}$ and $P_\mathrm{G}$, respectively). As this classification only comprises two kind of sources, it is trivially deduced that $P_\mathrm{G} = 1 - P_\mathrm{\star}$. We find that the distribution of probabilities is strongly bimodal, where the vast majority of the miniJPAS sources present $P_\mathrm{G}$ values around unity or null values. For instance, at magnitudes brighter than $r_\mathrm{SDSS} = 22$, only a $3$~\% fraction of miniJPAs sources show probabilities in the range $0.1 < P_\mathrm{G} < 0.99$. Intermediate values of $P_\mathrm{G}$ mainly comprises faint sources. As the majority of the miniJPAS value-added catalogues, the star and galaxy classification assumed in this work can be found at the CEFCA web portal, more precisely, in the table \texttt{minijpas.StarGalClass} and keyword \texttt{total\_prob\_star}.

It is worth mentioning that additional galaxy and star classifications, such as \citet[][]{Baqui2021}, have been added recently. Machine Learning (ML) methods as those explored in \citet[][]{Baqui2021} demonstrated to perform star and galaxy classifications with a high success rate, specially at faint magnitudes. However, we find that there are little discrepancies between performances of the best ML algorithms from \citet[][]{Baqui2021} and the `default' miniJPAS star and galaxy classification at magnitudes $r_\mathrm{SDSS} < 22.5$ \citep[see Figs.~11 and 12 in][]{Baqui2021}. Therefore, we do not expect that the method chosen to perform the star and galaxy classification (default versus ML) to play a role on the determination of the LMFs up to $r_\mathrm{SDSS} \le 22$ at all. However, as mentioned in previous sections, we do not intend to investigate the impact of utilising alternative classifications for determining the miniJPAS LMFs since it falls outside the scope of this study.


\subsection{Photometry and definition of our miniJPAS sample}\label{sec:minijpas:sample}

After the reduction of the miniJPAS images, the source detection and built of photometric catalogues was performed by the well-known Source-Extractor program \citep[SExtractor][]{Bertin1996}. As miniJPAS is a survey involving several astronomical topics, catalogues include magnitudes and fluxes for many different aperture definitions, such as fixed-circular apertures, Petrosian \citep[\texttt{PETRO},][]{Petrosian1976}, automatic \citep[\texttt{AUTO}, inspired by Kron's `first moment' algorithm][]{Kron1980}, and PSF-corrected apertures \citep[\texttt{PSFCOR}, see e.g.][]{Molino2014,Molino2019}. Except for \texttt{PSFCOR}, all of them were computed in both single- and dual-mode. For the latter, the detection and aperture definition of sources were done with the $r_\mathrm{SDSS}$-band images to subsequently perform forced photometry in the rest of band images. 

 For our aims, the choice of aperture photometry to be used primarily relies on the compromise of including or approaching the total flux of galaxies with the highest possible signal-to-noise ratio. With this in mind, the use of dual-mode photometry within the \texttt{AUTO} apertures from SExtractor is specially motivated. This kind of photometry comprises fluxes or magnitudes within elliptical apertures, which are scaled and defined according to the order moments of the light distribution of each source. In fact, these are intended to estimate the total flux of galaxies where the mean fraction of flux lost is around $6$~\% \citep[][]{Bertin1996,Graham2005}. The \texttt{PSFCOR} and \texttt{PETRO} magnitudes were mainly discarded for the LMF analysis addressed in this work because (i) the former commonly yield lower fluxes by $~0.5$ magnitudes owing to smaller apertures than \texttt{AUTO} \citep[][]{GonzalezDelgado2021}, meaning that the total luminosity and galaxy stellar mass is underestimated or biased; and (ii) \texttt{PETRO} uncertainties ($r_\mathrm{SDSS}$-band magnitude) are on average $\sim35$~\% larger as a consequence of larger apertures than \texttt{AUTO}, which results particularly troublesome in determining the luminosity and stellar mass of fainter miniJPAS galaxies and/or at high redshift. From now on, unless otherwise stated, all the magnitudes in this work refer to \texttt{AUTO} magnitudes.
 
The parent catalogue for this work is the miniJPAS Public Data Release (PDR201912)\footnote{\url{https://j-pas.org/datareleases/minijpas_public_data_release_pdr201912}}. As previously stated in \citet{Bonoli2021}, the primary miniJPAS photometric catalogue (table \texttt{minijpas.MagABDualObj}) is complete in detection at $r_\mathrm{SDSS} = 23.6$ and $r_\mathrm{SDSS} = 22.7$ for point-like and extended sources, respectively. As galaxies are in general extended sources and magnitudes close to the detection limit are noisy, we define a flux-limited initial sample at $r_\mathrm{SDSS} \le 22.5$. Moreover, we add extra constraints to guarantee aperture photometry quality: (i) 
sources must be detected within the image window frame and outside of masked regions (bright stars or artefacts), (ii) photometry in the detection band ($r_\mathrm{SDSS}$) can only include SExtractor flags equal to $\texttt{FLAGS} = 0$ and $2$, which removes saturated and truncated sources
among others. We note that at this point we have not imposed constraints on point-like sources or redshift, these issues will be statistically addressed during the LMF methodology and posterior analysis (Sect.~\ref{sec:zsty}). As a result, the flux-limited parent catalogue is composed of $N_\mathrm{S} \sim 12\,600$ sources. As we detail in Sect.~\ref{sec:sedfit}, we impose extra conditions to perform a proper SED-fitting analysis of this parent catalogue, which will slightly reduce the number of sources. Even tough this is the parent sample that we use for the present statistical research, it is worth remarking at this point that we base the determination of the miniJPAS LMFs on the cumulative probability of the sample at $0.05 \le z \le 0.7$ and $r_\mathrm{SDSS} \le 22$ (largely detailed in Sect.~\ref{sec:zsty}).


\section{SED-fitting analysis of the miniJPAS sources}\label{sec:sedfit}


Along with the parameters obtained from the value-added catalogues, we require more parameters to determine the LMFs of galaxies in the miniJPAS field. We obtain the rest of parameters from SED-fitting techniques (Sect.~\ref{sec:sedfit:muffit}), and more precisely, we base our method on the PDFs (Sect.~\ref{sec:sedfit:pdf}) of the parameters involved in order to perform a statistical analysis making the most of the available data. As we show below, SED-fitting results not only allow us to determine rest-frame luminosities and stellar mass, but also rest-frame colours corrected for extinction for spectral-type classification (Sect.~\ref{sec:qprob}) for each of the sources in our parent sample. Thanks to all these parameters and PDFs, we are also able to determine the stellar mass and luminosity completeness of our flux-limited sample (Sect.~\ref{sec:completeness}), which is also needed to set the limitations of the miniJPAS sample for the determination of the LMFs at certain stellar mass and luminosity ranges.


\subsection{Updated version of MUFFIT and input ingredients for SED-fitting analysis}\label{sec:sedfit:muffit}

The stellar population properties of galaxies needed to perform our analysis are constrained using an updated version of the code MUFFIT \citep[MUlti-Filter FITting for stellar population diagnostics;][]{DiazGarcia2015}. Briefly, MUFFIT is a SED-fitting code particularly developed and optimised to deal with multi-band photometric data. MUFFIT is based on an error-weighted $\chi^2$-test and includes composite models of stellar populations (CSP; a non-parametric star formation history based on mixtures of two simple stellar population models or SSPs) in order to constrain the stellar population properties of galaxies. MUFFIT has proven to be a tool for this aim and easily adaptable to tackle the typical peculiarities that this kind of surveys entail \citep[see e.g.][]{DiazGarcia2015,DiazGarcia2019a,DiazGarcia2019b,DiazGarcia2019c,GonzalezDelgado2021}.

All the properties of the stellar content of galaxies are established solely on the basis of the stellar continuum of galaxies or colours. For miniJPAS and J-PAS, this is specially relevant since their filter sets mainly include narrow bands that are sensitive to strong nebular or AGN emission lines, which are in turn not accounted by SSP models. In this regard, MUFFIT is able to remove bands affected by strong emission lines (e.g.~[\ion{O}{ii}], [\ion{O}{iii}], H$\beta$, H$\alpha$+[\ion{N}{ii}], and [\ion{S}{ii}]) from the SED-fitting analysis, hence getting a proper determination of galaxy properties via colours. Errors and degeneracies in the parameters due to photon-noise uncertainties are approached via Monte Carlo simulations through the flux uncertainties of each of the bands involved in the SED-fitting analysis. The number of Monte Carlo realisations is set to $N_\mathrm{MC} = 100$, although this number can be increased in a general case at the cost of a longer computational time. From our experience with SED-fitting techniques and multi-filter data, this number is large enough as to perform a statistical analysis of the errors, given the number of sources included in our sample.

Throughout this work, we choose a recent version of the \citet{Bruzual2003} SSP models (hereafter CB17) in order to build our CSP set. In particular, we selected CB17 models for $18$ ages ranging from $0.001$ to $13.5$~Gyr and PARSEC evolutionary stellar tracks with metallicities $\log_{10}\left(Z/Z_\odot\right) = -1.93$, $-1.53$, $-0.93$, $-0.63$, $-0.33$, $0.00$, $0.25$, and $0.55$. In addition, we assumed a universal \citet{Chabrier2003} IMF. The CB17 spectral coverage, $\lambda\lambda~15$~\AA--$36000$~$\mu\mathrm{m}$, is sufficiently wide as to carry out the SED-fitting analysis of galaxies at the highest redshift of the parent sample ($z=1.5$). For this reason, the redshift of our CSPs ranges from $z=0$ to $z=1.5$, which are equally spaced by $\Delta z = 0.01$. Moreover, we add cosmological constraints on the age of the SSP components of our CSPs, in the sense that the age of any component cannot be much older than the age of the Universe at the CSP redshift. Extinctions were added to the CSPs as a foreground screen assuming the same dust attenuation in each of the CSP components. For this purpose, we follow the attenuation law by \citet{Calzetti2000} with values in the range $A_V=0.0$--$2.5$ and assuming a constant $R_V=3.1$. We note that we assume that there is no distinction between extinction and attenuation law. It is worth mentioning that we do not assess potential systematics from the use of a given population synthesis model \citep[see e.g.][]{DiazGarcia2019b}, hence we adopt a unique CSP set and/or star formation history.

\begin{figure}
\centering
\includegraphics[width=\hsize]{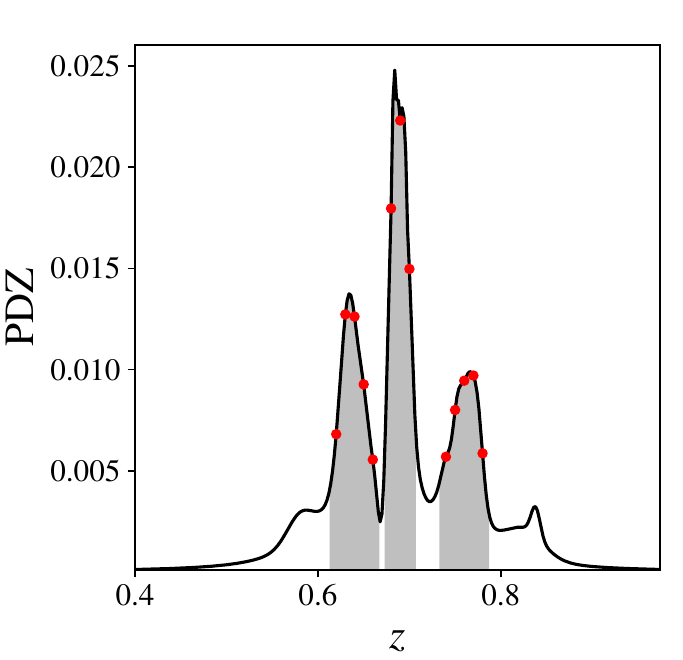}
\caption{Photo-$z$ probability distribution function (solid black line) of one noisy galaxy in miniJPAS (ID $2406$-$10239$ and $odds = 0.52$). The shaded area illustrates the photo-$z$ values defining the confidence level of $70$~\% of probability. The red dots show the photo-$z$ values used by MUFFIT to perform the SED-fitting analysis.}
\label{fig:pdz}
\end{figure}

In the updated version of MUFFIT, photo-$z$ are treated in a more proper way than in previous versions. Even though it is expected that a large fraction of miniJPAS galaxies have a photo-$z$ precision of $\sigma_\mathrm{NMAD} \sim 0.003$ \citep[see Sect.~\ref{sec:minijpas:photoz} and][]{HernanCaballero2021}, which would have little impact in the stellar mass determination \citep[details in][]{DiazGarcia2015}, some galaxies in the parent sample exhibit very complex PDZs that we must manage properly (see Fig~\ref{fig:pdz}). In brief, the SED-fitting analysis is restricted to the photo-$z$ values defining the confidence level of $70$~\% of probability in the PDZ (see shaded area in Fig~\ref{fig:pdz}). Hence, photo-$z$ are managed as an extra parameter to determine during the MUFFIT analysis of miniJPAS galaxies. Afterwards, each of the Monte Carlo realisations are weighted according the PDZ probability (see red dots in Fig~\ref{fig:pdz}) in order to determine the average stellar population properties and uncertainties \citep[see Eqs.~18--20 in][]{DiazGarcia2015} of each miniJPAS galaxy.

 Owing to the low signal-to-noise ratio of some of the galaxies in the parent sample, the $\chi^2$ minimisation is performed using fluxes instead of magnitudes. Furthermore, to ensure a proper fit of the data, we exclude from the analysis bands flagged with values different to $\texttt{FLAGS}=0$, $2$ and $\texttt{MASK\_FLAGS}=0$ and we require a minimum of ten bands with a signal-to-noise ratio larger than $1.5$. These constraints result in rejecting a $4$~\% fraction of the sources from the SED-fitting analysis, so our final sample is composed of $12\,100$ sources. 
 
\begin{figure}
\centering
\includegraphics[width=\hsize]{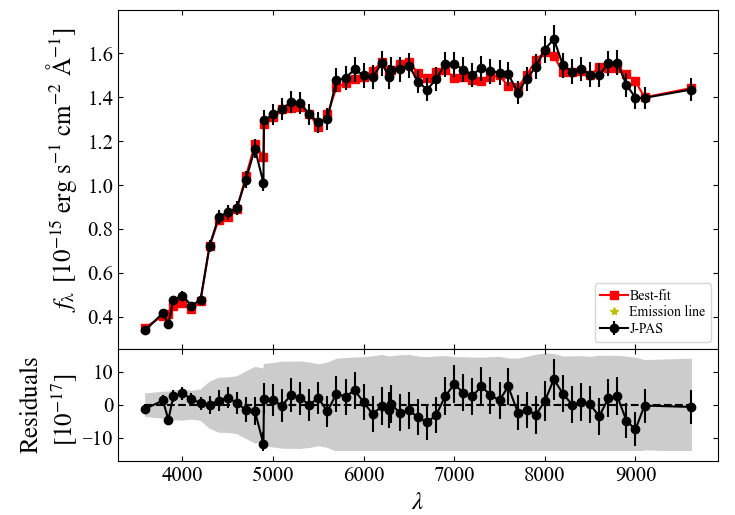}
\caption{SED-fitting analysis (\textit{top panel}) and residuals (\textit{bottom panel}) of a miniJPAS galaxy at $z=0.07$ (ID $2470$-$10239$). The black dots and vertical bars in the \textit{top panel} are the galaxy fluxes and errors as observed by miniJPAS, respectively. The red squares are the best-fitting CSP model. The shaded area shows $2.5$ times the photon-noise uncertainty of each band.}
\label{fig:sedfit}
\end{figure}

 As a result of the SED fitting analysis of the final sample, we obtain rest-frame luminosities, stellar mass, rest-frame colours corrected for extinction \citep[usually referred as intrinsic colours, see][]{DiazGarcia2019a}, and photo-$z$ (treated as another free parameter during the SED-fitting analysis with values in the PDZ $70$~\% confidence level). Indeed, other stellar population properties such as extinction, age, and metallicity (both luminosity- and mass-weighted) are constrained during the MUFFIT analysis and can be used in future works. We note that stars or point-like sources are included in the SED-fitting analysis, we statistically tackle this issue by making use of the star and galaxy classification ($P_\mathrm{G}$, see Sect.~\ref{sec:minijpas:starclass}) in subsequent sections. In Fig.~\ref{fig:sedfit}, we illustrate a SED-fitting case of one of the brightest galaxies in the miniJPAS survey. This galaxy exhibits an elliptical morphology with an evolved stellar content compatible with a mass-weighted age of $10$~Gyr and a sub-solar metallicity of $-0.2$~dex. It presents red colours that are barely produced by a high dust content ($A_V = 0.2$) with a stellar mass of $10^{11}$~$M_\odot$. There is no evidence of any presence of emission lines.

\subsection{Discretised PDF-analysis of galaxies and $B$-band luminosity}\label{sec:sedfit:pdf}

In the following, and due to the nature of the results yielded by our SED-fitting code MUFFIT, we distinguish two sets of results. Firstly, the set of best solutions obtained from the Monte Carlo realisations, which comprises $N_\mathrm{MC}$ stellar masses, luminosities or absolute magnitudes for a set of bands, photo-$z$, and rest-frame colours corrected for extinction for each of the sources in the final sample. At this point, we assume that the PDF of these stellar population properties is properly described by the Monte Carlo realisations, and therefore, this set of results is equivalent to a discretised version of the PDFs. Secondly, the average or most likely stellar population properties and their errors for each of the sources in the final sample. This second set results from the weighted mean and weighted standard deviation of the Monte Carlo realisations. These values are weighted according to the PDZ and $1/\chi^2$ values of each realisation.

As our aims comprise the comparison of the miniJPAS LMFs with results from the literature, we compute an extra set of luminosities or absolute magnitudes. In particular, we select the so-called $B$-broad band from the Classifying Objects by Medium-Band Observations survey\footnote{\url{https://www2.mpia-hd.mpg.de/COMBO/}} \citep[COMBO-17,][]{Wolf2003}. The choice of the $B$-band for this purpose is based on the fact that it has been used in several work for studying the luminosity functions of galaxies in a wide redshift range. Furthermore, given the observational wavelength range of the miniJPAS survey, rest-frame $B$-band luminosities (effective or pivot wavelength at $\lambda_\mathrm{pivot} \sim 4\,570$~\AA) are directly observed by the miniJPAS photometric system up to $z\sim1$. However, we are able to explore luminosity functions with redder bands thanks to our SED-fitting results, but depending on the redshift and band, luminosities would actually be an extrapolation of the SED-fitting analysis. For instance, luminosities for the miniJPAS $r_\mathrm{SDSS}$ band ($\lambda_\mathrm{pivot} \sim 6\,250$~\AA) are not directly observed at redshift higher than $z\sim0.5$, which set limits in the bands that can be used for determining luminosity functions up to $z\sim0.7$. The process for estimating the $B$-band luminosity of miniJPAS galaxies lies in the SED-fitting results. Briefly, for each of the sources in our final sample and thanks to the MUFFIT analysis, we have a set of SSP models reproducing the photometric SED or colours at the galaxy redshift. Therefore, from exactly the same combination of SSP models, we are able to reconstruct the rest-frame spectrum of each miniJPAS galaxy, which is subsequently convolved with the $B$-band and instrument transmission curves\footnote{\url{http://svo2.cab.inta-csic.es/theory/fps/index.php?id=LaSilla/COMBO17.B}}. We repeat this process for each of the Monte Carlo realisations, getting a discretised PDF of the $B$-band luminosity for each miniJPAS source. At this point, it is noteworthy that the SED-fitting analysis is performed without including emission lines, and therefore, all the luminosities reported in this work only contain predictions based on the stellar continuum and not about the nebular contribution.


\section{Spectral-type classification of galaxies}\label{sec:qprob}


During the last decades, many diagrams have been proposed to select or classify galaxies by star-formation activity or spectral type. These classifications mostly involve two main kinds of galaxies, which are unveiled through bimodal distributions of properties. Usually, these two types are referred to as star-forming or quiescent since the discrepancies between them are the result of high versus low levels of star formation activity, respectively \citep[see][and references therein]{DiazGarcia2019a}. For this aim, the most extended diagrams and modern methods confront sets of galaxy features that are sensitive to the spectral-type classification of galaxies such as rest-frame colours versus absolute magnitudes \citep[CMD, e.g.][]{Wyder2007,Faber2007}, empirically observed rest-frame colour--colour diagrams \citep[e.g.~the so-called $UVJ$ diagrams][]{Williams2009,Arnouts2013}, and others including stellar population properties \citep[star formation rate, stellar mass, etc.][]{Ilbert2010,Whitaker2012,Moustakas2013}. However, recent studies pointed out that samples of quiescent galaxies defined by these diagrams may include a remarkable fraction of dusty star-forming galaxies \citep[$5$--$40$~\% of galaxies depending on the stellar mass range and redshift][]{DiazGarcia2019a} and this selection can be significantly improved after correcting colours for extinction or dust reddening \citep[see also][]{Moresco2013,Schawinski2014,DiazGarcia2019a,DiazGarcia2019b,AntwiDanso2023}.

In the present work, we are particularly interested in basing our spectral-type classification of galaxies on the stellar mass--colour diagram corrected for extinction \citep[MCDE,][]{DiazGarcia2019a}. The MCDE was proposed as an efficient method in order to build non-biased samples of quiescent galaxies \citep[][]{DiazGarcia2019a} with respect to some previous methods, where the dust correction of the involved rest-frame colours is actually a key step in the process. In the following, we remark some of the advantages that motivated the use of this diagram for the spectral-type classification of galaxies. Firstly, the MCDE diagram includes multiple observables in order to perform the galaxy classification, such as stellar mass, redshift, and rest-frame colours. These aspects will account for the evolution of colours owing to galaxy ageing, as well as low-mass quiescent galaxies exhibit bluer colours on average than the more massive ones. Secondly, these diagrams do not present a strong dependency on the SSP model set employed for the SED-fitting analysis \citep[see][]{DiazGarcia2019a} and/or the star formation history assumed (e.g.~parametric or non-parametric) since it is primarily based on rest-frame colours rather than model-dependent stellar population properties such as star formation rates. In fact, the MCDE diagram demonstrated to yield results $98$~\% compatible with the stellar mass versus star formation rate diagram up to $z=1$ \citep[see Sect.~5 in][]{DiazGarcia2019a}. Thirdly, colours corrected for extinction mildly depend on the choice of the dust extinction law. As showed by \citet{DiazGarcia2019a}, the sample of quiescent galaxies differs by around $3$~\% in number when the attenuation law by \citet{Calzetti2000} is used instead of the \citet{Fitzpatrick1999} extinction law. Finally, and partly due to the above-mentioned reasons, the limiting values obtained through the MCDE diagram can be easily extended to current and future results from other SED-fitting codes in the J-PAS collaboration \citep[see][]{GonzalezDelgado2021}. It is worth mentioning that we also explored the possibility of using $UVJ$-like diagrams for the spectral classification of galaxies. However, we found out that rest-frame magnitudes of the near-infrared J-PAS bands (e.g.~$J1007$, which is close to the $Y$ band) were actually extrapolations of the SED-fitting analysis at redshift $z>0.1$ as a consequence of the J-PAS photometric system, which in turn are tightly related to the star formation history assumption leading to disparate results. For this reason, in this work we discarded $UVJ$ diagrams for the spectral classification of galaxies.

Nevertheless, we found that the sample of miniJPAS sources presented some difficulties in order to set limiting values in the MCDE diagram for the spectral classification of galaxies. This is mainly due to two factors. On the one hand, the volume or area imaged in the miniJPAS survey ($\Omega = 0.895~\mathrm{deg}^2$) limits the number of detected sources, especially in the nearby Universe. On the other hand, the depth of this survey does not allow to build samples including low-mass quiescent galaxies that are complete in stellar mass at intermediate redshift. For instance, the sample of quiescent galaxies is $95$~\% complete for stellar mass values of $\log_{10} M_\star \gtrsim 10$~dex at $z=0.3$ (see Sect.~\ref{sec:completeness}). This makes difficult a robust determination of the limiting values as a function of stellar mass and redshift. In other words, the slope of the stellar mass versus the rest-frame colour corrected for extinction show large uncertainties that complicates the galaxy classification. To overcome this problem, we make use of the sample that we also analysed with our SED-fitting code MUFFIT in \citet{DiazGarcia2019a} using galaxies from the ALHAMBRA survey, which comprises deeper observations than miniJPAS with $F814W < 24.5$ across $2.8$~deg$^2$ of the northern hemisphere \citep[further details in][]{Molino2014}. For this aim, we take all the galaxies from \citet{DiazGarcia2019a} and we reconstruct the MCDE diagram using the fittings obtained from ALHAMBRA but for a rest-frame colour compatible with the miniJPAS photometric system. In particular, we choose the rest-frame colour $(u_\mathrm{JAVA}-r_\mathrm{SDSS})$ corrected for extinction, which we refer as intrinsic colour or  $(u_\mathrm{JAVA}-r_\mathrm{SDSS})_\mathrm{int}$ in the following. For recomputing this colour for the ALHAMBRA galaxies, we follow a similar process than the detailed in \citet{DiazGarcia2019a}, which in turn is the process internally done by MUFFIT for performing the $k$-corrections of the involved photometric bands \citep[][]{DiazGarcia2015}. In brief, we take all the best-fitting models or the same combinations of SSP spectra (mixtures of two SSPs in our case) obtained from the Monte Carlo realisations that reproduces the SED of each galaxy in the sample, but at $z=0$ and null extinction ($A_V = 0$). These sets of rest-frame spectra with null extinction are subsequently convolved with the miniJPAS photometric system (i.e.~synthetic photometry) in order to determine the $(u_\mathrm{JAVA}-r_\mathrm{SDSS})_\mathrm{int}$ colours for each of the ALHAMBRA galaxies, and hence a MCDE diagram that can be easily extended to
miniJPAS galaxies. Once we reconstruct the MCDE for the miniJPAS case via ALHAMBRA galaxies, the next step is to set limiting values for the $(u_\mathrm{JAVA}-r_\mathrm{SDSS})_\mathrm{int}$ colours according to the stellar mass and redshift of this sample. For this goal, we repeat again the same process detailed in \citet{DiazGarcia2019a}: (i) after close inspection of the distribution of points, we set straight lines at different redshift bins that split the distribution of intrinsically red and blue galaxies in the MCDE (actually quiescent and star-forming galaxies, respectively); (ii) for each of the redshift bins, we compute the representative intrinsic colour of quiescent galaxies as a function of stellar mass and redshift; (iii) after removing uncertainty effects of the distributions of colours of quiescent galaxies via a maximum likelihood method \citep[MLE, further details in][]{LopezSanjuan2014,DiazGarcia2019a}, we set the limiting value of quiescent galaxies, $(u_\mathrm{JAVA}-r_\mathrm{SDSS})_\mathrm{int}^\mathrm{lim}$, as the $3\sigma$ limit of the distribution of colours fitted in the previous step.

As a result, the limiting values for the spectral-type classification of galaxies is expressed by an equation of the form
\begin{equation}\label{eq:mcde}
(u_\mathrm{JAVA}-r_\mathrm{SDSS})_\mathrm{int}^\mathrm{lim} = a \times (\log_{10}M_\star - 10) + b \times (z - 0.1) + c,
\end{equation}
where $a$, $b$, and $c$ are the constants determined through the ALHAMBRA galaxies and key for the miniJPAS galaxy classification (see values and uncertainties in Table~\ref{tab:mcde}). Nevertheless, these coefficients depend on the photometric apertures or magnitude system assumed. In this regard, the miniJPAS PSFCOR magnitudes were computed following the approach detailed in \citet[][i.e.~based on the same process carried out to perform the ALHAMBRA photometry]{Molino2014,Molino2019}, meaning that the $a$, $b$, and $c$ values obtained for the spectral-type classification of galaxies via ALHAMBRA data are `scaled' according to the rest-frame colours and stellar masses resulted from the miniJPAS \texttt{PSFCOR} photometry. This is important since we are interested in using \texttt{AUTO} magnitudes to constraint the miniJPAS LMFs, which are computed in larger apertures than \texttt{PSFCOR}. As a consequence, stellar masses obtained from SED-fitting analysis by using \texttt{AUTO} magnitudes are systematically larger than the \texttt{PSFCOR} ones and this fact can be also extended to the $(u_\mathrm{JAVA}-r_\mathrm{SDSS})_\mathrm{int}$ colours. With a view to future results of the J-PAS collaboration, we apply a second order correction to adapt the $a$, $b$, and $c$ values to the \texttt{AUTO} magnitudes. Therefore, we provide two sets of $a$, $b$, and $c$ values (see Table~\ref{tab:mcde}) that can be used to perform the classification of miniJPAS galaxies according to results based on \texttt{PSFCOR} or \texttt{AUTO} magnitudes. After detailed analysis of the results obtained from MUFFIT for \texttt{AUTO} apertures (this work) and those from \texttt{PSFCOR} magnitudes \citep[briefly introduced in][]{GonzalezDelgado2021}, we conclude that SED-fitting results of miniJPAS sources with \texttt{AUTO} photometry are in average shifted towards higher stellar mass values and bluer colours by $0.21$~dex and $0.07$ magnitudes, respectively, with respect to \texttt{PSFCOR} \citep[also in agreement with Figs.~8 and 9 in][]{GonzalezDelgado2021}. In addition, we find that these systematic differences are independent of the stellar mass range and galaxy colour, and hence of the spectral-type. Consequently, the second order correction of the relation obtained from ALHAMBRA galaxies for the classification of galaxies only comprises the intercept term, $c$, in Eq.~\ref{eq:mcde}. After accounting for the previous offsets, the intercept for galaxy classification for \texttt{AUTO} photometry is $\sim0.1$ smaller than for \texttt{PSFCOR} (see Table~\ref{tab:mcde}).

\begin{table}
\caption{Coefficients determining the limiting values of intrinsic colours for the spectral-type classification of miniJPAS galaxies (see Eq.~\ref{eq:mcde}) for \texttt{AUTO} and \texttt{PSFCOR} magnitudes.}
\label{tab:mcde}
\centering 
\begin{tabular}{c c c c}
\hline\hline
 & $a$ & $b$ & $c$ \\
\hline
\texttt{AUTO}   & $0.160 \pm 0.005$ & $-0.254 \pm 0.007$ & $1.689 \pm 0.005$ \\
\texttt{PSFCOR} & $0.160 \pm 0.005$ & $-0.254 \pm 0.007$ & $1.793 \pm 0.005$ \\
\hline
\end{tabular}
\end{table}

Once we have defined the limiting values for the galaxy classification, the next step is to assign probabilities to each of the sources in our sample. Sources with redder colours than the limiting values obtained by Eq.~\ref{eq:mcde} (whose parameters are detailed in Table~\ref{tab:mcde}) are labelled as quiescent galaxies, otherwise these are classified as star-forming galaxies. We note that in this work the spectral-type classification of galaxies is only composed of two galaxy categories and it must treated as a bimodal classification. As one of our aims is to perform a robust and statistical study for determining the miniJPAS LMFs, we make the most of the Monte Carlo realisations obtained from the SED-fitting analysis performed by the MUFFIT code. Essentially, we count the fraction of realisations presenting redder colours than the limiting values previously established for \texttt{AUTO} magnitudes, denoted as $P_\mathrm{Q}$, for each of the sources in the sample. Actually, $P_\mathrm{Q} \times P_\mathrm{G}$ can be assumed as the probability of a source in our sample of being a quiescent galaxy and this parameter accounts for parameter uncertainties, degeneracies and correlations amongst SED-fitting parameters (e.g.~extinction and rest-frame colours, stellar mass and redshift, etc.), the miniJPAS star and galaxy classification, and the redshift and stellar mass dependency of the limiting values for galaxy classification. Likewise, the probability of a source of being a star-forming galaxy can be established as $P_\mathrm{SF} \times  P_\mathrm{G}$, where $P_\mathrm{SF} = (1-P_\mathrm{Q})$. 

\begin{figure*}
\centering
\includegraphics[width=0.9\hsize]{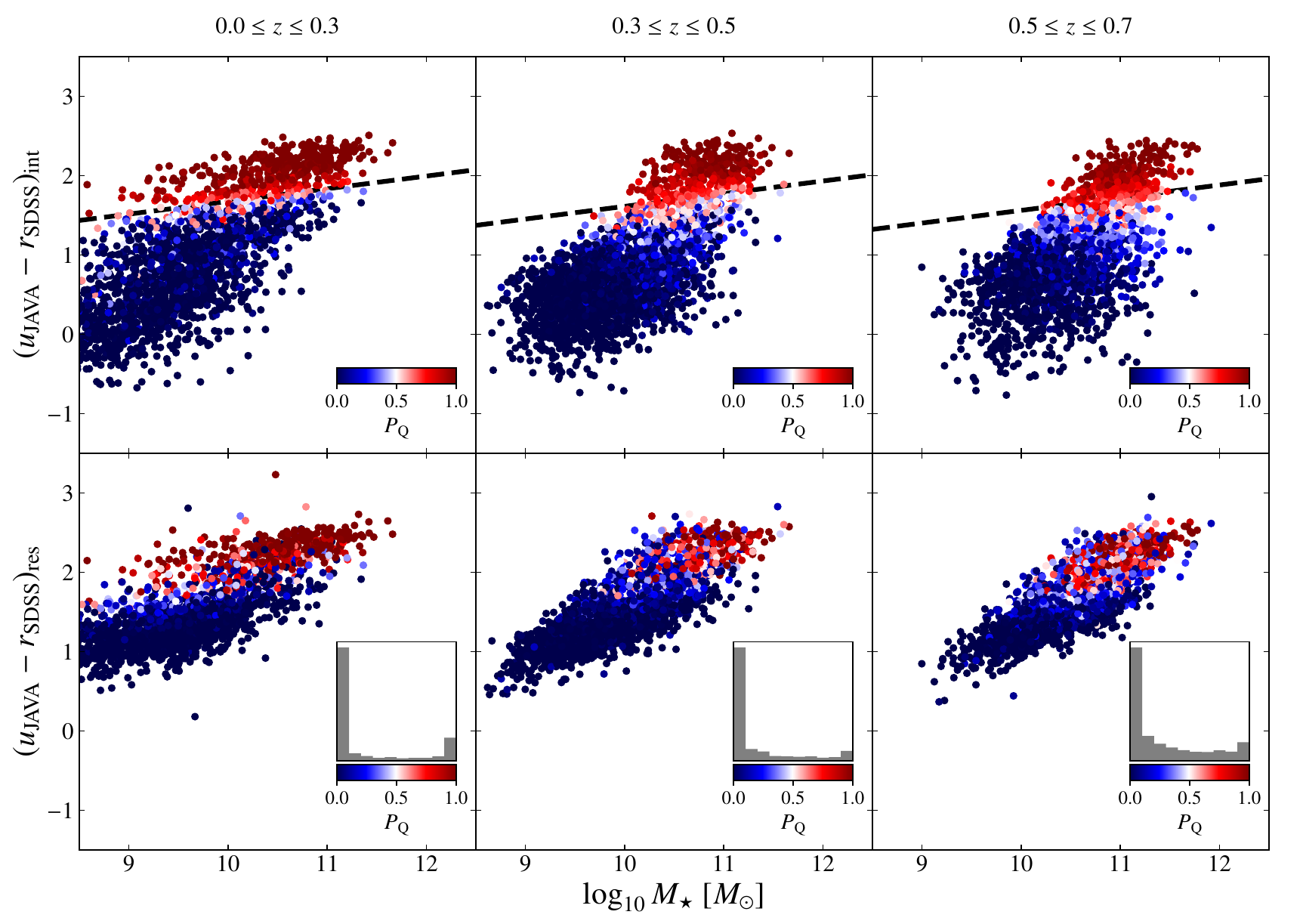}
\caption{Distribution of stellar mass versus rest-frame colour $(u_\mathrm{JAVA}-r_\mathrm{SDSS})$ obtained from miniJPAS sources with $r_\mathrm{SDSS} \le 22.5$ and $P_\mathrm{G} \ge 0.5$ at different redshift bins. \textit{Top}: diagram of the $(u_\mathrm{JAVA}-r_\mathrm{SDSS})$ colour after correcting for dust effects or intrinsic colour. \textit{Bottom}: diagrams without correcting the rest-frame $(u_\mathrm{JAVA}-r_\mathrm{SDSS})$ colour for extinction and normalised histograms of $P_\mathrm{Q}$ values for each redshift bin (see insets). The dashed black line illustrates the limiting relation for selecting quiescent galaxies and \texttt{AUTO} photometry at the central redshift of each bin (i.e.~$z=0.15$, $0.4$, and $0.6$; see Eq.~\ref{eq:mcde}). Dots are colour coded according to the quiescent and star-forming classification (redder and bluer colours, respectively) or quiescent probability ($P_\mathrm{Q}$).}
\label{fig:mcde}
\end{figure*}

As expected, the distribution of $P_\mathrm{Q}$ values strongly correlates with the positions that sources occupy in the MCDE (see Fig.~\ref{fig:mcde}). Galaxies with the intrinsic reddest colours (upper parts of the MCDE) present the highest $P_\mathrm{Q}$ values (quiescent galaxies, see red dots in Fig.~\ref{fig:mcde}), whereas the lower $P_\mathrm{Q}$ values lie on the lower parts of the MCDE (star-forming galaxies, see blue dots in Fig.~\ref{fig:mcde}). Sources with an uncertain spectral-type classification (i.e.~$P_\mathrm{Q} \sim 0.5$) commonly present intermediate rest-frame colours in the MCDE, which overlaps with the colour range of the so-called `green valley' galaxies. We remark that for this work we only classify galaxies as either quiescent or star-forming, that is, extra or alternative classifications such as green valley galaxies are not explored in the present paper. We also find that the shape of the distribution of $P_\mathrm{Q}$ values slightly depends on redshift. At $z \le 0.3$, the spectral-type classification of galaxies is highly bimodal where there is a $83$~\% fraction of galaxies showing values of $P_\mathrm{Q} \le 0.1$ or $P_\mathrm{Q} \ge 0.9$. However, only half of the sample at $0.5 \le z \le 0.7$ show values of $P_\mathrm{Q} \le 0.1$ or $P_\mathrm{Q} \ge 0.9$ (see insets in Fig.~\ref{fig:mcde}). Similarly, the fraction of galaxies with an uncertain classification, $0.4 \le P_\mathrm{Q} \le 0.6$, also depends on redshift. Thus, there is a $3$~\% fraction of galaxies with $0.4 \le P_\mathrm{Q} \le 0.6$ at $z \le 0.3$, which increases up to $12$~\% for the galaxies at the highest redshift of our sample. In fact, this is not surprising since this is a direct consequence of the higher photon-noise uncertainties in the observation of galaxies at higher redshifts, which complicates the determination of the stellar population properties involved in the MCDE. Finally, we find that there also are correlations between the $P_\mathrm{Q}$ values and their positions across the classical stellar mass versus colour diagram, that is, without correcting for dust reddening (see bottom panels in Fig.~\ref{fig:mcde}). In fact, the higher (lower) $P_\mathrm{Q}$ also lie on the upper (lower) parts of this diagram, but as also found in \citet{DiazGarcia2019a}, there is a non-negligible fraction of star-forming galaxies ($P_\mathrm{Q} < 0.5$) that present red colours owing to a large dust content or extinction. Moreover, galaxies with $0.4 \le P_\mathrm{Q} \le 0.6$ are scattered in the colour range of quiescent galaxies in this diagram.

\section{Luminosity and stellar mass completeness}\label{sec:completeness}


Completeness is one of the key ingredients that are needed for determining LMFs in any survey since it sets the limitations of our observations and sample definition. In brief, the completeness points out how complete the sample is or the part of the whole galaxy population that we are missing or not including as part of our sample owing to different reasons such as target selection, survey depth, data quality constraints, etc. Owing to the nature of multi-filter surveys like J-PAS, there is not any pre-selection of galaxy candidates because all the sources in each of the survey pointings are imaged and included in the photometric catalogues. Nevertheless, we distinguish three types of completeness limits that can affect the determination of the miniJPAS LMFs. Firstly, the image depth sets a first unavoidable limitation for defining our galaxy sample. In miniJPAS, it was established that the sample of extended sources is complete up to $r_\mathrm{SDSS} = 22.7$, that is, all the galaxies in the field that are equal or brighter than this magnitude limit are detected in the miniJPAS images and included in the photometric catalogue. As our sample is brighter than this limit, our galaxy sample is not affected by this selection effect. On the other hand, our sample of miniJPAS sources is defined according to a magnitude limit that in turns triggers limitations in stellar mass and $B$-band luminosity completeness. This would imply that, at a given redshift and close to the magnitude limit of the sample, part of the galaxies below a certain stellar mass and luminosity limit are not going to be observed or included in the photometric catalogue.

It is well-known that stellar mass and luminosity completeness of flux-limited samples are properly described by Fermi-Dirac distribution functions \citep[see e.g.][]{Sandage1979,Ilbert2010,DiazGarcia2019a}. For our aims, we therefore determine two Fermi-Dirac distribution functions for the stellar mass and $B$-band luminosity completeness ($\mathcal{C}$, hereafter), where each of them is redshift dependent and formally parametrised by two parameters as following
\begin{equation}\label{eq:fermidirac_mass}
\mathcal{C}(z,M_\star) = \frac{1}{\exp\left[\left( M_{\mathrm{F},M_\star}(z) - \log_{10}M_\star \right) / \Delta_{\mathrm{F},M_\star}(z)\right] + 1} \ ,
\end{equation}
\begin{equation}\label{eq:fermidirac_mabs}
\mathcal{C}(z,M_B) = \frac{1}{\exp\left[\left( M_B - M_{\mathrm{F},M_B}(z) \right) / \Delta_{\mathrm{F},M_B}(z) \right] + 1} \ ,
\end{equation}
where $0 \le \mathcal{C} \le 1$ (i.e.~the completeness of the sample at certain stellar mass or luminosity at a given redshift), $M_\mathrm{F}$ is the stellar mass in dex units or absolute magnitude in the $B$-band for which the completeness reaches $50$~\% ($\mathcal{C}=0.5$), and $\Delta_\mathrm{F}$ the decrease rate on the fraction of galaxies. It is also of note that both $M_\mathrm{F}$ and $\Delta_\mathrm{F}$ are functions that depend on redshift in a general case, meaning that we do not assume a constant value for $\Delta_\mathrm{F}$. From Eqs.~\ref{eq:fermidirac_mass} and \ref{eq:fermidirac_mabs}, it is easy to obtain the limiting values of stellar mass and absolute magnitude for a given completeness level ($\log_{10}M_\star^\mathcal{C}$ and $M_B^\mathcal{C}$, respectively) as  
\begin{equation}\label{eq:completeness_mstar}
\log_{10}M_\star^\mathcal{C}(z) = M_{\mathrm{F},M_\star}(z) + \Delta_{\mathrm{F},M_\star}(z) \times \ln \left( \frac{\mathcal{C}}{1-\mathcal{C}} \right)\ ,
\end{equation}
and
\begin{equation}\label{eq:completeness_mabs}
M_B^\mathcal{C}(z) = M_{\mathrm{F},M_B}(z) - \Delta_{\mathrm{F},M_B}(z) \times \ln \left( \frac{\mathcal{C}}{1-\mathcal{C}} \right)\ .
\end{equation}
In addition, we can conclude from Eqs.~\ref{eq:completeness_mstar} and \ref{eq:completeness_mabs} that $M_\mathrm{F}$ and $\Delta_F$ can be determined or constrained as long as we knew two stellar mass and $B$-band luminosity values at two certain completeness levels.

For the determination of the stellar mass completeness of the miniJPAS flux-limited sample, our starting point is the method proposed by \citet{Pozzetti2010} for this aim. This method relies on the assumption that, at a given redshift, the distribution of the mass-to-light ratio of the fainter galaxies in the sample ($M_\star/L_r$) should be similar to the one at the magnitude limit of the sample ($M_\mathrm{\star}^\mathrm{lim}/L_r^\mathrm{lim}$). Keeping this in mind, the distribution of stellar mass at a fixed magnitude limit can be reconstructed from the above assumption as 
\begin{equation}\label{eq:lim_mstar}
\log_{10}M_\star^\mathrm{lim} = \log_{10}M_\star - 0.4 \times (r_\mathrm{SDSS}^\mathrm{lim} - r_\mathrm{SDSS})\ ,
\end{equation}
where $r_\mathrm{SDSS}$ is the distribution of observed magnitudes for the galaxies with the fainter magnitudes at certain redshift, $\log_{10}M_\star$ are the stellar mass values for each of these faint galaxies, and $r_\mathrm{SDSS}^\mathrm{lim}$ the limiting magnitude imposed for the definition of the sample. If our assumption is correct, the distribution of $\log_{10}M_\star^\mathrm{lim}$ values illustrates the range and frequency of stellar mass values of galaxies close to the magnitude limit. Consequently, all the galaxies with a stellar mass higher than the $95^\mathrm{th}$ percentile of the $\log_{10}M_\star^\mathrm{lim}$ distribution are roughly included in the sample (i.e.~$\mathcal{C}=0.95$), whereas galaxies with stellar mass under the $5^\mathrm{th}$ percentile of the distribution are not going to be included in the sample (i.e.~$\mathcal{C}=0.05$).

Regarding the $B$-band luminosity completeness, we follow a similar approach, but we assume that the distribution of rest-frame colours of the fainter galaxies in the flux-limited sample at certain redshift is similar to those at the magnitude limit instead. As for the stellar mass completeness case, this results in a distribution of $B$-band absolute magnitudes at the magnitude limit of the sample formally expressed as 
\begin{equation}\label{eq:lim_mabs}
M_B^\mathrm{lim} = M_B + (r_\mathrm{SDSS}^\mathrm{lim} - r_\mathrm{SDSS})\ .
\end{equation}
For this case, the sample would be complete in terms of the $B$-band absolute magnitude for galaxies brighter than the $5^\mathrm{th}$ percentile of the $M_B^\mathrm{lim}$ distribution and incomplete for galaxies fainter than the $95^\mathrm{th}$ percentile, which we assume that roughly correspond to $\mathcal{C}=0.95$ and $0.05$, respectively. 

For the determination of the $\mathcal{C}=0.05$ and $0.95$ limits of the $\log_{10}M_\star^\mathrm{lim}$ and $M_B^\mathrm{lim}$ distributions, we make use of the average results and uncertainties obtained by MUFFIT. We split the sample in redshift bins of width $\Delta z = 0.1$ from $z=0.05$ to $0.75$ for all the miniJPAS sources with $P_\mathrm{G} \ge 0.5$. Moreover, we split this subsample in quiescent and star-forming galaxies, meaning that sources in the subsample with $P_\mathrm{Q} \ge 0.5$ ($P_\mathrm{Q} < 0.5$) are classified as quiescent (star-forming) for setting constraints in the galaxy sample completeness. For each of the redshift bins in our subsamples, we select galaxies with observed magnitudes in the range $r_\mathrm{SDSS}^\mathrm{lim} - 0.5 \le r_\mathrm{SDSS} \le r_\mathrm{SDSS}^\mathrm{lim}$ for building our distributions of limiting parameters (see Eqs.~\ref{eq:lim_mstar} and \ref{eq:lim_mabs}). However, for the cases for which the number of galaxies is lower than $30$, we extend the bright magnitude limit until this minimum number of galaxies is fulfilled. This particularly happens for quiescent galaxies in low redshift bins since the observed volume and number density of this kind of galaxies is lower. Finally, we account for the fact that the $\log_{10}M_\star^\mathrm{lim}$ and $M_B^\mathrm{lim}$ distributions are built from samples that were determined via SED-fitting techniques and these are affected by uncertainties. If this issue were not addressed, the distribution of limiting magnitudes would be broader than it actually is. Consequently, the upper completeness limit would be overestimated, which would greatly affect and restrict the number of sources during the definition of complete samples, as well as the lower limit would be underestimated. In fact, as detailed in Sect.~\ref{sec:results:errors}, the errors in stellar mass and luminosity are large enough at the fainter magnitudes of the sample as to significantly alter the $\log_{10}M_\star^\mathrm{lim}$ and $M_B^\mathrm{lim}$ distributions. To overcome this issue, we apply the MLE method previously mentioned for deconvolving uncertainty effects in the $\log_{10}M_\star^\mathrm{lim}$ and $M_B^\mathrm{lim}$ distributions. Interestingly, the $\log_{10}M_\star^\mathrm{lim}$ and $M_B^\mathrm{lim}$ distributions of values are properly fitted by Gaussian and/or log-normal distributions, and consequently, the MLE method described in \citet{DiazGarcia2019a,DiazGarcia2019b} can be easily and analytically applied here. Then, the $\mathcal{C}=0.05$ and $0.95$ limits are obtained from the analytic functions obtained through the MLE (i.e.~the intrinsic distributions of  $\log_{10}M_\star^\mathrm{lim}$ and $M_B^\mathrm{lim}$ after deconvolving the errors in stellar mass and $B$-band absolute magnitude provided by MUFFIT), which are used to subsequently determine $M_\mathrm{F}$ and $\Delta_\mathrm{F}$ by Eqs.~\ref{eq:completeness_mstar} and \ref{eq:completeness_mabs}. In this regard, $M_\mathrm{F}$ matches the intermediate value between those for $\mathcal{C}=0.05$ and $0.95$ (see Eqs.~\ref{eq:completeness_mstar} and \ref{eq:completeness_mabs}). In fact, as the $\log_{10}M_\star^\mathrm{lim}$ and $M_B^\mathrm{lim}$ distributions are properly fitted by Gaussian-like functions, it turns out to be a good approach that $M_\mathrm{F}$ corresponds to the medians since resulted from the MLE method. We note that as our plan is to study the stellar mass and $B$-band luminosity completeness for a set of magnitude limit values, we repeat the method detailed above for each of these sample constraints, more precisely, we perform this method for $r_\mathrm{SDSS}^\mathrm{lim} = 21.5$, $21.75$, $22$, $22.25$, and $22.5$.

\begin{figure*}
\centering
\includegraphics[width=0.8\hsize]{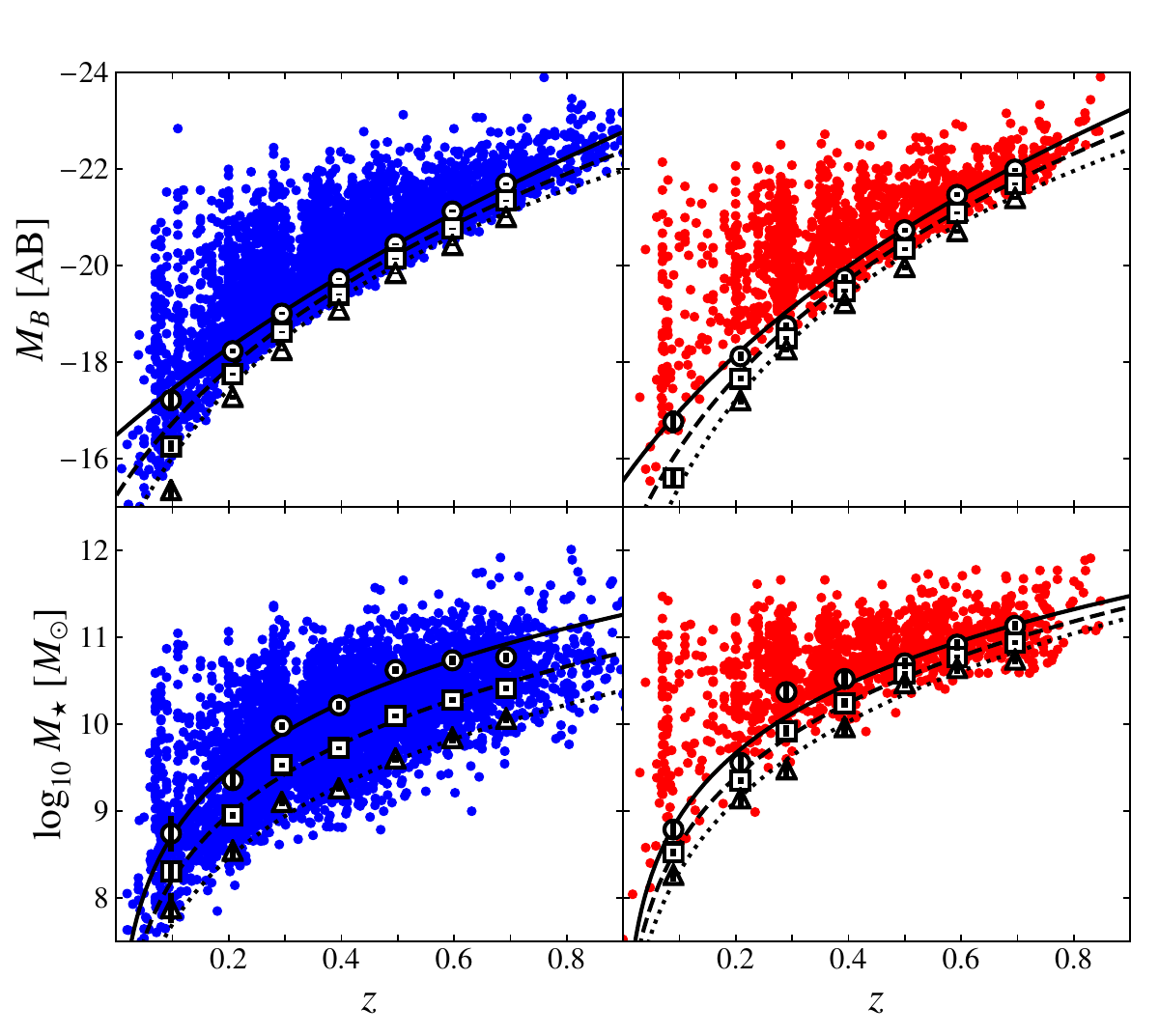}
\caption{$B$-band luminosity and stellar mass completeness as a function of redshift (\textit{top and bottom panels}, respectively) for star-forming and quiescent miniJPAS galaxies (\textit{left and right panels}, respectively) from the flux-limited sample at $r_\mathrm{SDSS}\le 22$ magnitudes (i.e.~$r_\mathrm{SDSS}^\mathrm{lim} = 22$). The dotted, dashed, and solid black lines illustrate the fit of the $5$, $50$, and $95$~\% completeness levels (i.e.~$\mathcal{C} = 0.05$, $0.5$, and $0.95$) obtained from the completeness methodology detailed in Sect.~\ref{sec:completeness} (triangle, square, and circle shape markers), respectively.}
\label{fig:completeness}
\end{figure*}

With the ultimate goal of obtaining an analytic expression of the $M_\mathrm{F}$ and $\Delta_\mathrm{F}$ parameters as a function of redshift, we fit the $\log_{10}M^{\mathcal{C}}_\star$ and $M_B^\mathrm{C}$ values obtained in the previous step to a power-law function of the form $\mu\times(z+\gamma)^\nu$ and $\delta \times (z+\zeta)^\epsilon$, respectively. This process is repeated for the completeness values of $\mathcal{C} = 0.05$, $0.5$, and $0.95$. In fact, these directly determine $M_\mathrm{F}(z)$ since $\log_{10}M^{\mathcal{C}=0.5}_\star(z) = M_{\mathrm{F}, M_\star}(z)$ and $M^{\mathcal{C}=0.5}_B(z) = M_{\mathrm{F}, M_B}(z)$. On the other hand, $\Delta_{\mathrm{F},M_\star}(z)$ and $\Delta_{\mathrm{F},M_B}(z)$ can be obtained by Eqs.~\ref{eq:completeness_mstar} and \ref{eq:completeness_mabs} and the fittings just mentioned. The $\mu$, $\nu$, $\gamma$, $\delta$, $\epsilon$, and $\zeta$ values for the completeness levels $\mathcal{C}=0.05$, $0.5$, and $0.95$ and a flux-limited sample of $r_\mathrm{SDSS} \le 22$ (i.e.~$r_\mathrm{SDSS}^\mathrm{lim}=22$) are shown in Table~\ref{tab:completeness}. In Appendix~\ref{sec:appendix:completeness}, we extend our results for alternative magnitude-limited samples (see Tables~\ref{tab:completeness_r215}--\ref{tab:completeness_r225}), which can be used in future J-PAS studies. Overall, all the values that we obtain for each of the redshift bins are properly fitted by power-law functions independently of the spectral-type of galaxies (see Fig.~\ref{fig:completeness}) and the sample is complete ($\mathcal{C}=0.95$) for higher luminosity and stellar mass limits at increasing redshift. As previously observed in similar works, samples of quiescent galaxies are complete at higher luminosity and stellar mass limits than star-forming galaxies at same redshift, which is reflected in larger $M_\mathrm{F}$ values. On the other hand, we find that $\Delta_{\mathrm{F},M_\star}$ is smaller for quiescent galaxies than for star-forming galaxies as a consequence of a larger diversity of mass-to-light ratios in star-forming galaxies. In fact, $\Delta_{\mathrm{F},M_\star}$ is roughly constant for our star-forming galaxy sample with $r_\mathrm{SDSS} \le 22$ with a value of $0.16$, whereas for quiescent galaxies this value is always lower and mildly decreases from $0.11$ at $z \sim 0$ to $0.06$ at $z \sim 0.7$. Regarding $\Delta_{\mathrm{F},M_B}$, our results point out that this value present larger changes with redshift than for the stellar mass case, although this value is approximately constant at $z \gtrsim 0.3$ ($\Delta_{\mathrm{F},M_B}\sim 0.10$) for both star-forming and quiescent galaxies. At lower redshift,  $\Delta_{\mathrm{F},M_B}$ increases up to $\sim 0.3$ at $z\sim0.1$, independently of the galaxy spectral-type. Consequently, we find that amongst the four parameter defining the completeness limits of the sample, $\Delta_{\mathrm{F},M_B}$ and $\Delta_{\mathrm{F},M_\star}$ are almost independent parameters on the galaxy spectral-type and redshift, respectively. Finally, it is worth mentioning that given the volume observed in miniJPAS, the completeness limits constrained in this work are valid at $0.1 \lesssim z \lesssim 0.7$, and out of this redshift range these should be managed as an extrapolation of the fittings. 

Even though this is not an ideal method for determining the completeness curves at the magnitude limit, we can set rough estimations of them that can be used to set limits for the definition of galaxy samples in the miniJPAS survey. An ideal methodology for the determination of the completeness limits and analytical functions would be a direct comparison with respect to the LMFs functions of a similar and deeper survey in an overlapping area \citep[see e.g.][]{DiazGarcia2019a}. Other methods involve the injection of synthetic sources or mock catalogues built from simulations in order to check the effect of the introduction of flux limits in the sample. However, the use of mock catalogues involve the inclusion of population synthesis models (as well as diverse star formation histories) and a large set of assumptions or distributions (e.g.~involving number densities and morphology) typically based on deeper surveys, which may not be trivially accounted for. Nevertheless, we find this method turns out to be a good approach for the distribution of stellar mass and luminosity of galaxies at the magnitude limit of the sample, which somehow reflects the kind of bias introduced in the selection of the sample in terms of stellar mass and absolute magnitude \citep[as also concluded by][]{Meneux2009,Pozzetti2010,DiazGarcia2019a}. With the incoming of J-PAS in a close future, which will involve large overlapping areas with deeper surveys such as ALHAMBRA, we expect to confirm how much precise this method is for the determination of the completeness limits explored in this work.

\begin{table*}
\caption{Coefficients determining the stellar mass and $B$-band luminosity completeness (\textit{top} and \textit{bottom panels}, respectively) of the star-forming and quiescent galaxies from miniJPAS for our flux-limited sample at $r_\mathrm{SDSS} \le 22$.}
\label{tab:completeness}
\centering
\begin{tabular}{c c c c c c c}
\hline\hline
& \multicolumn{3}{c}{Star-forming} & \multicolumn{3}{c}{Quiescent}\\
\multirow{2}{*}{$\mathcal{C}$} & \multirow{2}{*}{$\mu$} & \multirow{2}{*}{$\nu$} & \multirow{2}{*}{$\gamma$} & \multirow{2}{*}{$\mu$} & \multirow{2}{*}{$\nu$} & \multirow{2}{*}{$\gamma$}\\
 & & & & & & \\
\hline
$0.05$ & $10.490 \pm 0.090$ & $0.134 \pm 0.032$ & $0.025 \pm 0.086$ & $11.368 \pm 0.087$ & $0.143 \pm 0.028$ & $0.016 \pm 0.049$\\
$0.50$ & $10.969 \pm 0.110$ & $0.126 \pm 0.033$ & $0.000 \pm 0.080$ & $11.499 \pm 0.052$ & $0.127 \pm 0.016$ & $0.003 \pm 0.029$\\
$0.95$ & $11.398 \pm 0.175$ & $0.116 \pm 0.051$ & $0.000 \pm 0.146$ & $11.614 \pm 0.108$ & $0.114 \pm 0.028$ & $0.000 \pm 0.056$\\
\hline
\multirow{2}{*}{$\mathcal{C}$} & \multirow{2}{*}{$\delta$} & \multirow{2}{*}{$\epsilon$} & \multirow{2}{*}{$\zeta$} & \multirow{2}{*}{$\delta$} & \multirow{2}{*}{$\epsilon$} & \multirow{2}{*}{$\zeta$}\\
 & & & & & & \\
\hline
$0.05$ & $-22.163 \pm 0.060$ & $0.171 \pm 0.010$ & $0.043 \pm 0.025$ & $-22.793 \pm 0.079$ & $0.176 \pm 0.014$ & $0.004 \pm 0.026$\\
$0.50$ & $-21.982 \pm 0.187$ & $0.216 \pm 0.016$ & $0.183 \pm 0.045$ & $-22.844 \pm 0.121$ & $0.210 \pm 0.015$ & $0.094 \pm 0.034$\\
$0.95$ & $-20.249 \pm 0.742$ & $0.326 \pm 0.037$ & $0.532 \pm 0.111$ & $-22.396 \pm 0.885$ & $0.262 \pm 0.064$ & $0.248 \pm 0.160$\\
\hline
\end{tabular}
\end{table*}

\section{Methodology for determining the LMF parameters}\label{sec:zsty}


In this research, we depart from the assumption in which the population of both star-forming and quiescent galaxies are properly described by a single Schechter function. This decision is mainly motivated by the fact that our sample at the low-mass galaxy regime is highly reduced in number and only present at the nearby Universe. Specifically, there are few quiescent galaxies of $\log_{10}M_\star<9.5$ at $z<0.2$, which makes largely difficult the proper determination of a second Schechter component at the low-mass/faint regime. This inconvenient is a direct consequence of the imaging depth and the small area imaged in miniJPAS, although in a close future with the arrival of J-PAS and its huge observing area, we expect to explore the LMFs via double Schechter functions. Secondly, in this research the LMFs of the whole galaxy population is the sum of the two galaxy spectral-types, meaning that the LMFs of all the galaxies in miniJPAS are actually double Schechter functions. Having said this, the LMFs of a single Schechter function at a given redshift are formally expressed as 
\begin{equation}\label{eq:schechter_lum}
\Phi (L_B)~\mathrm{d}L_B = \Phi_{L_B}^{\star}~\left( \frac{L_B}{\mathcal{L}_B} \right)^\beta~\exp{\left( -\frac{L_B}{\mathcal{L}_B}\right)}~\frac{\mathrm{d}L_B}{L_B}\ ,
\end{equation}
\begin{equation}\label{eq:schechter_mass}
\Phi (M_\star)~\mathrm{d}M_\star = \Phi_{M_\star}^{\star}~\left( \frac{M_\star}{\mathcal{M}_\star} \right)^\alpha~\exp{\left( -\frac{M_\star}{\mathcal{M}_\star}\right)}~\frac{\mathrm{d}M_\star}{M_\star}\ ,
\end{equation}
where $\mathcal{L}_B$ and $\mathcal{M}_\star$ are the so-called characteristic $B$-band luminosity and characteristic stellar mass, respectively, $(\beta+1)$ and $(\alpha+1)$ are referred as the slopes of the faint-end of the LMFs (see also Eqs.~\ref{eq:schechter_mabs} and \ref{eq:schechter_logmass}), and $\Phi_{L_B}^\star$ and $\Phi_{M_\star}^\star$ their normalizations in number density units. Alternatively, these functions can be also found in their logarithmic form as
\begin{equation}\label{eq:schechter_mabs}
\begin{split}
\Phi (M_B)~\mathrm{d}M_B =\ & 0.4~\ln{10}~\Phi_{L_B}^\star~10^{0.4(\mathcal{M}_B-M_B)~(\beta+1)} \cdot \\
& \cdot \exp\left[ -10^{0.4(\mathcal{M}_B-M_B)}\right]~\mathrm{d}M_B\ ,
\end{split}
\end{equation}
\begin{equation}\label{eq:schechter_logmass}
\begin{split}
\Phi (\bar{M}_\star)~\mathrm{d}\bar{M}_\star =\ & \ln{10}~\Phi_{M_\star}^\star~10^{(\bar{M}_\star-\bar{\mathcal{M}}_\star)~(\alpha+1)} \cdot \\
& \cdot \exp\left[ -10^{(\bar{M}_\star-\bar{\mathcal{M}}_\star)}\right]~\mathrm{d}\bar{M}_\star\ ,
\end{split}
\end{equation}
where $M_B$ is the $B$-band absolute magnitude, $\mathcal{M}_B$ is the absolute magnitude of the characteristic $B$-band luminosity, $\bar{M}_\star$ the stellar mass logarithm (i.e.~$\bar{M}_\star=\log_{10}M_\star$), and $\bar{\mathcal{M}}_\star$ the logarithm of the characteristic stellar mass.

Accordingly, for studying the evolution of the miniJPAS LMFs it is needed to determine the two sets of parameters including the characteristic parameters, faint-end slopes, and normalisations at different redshift, meaning that \{$\mathcal{M_\star}(z)$, $\alpha(z)$, $\Phi_{M_\star}^\star(z)$\} and \{$\mathcal{L}_B(z)$, $\beta(z)$, $\Phi_{L_B}^\star(z)$\}. As we detail below, we also constrain these parameters according to the galaxy spectral-type.


\subsection{Characteristic parameters and slopes of the faint/low-mass ends of LMFs}\label{sec:zsty:par}

For a robust and statistical determination of the miniJPAS LMFs, as well as for making the most of the miniJPAS sample and statistical results from our SED-fitting codes \citep[][]{HernanCaballero2021,GonzalezDelgado2021}, we develop and base our techniques on a maximum likelihood method in which we include the probability distribution functions of the involved parameters (see Sect.~\ref{sec:sedfit:pdf}). In a sense, our methodology follows the guidelines and includes some aspects of the works by \citet{Sandage1979,Efstathiou1988,Zucca1994,Ilbert2005,LopezSanjuan2017} and is adapted for the inclusion of all the peculiarities of the miniJPAS survey, the results from MUFFIT and similar codes in the J-PAS collaboration, the statistical spectral-type classification of galaxies, and the completeness of our flux-limited sample (details in Sects.~\ref{sec:sedfit}--\ref{sec:completeness}). Based on the idea that the probability of observing a galaxy of stellar mass $M_\star'$ at a given redshift $z'$ from a sample complete in stellar mass is $\Phi(M_\star',z')/\int_{M_\star^\mathrm{f}(z')}^{M_\star^\mathrm{b}(z')}{\Phi(M_\star,z')\ \mathrm{d}M_\star}$, where $\Phi(M_\star,z')$ is the stellar mass function (as in Eq.~\ref{eq:schechter_mass}) and $M_\star^\mathrm{f}(z')$ and $M_\star^\mathrm{b}(z')$ the low and high stellar mass limits of the sample at the source redshift, respectively; the likelihood ($\mathcal{L}$) of our method can be defined as the joint probability of observing our sample that is expressed as:
\begin{equation}\label{eq:likelihood}
\mathcal{L} =  \prod_{i=1}^{N_\mathrm{S}} \left( \frac{\Phi(M_{\star,i},z_i)} {\int_{M_{\star}^\mathrm{f}(z_i)}^{M_{\star}^\mathrm{b}(z_i)}{\Phi(M_\star,z_i)\ \mathrm{d}M_\star}} \right)^{\omega_i}\ ,
\end{equation}
where $N_\mathrm{S}$ is the number of sources in the sample and $\omega_i$ is a weight accounting for different factors such as the probability of a source to be a star, the completeness of the sample at the $i^\mathrm{th}$ source stellar mass and redshift, priors, etc. (further details in Sect.~\ref{sec:zsty:weight}). The maximisation of this likelihood makes possible the determination of the \{$\mathcal{M_\star}(z)$, $\alpha(z)$, $\Phi_{M_\star}^\star(z)$\} parameters that characterise the most likely LMFs describing the observations of our sample. As in \citet{LopezSanjuan2017}, we assume redshift dependent functions for these parameters of the form: $\alpha(z) = \alpha_1 + \alpha_2 \times z$, $\beta(z) = \beta_1 + \beta_2 \times z$, $\log_{10} \mathcal{M}_\star (z) = \bar{\mathcal{M}}_{\star,1} + \bar{\mathcal{M}}_{\star,2}\times z$, $\mathcal{M}_B = \mathcal{M}_{B,1} + \mathcal{M}_{B,2}\times z$, $\log_{10}\Phi_{L_B}^{\star} = \bar{\Phi}_{L_B,1}^\star + \bar{\Phi}_{L_B,2}^\star \times  z$, and $\log_{10}\Phi_{M_\star}^{\star} = \bar{\Phi}_{M_\star,1}^\star + \bar{\Phi}_{M_\star,2}^\star \times  z$. Actually, we can directly include these analytical terms in the process of maximisation. Hence, we only need to constrain the sets of \{$\alpha_1$, $\alpha_2$, $\bar{\mathcal{M}}_{\star,1}$, $\bar{\mathcal{M}}_{\star,2}$, $\bar{\Phi}_{M_\star,1}^\star$, $\bar{\Phi}_{M_\star,2}^\star$\} and \{$\beta_1$, $\beta_2$, $\mathcal{M}_{B,1}$, $\mathcal{M}_{B,2}$, $\bar{\Phi}_{L_B,1}^\star$, $\bar{\Phi}_{L_B,2}^\star$\} values for both star-forming and quiescent galaxies in a general case. In this regard, we note that this method requires no photo-$z$ binning of the data other than the upper and lower redshift limits of the sample (see Sect.~\ref{sec:zsty:weight}). As part of our goals, we also include the PDFs of the parameters involved. This is done by the Monte Carlo realisations obtained during the SED-fitting analysis performed by MUFFIT and the weights included in Eq.~\ref{eq:likelihood} (further details in Sect.~\ref{sec:zsty:weight}). In practise, this is equivalent to have a sample of $N_\mathrm{S} \times N_\mathrm{MC}$ sources, where the weight in number of each Monte Carlo realisation is $1/N_\mathrm{MC}$. After the inclusion of these redshift dependent functions of the parameters and the Monte Carlo realisations, the likelihood to be maximised for determining the stellar mass functions is expressed as:
\begin{align}\label{eq:likelihood_z}
\ln{\mathcal{L}} = & 
\sum_{i=1}^{N_\mathrm{S} \times N_\mathrm{MC}}
\alpha_1 \omega_i \ln M_{\star,i} + \alpha_2 \omega_i z_i  \ln M_{\star,i}
- \frac{\omega_i M_{\star,i}}{10^{\bar{\mathcal{M}}_{\star,1} + \bar{\mathcal{M}}_{\star,2} z_i}} \nonumber\\
& - \frac{1}{\log_{10}e} [(\alpha_1 \bar{\mathcal{M}}_{\star,1} + \bar{\mathcal{M}}_{\star,1})\omega_i \nonumber\\
& + (\alpha_1 \bar{\mathcal{M}}_{\star,2} + \alpha_2 \bar{\mathcal{M}}_{\star,1} + \bar{\mathcal{M}}_{\star,2})\omega_i z_i + \alpha_2 \bar{\mathcal{M}}_{\star,2} \omega_i z_i^2] \\
& - \omega_i \ln \left[\Gamma\left(\alpha_1 + \alpha_2 z_i + 1, \frac{M_{\star}^\mathrm{f}(z_i)}{10^{\bar{\mathcal{M}}_{\star,1} + \bar{\mathcal{M}}_{\star,2} z_i}} \right) \right. \nonumber\\
& \left. - \Gamma\left(\alpha_1 + \alpha_2 z_i + 1, \frac{M_{\star}^\mathrm{b}(z_i)}{10^{\bar{\mathcal{M}}_{\star,1} + \bar{\mathcal{M}}_{\star,2} z_i}}\right)\right] \nonumber\ ,
\end{align}
where $\Gamma$ is the incomplete Euler gamma function (i.e.~$\Gamma(s,x) = \int_{x}^{\infty} t^{s-1}~e^{-t}~\mathrm{d}t$), and $M_{\star}^\mathrm{f}(z_i)$ and $M_{\star}^\mathrm{b}(z_i)$ are the lower and upper stellar mass limits (a.k.a.~the faint and bright limits) of the sample at the redshift of the $i^\mathrm{th}$ realisation ($z_i$). A similar expression is found for the likelihood of the $B$-band luminosity function. As described in Sect.~\ref{sec:zsty:weight}, the faint limits $M_{\star}^\mathrm{f}(z)$ and $M_{B}^\mathrm{f}(z)$ in this research are defined according to the completeness level $\mathcal{C}$ of the sample employed for the determination of the LMFs, whose expressions and values are studied in detail above in Sect.~\ref{sec:completeness}. We note that our estimator loses the normalisation of the LMFs during the maximisation of Eq.~\ref{eq:likelihood_z}. Therefore, once the \{$\alpha_1$, $\alpha_2$, $\bar{\mathcal{M}}_{\star,1}$, $\bar{\mathcal{M}}_{\star,2}$\} and \{$\beta_1$, $\beta_2$, $\mathcal{M}_{B,1}$, $\mathcal{M}_{B,2}$\} are constrained, the LMF normalisations can be recovered \textit{a posteriori} from the number density of sources in the sample (details in Sect.~\ref{sec:zsty:norm}). 

For the maximisation and sampling of the posterior distribution of the likelihood detailed in Eq.~\ref{eq:likelihood_z}, we use the \texttt{Python} implementation of the affine-invariant ensemble sampler for Markov chain Monte Carlo code known as \textit{emcee}\footnote{\url{https://emcee.readthedocs.io}} \citep[][]{ForemanMackey2013}. For this aim, the \textit{emcee} analysis is performed with $100$ walkers (i.e.~$25$ walkers per degree of freedom) and $2\,000$ steps per walker for both the `burn-in' phase and sampling the posterior distribution. We check that there is negligible or small impact in the posterior distribution retrieved during the \textit{emcee} analysis by the use of a larger number of steps. In fact, we find that after a few dozens of steps, \textit{emcee} promptly converges to a stable solution that maximises Eq.~\ref{eq:likelihood_z} and that the posterior distribution can be properly sampled with only $\sim20$ walkers and after $\sim 500$ steps. It is worth mentioning that a too large number of steps rapidly increases the computational time of this \textit{emcee} analysis, which may be especially long owing to the calculation of the $\Gamma$ terms for each of the realisations of each of the sources. In view of the future J-PAS project, we may therefore reduce the number of walkers and steps, where this decision would be based on a consensus to determine the LMFs in a short enough period of time without renouncing to the precision of the LMF parameters. 


\subsection{Weights and priors}\label{sec:zsty:weight}

In this research, the weights $\omega_i$ in Eq.~\ref{eq:likelihood_z} are one of the key parameters for the determination of the miniJPAS LMFs. These weights can account for, among others, the probability of each of the sources for being a galaxy, the spectral classification of galaxies, the completeness of the sample at certain redshift and stellar mass/absolute magnitude, how relevant is a source owing to the quality of the determination of the parameters involved, and even the statistical number of galaxies in the sample, which is also needed for the determination of the LMF normalisation (see Sect.~\ref{sec:zsty:norm}). Moreover, as we base the determination of the miniJPAS LMFs on the cumulative probability of the sample at $0.05 \le z \le 0.7$ and $r_\mathrm{SDSS} \le 22$, these constraints on the parent sample can be easily included in a statistically rigorous manner through the mentioned weights.

In the following, we distinguish two kind of weights in order to facilitate the understanding of the role of the different key aspects included in the weights. For this reason, the weight $\omega_i$ included in Eqs.~\ref{eq:likelihood} and \ref{eq:likelihood_z} can be decomposed in two terms or contributions of the form
\begin{equation}\label{eq:weight}
\omega_{i} = \kappa_i \times \frac{\tau_i}{\langle\tau \rangle}\ ,
\end{equation}
where $\kappa_i$ accounts for all the terms that contribute to the number of galaxies in the sample and $\tau_i$ for any kind of statistical significance meaning that some sources or realisations should have a higher consideration during the maximisation of Eq.~\ref{eq:likelihood_z}. As in \citet{Zucca1994}, we introduce the median of the distribution of the $\tau_i$ weights, $\langle\tau \rangle$, to avoid artificially increasing the error contours of the LMF parameters during the \textit{emcee} analysis in a general case. In fact, $\tau_i / \langle\tau \rangle$ can act as a Bayesian prior during the sampling of the posterior distribution.

For this work and owing to the selection of our sample, $\kappa_i$ have to include the probability of a source to be a galaxy or $P_\mathrm{G}$. This way, those sources with null probability to be a galaxy are removed from the analysis, whereas those sources with an uncertain star/galaxy classification are weighted according to their probability. Secondly, as we plan to derive the LMFs according to the galaxy spectral-type, $\kappa_i$ must reflect the number and stellar population parameters of quiescent or star-forming galaxies. For this aim, we include the $P_{\mathrm{Q},i}$ and $P_{\mathrm{SF},i}$ estimated for each of the Monte Carlo realisations ($1$ or $0$ depending whether the realisation is above or below Eq.~\ref{eq:mcde}, respectively; see Sect.~\ref{sec:qprob}). Hence, galaxies with $P_\mathrm{Q}$ values for the spectral classification different than unity or zero partly contribute to both star-forming and quiescent LMFs with those realisations that are compatible with these galaxy spectral-types. Thirdly, we also include the completeness level of each of the realisations. This allows us to include in the analysis those sources that were observed in miniJPAS and that are below the $\mathcal{C}=0.95$ completeness level, thus making the most of our sample. For instance, if one of the realisations correspond to an observation for a completeness level of $\mathcal{C}=0.5$, this would mean that only half of the sources were observed or included in the sample, and therefore, the LMFs must account for the double of sources at that redshift and stellar mass/absolute magnitude. Fourthly, as we are using the $N_\mathrm{MC}$ Monte Carlo realisations for the inclusion of the PDFs in our analysis, we have to take this into consideration to avoid overestimating the number of sources in the sample. Finally, for the determination of the LMFs we employ a flux-limited sample down to $r_\mathrm{SDSS}^\mathrm{lim}=22$ (i.e.~$r_\mathrm{SDSS} \le 22$), in order to avoid the well-known border effects owing to uncertainties in our $r_\mathrm{SDSS}$ detection band. In this regard, we also include fainter sources (down to $r_\mathrm{SDSS}=22.5$) than this magnitude limit with uncertainties that make them  compatible with $r_\mathrm{SDSS} \le 22$ and we weight its probability ($P_{r}$) according to the following equation:
\begin{equation}\label{eq:weightr}
P_{r,i} = \frac{1}{2}-\frac{1}{2} \mathrm{erf} \left \{ \frac{2.5}{\sqrt{2}~\ln{10}~\sigma_{r,i}}~\left[ 10^{0.4~(r_{\mathrm{SDSS},i}-r_{\mathrm{SDSS}}^\mathrm{lim})}-1 \right] 
\right\}\ ,
\end{equation}
where $\sigma_{r,i}$ are the magnitude uncertainty included in the miniJPAS photometric catalogue, $r_\mathrm{SDSS}$ its magnitude, and erf is the error function, which is defined as $\mathrm{erf}(x) = 2~\pi^{-1/2}~\int_0^{x} e^{-t^2}~\mathrm{d}t$. Considering all these factors, the $\kappa_i$ weight used for deriving the stellar mass functions of miniJPAS quiescent galaxies would be the following:
\begin{equation}\label{eq:weightq}
\kappa_{i} = \frac{P_{\mathrm{Q},i} \times P_{\mathrm{G},i} \times P_{r,i} }{N_{\mathrm{MC},i} \times \mathcal{C}(z_i,M_{\star,i})}\ .
\end{equation}
For the stellar mass functions of star-forming galaxies, $\kappa_i$ is the same than in Eq.~\ref{eq:weightq}, except for changing the $P_\mathrm{Q,i}$ for $P_\mathrm{SF,i}$.
Furthermore, we note that for the luminosity function case, the completeness of the $B$-band absolute magnitude must be used instead of the stellar mass completeness one, meaning that $\mathcal{C}(z_i,M_{B,i})$ (see Eq.~\ref{eq:completeness_mabs}). As $N_{\mathrm{MC},i}$ is the same number for all the sources in the sample, we note that the number of realisations is not relevant for the maximisation of Eq.~\ref{eq:likelihood_z}, but it is needed for the LMF normalisation (see Sect.~\ref{sec:zsty:norm}).

As we mention in Sect.~\ref{sec:minijpas:photoz}, \textit{odds} is a parameter that correlates with the quality of the PDZs of each of the miniJPAS sources and is sensitive to outliers. In fact, \citet{HernanCaballero2021,HernanCaballero2023} showed that J-PAS-like sources with low \textit{odds} values show larger discrepancies with respect to spectroscopic redshifts and more frequently present larger values than these ones. However, we cannot restrict our sample to only sources with high \textit{odds} values because this would bias our sample and the LMFs are sensitive to the total number density of galaxies. In order to circumvent this issue, we also include the \textit{odds} parameter as an extra weight during the maximisation of Eq.~\ref{eq:likelihood_z}. We find that including \textit{odds} values in the weight budget improves the constraints on the faint-end slopes of miniJPAS LMFs. This is a consequence of the fact that faint sources in the sample exhibit lower \textit{odds} in average, and hence overestimated photo-$z$, that induces higher $\alpha$ and $\beta$ values. It is also of note that we do not adopt the photo-$z$ probability associated to PDZs, since the JPHOTOZ package used to constrain these probability distributions includes priors based on distributions of the $i$ band from deeper surveys such as VVDS, meaning that its inclusion may require our LMF results to be similar to the VVDS luminosity functions. In addition and owing to the depth of miniJPAS, the faint-end slopes of LMFs are mainly constrained by galaxies at lower redshifts (i.e.~$z\lesssim 0.25$), whereas the bulk of the galaxy sample is concentrated around $z \sim 0.5$. This means that the maximisation of Eq.~\ref{eq:likelihood_z}, and hence $\alpha$ and $\beta$, is dominated in number by sources with stellar masses and luminosities above the LMF faint-end limits (i.e.~around $\mathcal{M}_\star(z)$ and $\mathcal{M}_{B}(z)$). In order to mitigate this fact and balance the significance of the nearby galaxies for the determination of the LMF faint-end slopes, we increase the weight of nearby galaxies for the maximisation of Eq.~\ref{eq:likelihood_z} by accounting for the comoving volume ($V_\mathrm{C}$). Therefore, for this work we only account for \textit{odds} and comoving volume in the $\tau_i$ weight, that is, $\tau_i = {odds_i}^2 \times (\frac{\mathrm{d}^2 V_\mathrm{C} (z_i)}{\mathrm{d}z\ \mathrm{d}\Omega})^{-1}$ and $\langle\tau \rangle$ the median of the $\tau_i$ distribution of the sources with non-null $\omega_i$ weights.

Regarding other priors and constraints, we set lower limits for the stellar mass and luminosity completeness levels that are going to be included in the maximisation of Eq.~\ref{eq:likelihood_z}, which directly affects to the definition of $M_{\star,i}^\mathrm{f}$ and $M_{B,i}^\mathrm{f}$. This implies that the weight of a realisation with a stellar mass or absolute magnitude under this completeness limit is null (i.e.~$\omega_i = 0$) and can be removed from the analysis process. Regardless of the galaxy spectral-type, we assume that the minimum completeness limit imposed for the determination of the $B$-band luminosity function is $\mathcal{C}=0.25$, whereas for the stellar mass case is $\mathcal{C}=0.05$. Consequently, $M_{B,i}^\mathrm{f} = M_B^{\mathcal{C}=0.25}(z_i)$ and $\log_{10}M_{\star,i}^\mathrm{f}=\log_{10}M_\star^{\mathcal{C}=0.05}(z_i)$ (see Eqs.~\ref{eq:completeness_mstar} and \ref{eq:completeness_mabs}). The former limit is set to $\mathcal{C}=0.25$ because this helps to prevent overestimation of the $\alpha$ and $\beta$ values since the miniJPAS sources with the lowest $B$-band luminosities also exhibit the lowest $odds$ values. This is not the case for stellar mass functions since low stellar mass galaxies do not preferentially show low $odds$ values, but they are more homogeneously distributed and weighted accordingly (see $\tau_i$ above). Furthermore, we find that the assumption of a lower limit for completeness prevents the inclusion of some outliers without significantly altering the results. In the same sense, as the Fermi-Dirac function used for the parametrisation of the completeness (see Eqs.~\ref{eq:fermidirac_mass} and \ref{eq:fermidirac_mabs}) is a divergent function, we assume that the sample is complete (i.e.~$\mathcal{C}(z_i, M_{\star,i})=1$ in Eq.~\ref{eq:weightq}) for values above $\mathcal{C}=0.975$.

During the \textit{emcee} analysis, we also impose limits to characteristic parameters and the faint-end slopes of the LMFs. In particular, we impose that within our redshift limits these parameters must accomplish that: $-4 \le \alpha(z) \le 4$, $-4 \le \beta(z) \le 4$, $9.5 \le \log_{10}\mathcal{M}_\star(z) \le 12.5$, and $-24 \le \mathcal{M}_B(z) \le -17$. As previously commented, we restrict our analysis to those realisations within a redshift interval of $0.05 \le z \le 0.7$, with the goal of restricting our analysis to a non-biased sample of quiescent galaxies owing to the use of $r_\mathrm{SDSS}$ as detection band and removing biased or systematic results related to the parameters involved in the determination of the LMFs. Owing to the low number of galaxies characterising the low mass or faint end slopes of the LMFs (with large uncertainties as well), as well as for decreasing the uncertainties in the parameters involved, for this work we opted for assuming $\alpha_2 = \beta_2 = 0$ for the stellar mass function of star-forming galaxies and both $B$-band luminosity functions. Hence, there is no constraint on $\alpha_2$ for the stellar mass function of quiescent galaxies. This is motivated by the fact that we are using single-Schechter functions for the LMF parametrisation and the number density of low-mass quiescent galaxies rapidly increases at decreasing redshift, while the number of massive quiescent galaxies does not change significantly \citep[see e.g.][and references therein]{DiazGarcia2019b}. On the other hand, the assumption of fixed or constant $\alpha$ and $\beta$ values is frequently used and based on previous works in deeper surveys, where these parameters show little variation with redshift at least since $z = 0.7$ \citep[see e.g.][]{Faber2007,LopezSanjuan2017}.

As a result of all these sample restrictions and probabilities, the effective number of galaxies down to $r_\mathrm{SDSS}=22$ (i.e.~$\sum_{i} \frac{P_{\mathrm{G},i} \times P_{r,i}}{N_{\mathrm{MC},i}}$) employed in this work for determining the LMFs of miniJPAS galaxies at $0.05 \le z \le 0.7$ amounts to $\sim 5\,000$ galaxies.  


\subsection{Normalisation of the LMFs}\label{sec:zsty:norm}

Even though the normalisation is lost during the maximisation of the likelihood in our method, this one can be retrieved by the characteristic parameters, faint-end slopes, and the number densities of our galaxy sample. By definition, the LMFs describes the number density of galaxies per unit of comoving volume and luminosity or stellar mass. In other words, if we integrate the LMFs in a certain comoving volume, we must obtain the number density of galaxies in it. With this in mind and taking the parameters obtained during the maximisation of Eq.~\ref{eq:likelihood_z} into account, as well as the redshift dependence of these parameters, the normalisation of our stellar mass function at a redshift bin of width $\Delta z'$ and centred at $z'$ is 
\begin{equation}\label{eq:normalisation}
\Phi_{M_\star}^{\star}(z',\Delta z') = \frac{\sum_{i=1}^{N_\mathrm{S} \times N_\mathrm{MC}} \kappa_i W(z' - z_i)}{\int_\Omega \int_{z'-\Delta z'/2}^{z'+\Delta z'/2} \frac{\mathrm{d}^2 V_\mathrm{C}(z)}{\mathrm{d}z~\mathrm{d}\Omega}~\Delta\Gamma(z)~\mathrm{d}z~\mathrm{d}\Omega}\ ,
\end{equation}
where $\Omega$ is the solid angle or the survey footprint area ($\Omega = 0.895$~deg$^2$ for miniJPAS), $V_\mathrm{C}$ is the comoving volume, and $\Delta \Gamma$ is the function that results of integrating the stellar mass function in the stellar range of our sample, that is:
\begin{equation}\label{eq:deltagamma}
\Delta \Gamma (z) = \Gamma \left( \alpha(z)+1, \frac{M_{\star}^\mathrm{f}(z)}{\mathcal{M}_{\star}(z)}\right) - \Gamma \left( \alpha(z)+1, \frac{M_{\star}^\mathrm{b}(z)}{\mathcal{M}_{\star}(z)}\right)\ .
\end{equation}
$W$ is the window function for selecting galaxies or realisations in a redshift bin:
\begin{equation}\label{eq:window}
W(z'-z) = \left\{  \begin{array}{rcl}
1 & \mathrm{if} & -\frac{\Delta z'}{2} \le z'-z \le \frac{\Delta z'}{2} \\
0 & \mathrm{otherwise} & 
\end{array}\right . \ .
\end{equation}
The normalisation of the $B$-band luminosity functions in this work is as Eqs.~\ref{eq:normalisation} and \ref{eq:deltagamma}, but including $\beta(z)$ and $\mathcal{L}_B$ instead. We note that the normalisation of the LMFs as a function of the galaxy spectral-type is implicit in the $\kappa_i$ term. 

However, if we need the normalisation of the LMFs at a given redshift, this expression can result hard to compute. To make it easier to obtain, we compute Eq.~\ref{eq:normalisation} in $20$ redshift bins of equal width ranging from $z=0.05$ to $0.7$ to subsequently fit this set of values to functions of the form $\log_{10}\Phi_{M_\star}^{\star} = \bar{\Phi}_{M_\star,1}^\star + \bar{\Phi}_{M_\star,2}^\star \times  z$ and $\log_{10}\Phi_{L_B}^{\star} = \bar{\Phi}_{L_B,1}^\star + \bar{\Phi}_{L_B,2}^\star \times  z$. We explicitly checked that a linear function of this kind is enough to fairly reproduce the evolution of the normalisation with redshift since $z=0.7$, although quadratic and cubic equations can be also used for the fitting in a general case.


\subsection{Eddington bias}\label{sec:zsty:eddington}

As a consequence of uncertainties and the exponential-like shape of galaxy LMFs for values greater than the characteristic stellar mass or luminosity, these functions can be subject to the so-called Eddington bias. This effect can be especially relevant at the high mass and/or bright-end of the LMFs and mainly results in an overestimation of the number of massive or bright galaxies \citep[see e.g.][]{Ilbert2013,Obreschkow2018}. If we do not account for this effect, other parameters resulting from the LMFs (e.g.~the stellar mass and luminosity densities) would be affected in different ways. In our case, the dependency of the errors on redshift and magnitude (see Sect.~\ref{sec:results:errors} and Fig.~\ref{fig:errors}) makes that the cosmic evolution of the LMFs may be also affected by Eddington bias, and therefore, the cosmic evolution of the stellar mass and luminosity densities as well. The Eddington bias is mathematically expressed as the convolution of the error distribution with the proper distribution of interest. The LMFs determined following the methodology detailed in this paper are actually bidimensional density distributions in the redshift and stellar mass/luminosity space. Consequently, the Eddington bias effect is included in the LMFs according to 
\begin{equation}\label{eq:eddington}
\Phi_\mathrm{Edd} (\bar{M}_\star,z) = \int_{-\infty}^{+\infty}\int_{-\infty}^{+\infty} \Phi(x,y)~G\left( \bar{M}_\star-x, z-y\right) \mathrm{d}x~\mathrm{d}y\ ,
\end{equation}
where $G$ is the bivariate normal distribution of errors that includes the photo-$z$ uncertainties, the logarithm stellar mass or $B$-band absolute magnitude uncertainties, and parameter correlations.

We correct for the Eddington bias effect using the average values of parameters obtained during the SED-fitting analysis. For this aim, we split our galaxy sample in bins of redshift, stellar mass, and absolute magnitude for each of the galaxy spectral-types. For each of these bins, we compute the median values and average uncertainties, as well as the correlations between redshift and the other two stellar population parameters. From this map or distribution of values, we construct the bivariate normal distribution of errors included in Eq.~\ref{eq:eddington} as a function of redshift, stellar mass, and absolute magnitude. For each of the iterations performed during the sampling of the posterior, we find the LMF functions $\Phi(x,y)$ that minimises the difference $|\log_{10}\Phi'(M_\star,z)-\log_{10}\Phi_\mathrm{Edd}(M_\star,z)|$ and $|\log_{10}\Phi'(M_B,z)-\log_{10}\Phi_\mathrm{Edd}(M_B,z)|$, where $\Phi'(M_\star,z)$ and $\Phi'(M_B,z)$ are the LMFs obtained from the maximisation of Eq.~\ref{eq:likelihood_z} (Sects.~\ref{sec:zsty:par} and \ref{sec:zsty:weight}) and its subsequently normalization (Sect.~\ref{sec:zsty:norm}). Therefore, the Eddington bias correction is performed after the \textit{emcee} analysis and for each of the realisations sampling the posterior distribution.


\section{Results}\label{sec:results}


The combination of all the methodologies and analysis techniques detailed in previous sections, which naturally include a proper treatment of the stellar mass and luminosity uncertainties of miniJPAS galaxies (details in Sect.~\ref{sec:results:errors}),  result in the LMFs of the miniJPAS galaxies (Sect.~\ref{sec:results:lmf}). Furthermore, we can easily take advantage of these functions for additionally deriving the cosmic evolution of the stellar mass and luminosity densities (Sect.~\ref{sec:results:density}) and the fraction of quiescent galaxies at different redshift, stellar mass, and luminosity (Sect.~\ref{sec:results:fraction}). 


\subsection{Stellar mass and $B$-band absolute magnitude uncertainties of miniJPAS galaxies}\label{sec:results:errors}

\begin{figure}
\centering
\includegraphics[width=\hsize]{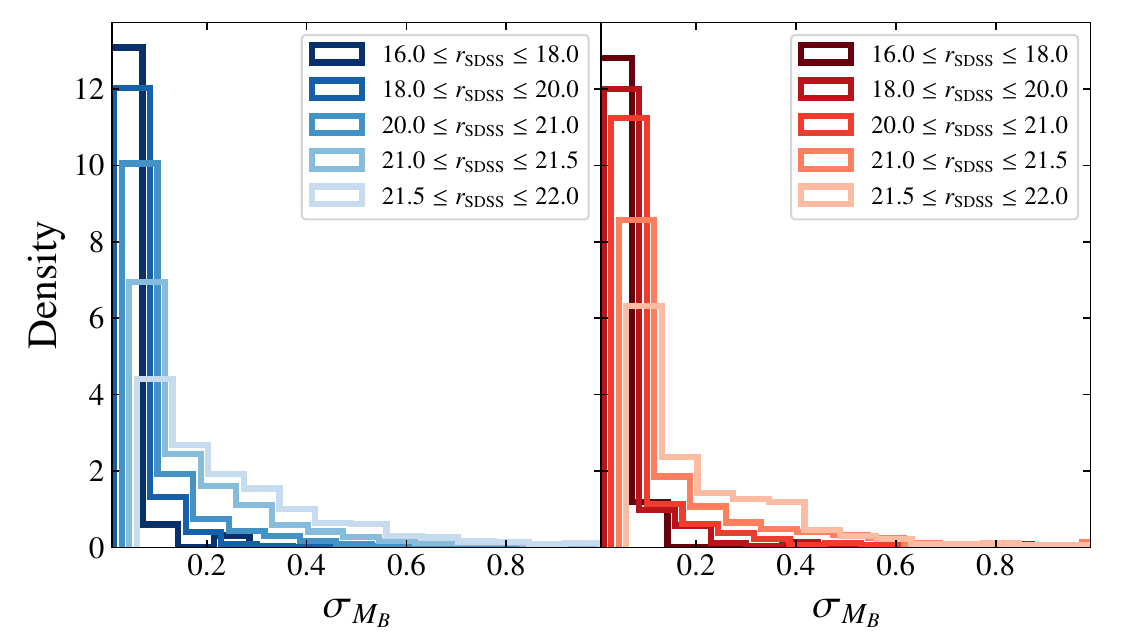}\\
\includegraphics[width=\hsize]{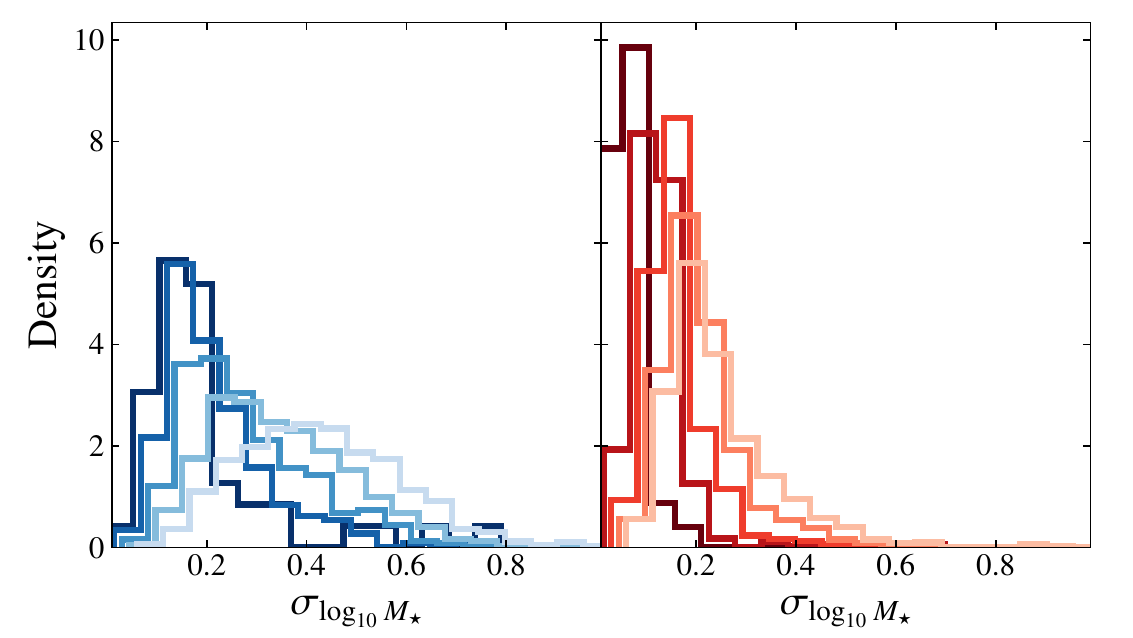}
\caption{Normalised distributions of errors of $B$-band luminosity (\textit{top panels}) and stellar mass (\textit{bottom panels}) at different $r_\mathrm{SDSS}$ magnitude bins (see insets) for star-forming and quiescent galaxies (\textit{left} and \textit{right panels}, respectively) in miniJPAS.}
\label{fig:errors}
\end{figure}

We find that uncertainties of stellar mass and $B$-band luminosity mainly correlate with magnitude (see Fig.~\ref{fig:errors}) and, by extension, with redshift. For our final sample, uncertainties in the $B$-band luminosity ($\sigma_{M_B}$, see upper panels in Fig.~\ref{fig:errors}) typically cover values of $\sigma_{M_B} < 0.5$, whereas for stellar mass this range increases up to $\sigma_{\log_{10}M_\star} < 0.7$~dex. Sources with the lowest luminosity uncertainties are usually brighter and appear more frequently at lower redshifts, while the more noisy estimations of the $B$-band luminosity are only present at fainter magnitudes (see top panels in Fig.~\ref{fig:errors}). Interestingly, there is a significant fraction of sources exhibiting typical values of $\sigma_{M_B} \lesssim 0.15$ ranging from $95$~\% for magnitudes $r_\mathrm{SDSS} \le 20$ to $50$~\% for $21.5 \le r_\mathrm{SDSS} \le 22$. We find that these fractions are $\sim5$~\% higher for quiescent galaxies. 

Concerning stellar mass uncertainties ($\sigma_{\log_{10}M_\star}$, see bottom panels in Fig.~\ref{fig:errors}), these present greater correlations with  magnitude than luminosity uncertainties and also rely on the galaxy spectral-type. In this regard, the stellar mass uncertainties of miniJPAS galaxies are systematically smaller for quiescent than for star-forming galaxies (right- and left-bottom panels in Fig.~\ref{fig:errors}, respectively). More precisely, the $\sim 85$~\% fraction of all the quiescent galaxies present stellar mass uncertainties of $\sigma_{\log_{10}M_\star} \lesssim 0.2$~dex down to magnitude $r_\mathrm{SDSS} = 21$, which can be treated as an upper limit for the stellar mass precision of the luminous red galaxies from miniJPAS. However, only a $70$~\% fraction of star-forming galaxies show uncertainties lower than $\sigma_{\log_{10}M_\star} = 0.3$~dex for magnitudes brighter than $r_\mathrm{SDSS} = 21$. At fainter magnitudes, $r_\mathrm{SDSS} > 21$, the stellar mass uncertainties of galaxies from miniJPAS rapidly increases, especially for the star-forming case. At this faint regime, the stellar mass uncertainties of star-forming galaxies ranges from $\sigma_{\log_{10}M_\star} \sim 0.25$ to $0.55$~dex, whereas for quiescent galaxies is from $\sigma_{\log_{10}M_\star} \sim 0.1$ to $0.4$~dex.

This highlights that stellar masses are in average poorly constrained than luminosity, especially at fainter magnitudes. This is a consequence of stellar mass relying on both the brightness-distance and mass-luminosity relation, which results more uncertain than only distance or photo-$z$. In fact, these results along with the complex PDZ exhibited by some miniJPAS sources motivate the use of the discretised PDF for constraining the miniJPAS LMFs, all this for making the most possible of our sample at fainter magnitudes.

\subsection{Stellar mass and luminosity functions up to $z=0.7$}\label{sec:results:lmf}

In the redshift range explored in this work, the values of the characteristic parameters, the low-mass and faint-end slopes, and normalisations of the miniJPAS LMFs (see Table~\ref{tab:lmf_results}) present median values and posterior distributions (see Figs.~\ref{fig:corner_LF} and \ref{fig:corner_MF}) that are properly sampled within the range of values imposed or priors (i.e.~$-4 \le \alpha(z) \le 4$, $-4 \le \beta(z) \le 4$, $9.5 \le \log_{10}\mathcal{M}_\star(z) \le 12.5$, and $-24 \le \mathcal{M}_B(z) \le -17$). In general, even though the assumption of $\alpha_2 = \beta_2 = 0$, the parameters describing our LMFs do not present a significant evolution. However, we find that the evolution of these parameters depend on the galaxy spectral-type, meaning that galaxies may follow different evolutionary paths according to their spectral-type. 

Secondly, it is well-known that during the process for obtaining the Schechter parameters, degeneracies and correlations amongst the parameters might arise. For instance, there is a strong degeneracy between the faint-end slope and the characteristic parameters. We find that these degeneracies are extended also to the coefficients describing the evolution of these parameters with redshift as shown by the posterior distributions (see Figs.~\ref{fig:corner_LF} and \ref{fig:corner_MF}). In this regard, more massive or brighter characteristic parameters lead to lower low-mass or faint-end slopes and viceversa. As a consequence of the direct correlation between the normalisations and the rest of Schechter parameters, see Eq.~\ref{eq:normalisation}, higher characteristic parameters involve lower normalization values. We also find that these degeneracies are independent of the galaxy spectral-type. 

\begin{figure*}
\centering
\includegraphics[width=0.45\hsize]{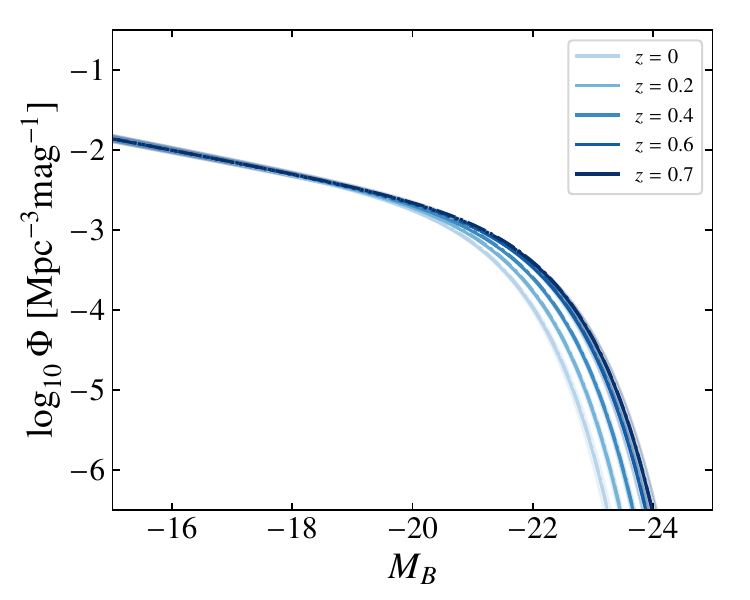}
\includegraphics[width=0.45\hsize]{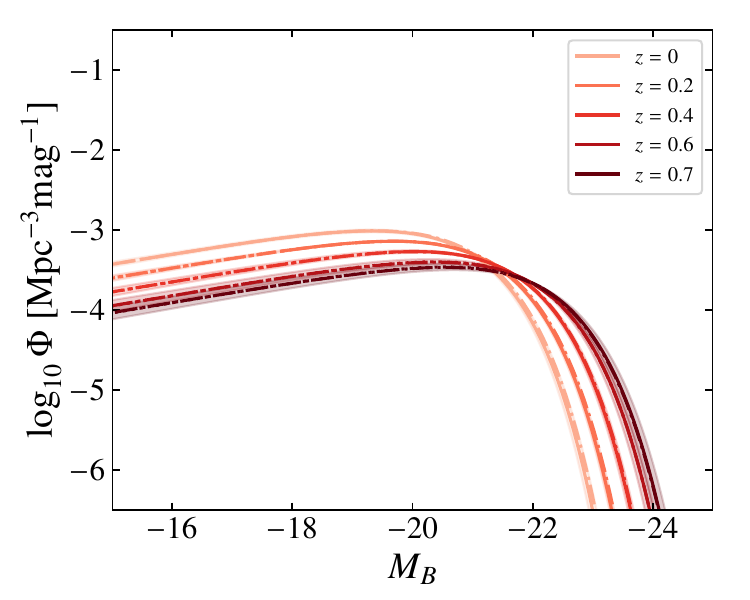}
\caption{Evolution with redshift of the $B$-band luminosity functions of star-forming and quiescent galaxies (solid lines in \textit{left and right panels}, respectively) from miniJPAS. The shaded areas exhibit the uncertainties in the luminosity functions obtained with our methodology. The dash-dot lines show the $B$-band luminosity functions before correcting of Eddington bias effects. The dashed and dotted lines point the absolute magnitude range in which our sample is affected by incompleteness after and before correcting for Eddington bias effects, respectively.}
\label{fig:LF}
\end{figure*}

The first aspect that we observe is that the $B$-band luminosity functions of miniJPAS galaxies present different shapes and evolution with redshift depending on the galaxy spectral-type (see Fig.~\ref{fig:LF}). Thus, the slope of the $B$-band faint-end (i.e.~$\beta_1 + 1$) of star-forming galaxies is negative down to $z=0.7$, while the one for quiescent galaxies is positive (see left and right panels in Fig.~\ref{fig:LF}, respectively). Regarding the bright-end of the luminosity functions, we find that star-forming and quiescent galaxies present similar values. At decreasing redshift, the number density of both bright star-forming and quiescent galaxies decreases. However, the number of faint quiescent galaxies ($M_B \gtrsim -21$) is increasing since $z=0.7$, as it is reflected in the evolution of the faint-end of its $B$-band luminosity function. On the other hand, the evolution in number of faint star-forming galaxies exhibit negligible evolution, or at least is compatible with the uncertainties of this function. As a consequence of the low uncertainties in the determination of the $B$-band absolute magnitudes of the brightest miniJPAS galaxies (see Fig.~\ref{fig:errors}), the effect of the Eddington bias is negligible for the miniJPAS luminosity functions (see Fig.~\ref{fig:LF}).

\begin{figure*}
\centering
\includegraphics[width=0.45\hsize]{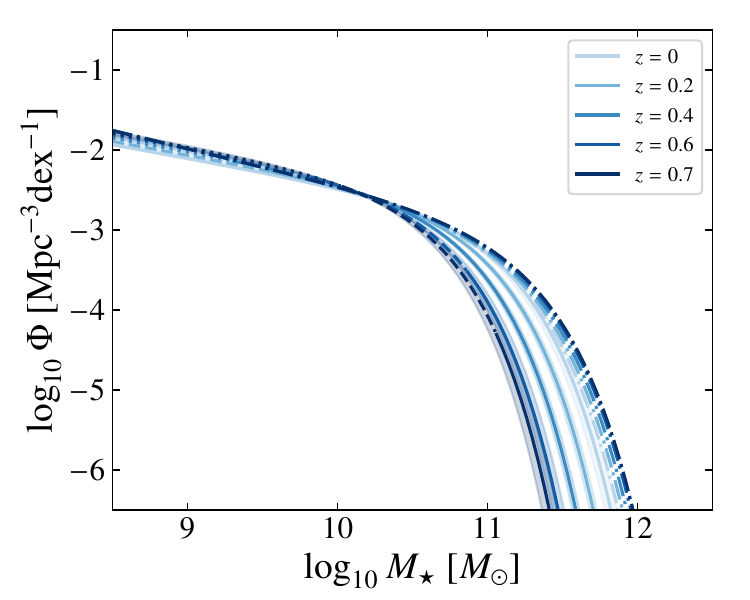}
\includegraphics[width=0.45\hsize]{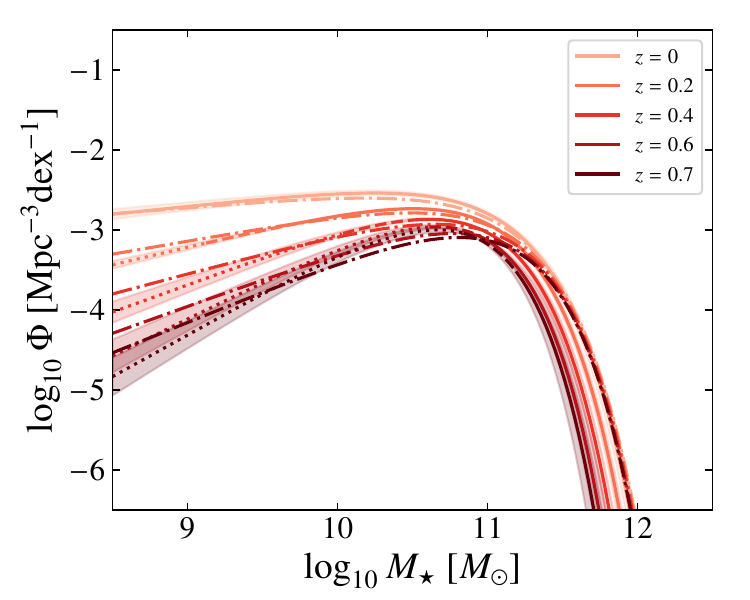}
\caption{As Fig.~\ref{fig:LF}, but for stellar mass functions.}
\label{fig:MF}
\end{figure*}

The stellar mass functions of star-forming and quiescent galaxies also show differences in shape and evolution depending on the spectral-type (see Fig.~\ref{fig:MF}). In addition, these functions show larger evolution than the luminosity ones, especially for quiescent galaxies. The high-mass end of star-forming galaxies show evolution since $z=0.7$, which points out that the most massive galaxies, both star-forming and mainly quiescent, are present at lower redshifts in our sample. However, we find little evolution in the number density of the most massive quiescent galaxies. The evolution of the stellar mass functions of quiescent galaxies mainly resides in the less massive counterpart. Thus, our results points out that there is a more significant increase in the number of intermediate and low mass quiescent galaxies (i.e.~$\log_{10}M_\star < 10.7$~dex). The effect of Eddington bias is more remarkable for the stellar mass functions than for the luminosity functions. This is expected since the uncertainties of the stellar masses are greater in our sample (see Fig.~\ref{fig:errors}) and mainly affecting the high-mass end of these functions. In particular, the Eddington bias seems to be more relevant for the star-forming case (see Fig.~\ref{fig:MF}). In fact, without the Eddington bias correction, the stellar masses of star-forming galaxies would be largely overestimated in the high-mass end. For quiescent galaxies, the Eddington bias effect is less prominent because the stellar mass uncertainties of quiescent galaxies are in average lower than for the star-forming ones. However, we still find necessary to perform the Eddington bias correction described in Sect.~\ref{sec:zsty:eddington}. Otherwise, the high-mass end of the stellar mass function of quiescent galaxies would agree with a null evolution since $z=0.7$. In general, the evolution of the LMFs show that the number density of quiescent galaxies increases at decreasing redshift, but this evolution in number is mainly due to the evolution in number of less massive galaxies, which confirms the results obtained in previous work involving the evolution of the number density of quiescent galaxies with redshift \citep[see e.g.][and references therein]{DiazGarcia2019a}.

\begin{table*}
\caption{Coefficients determining the stellar mass and $B$-band luminosity functions (top and bottom panels, respectively; details in Sect.~\ref{sec:results:lmf}) of the star-forming and quiescent galaxies from miniJPAS. Except for the stellar mass function of quiescent galaxies, a prior $\alpha_2 = \beta_2 = 0$ is assumed for getting these results.}
\label{tab:lmf_results}
\centering
\begin{tabular}{c c c c c c c}
\hline\hline
\multirow{2}{*}{Spectral-type} & \multirow{2}{*}{$\alpha_1$} & \multirow{2}{*}{$\alpha_2$} & \multirow{2}{*}{$\bar{\mathcal{M}}_{\star,1}$} & \multirow{2}{*}{$\bar{\mathcal{M}}_{\star,2}$} & \multirow{2}{*}{$\bar{\Phi}_{M_\star,1}^{\star}$} & \multirow{2}{*}{$\bar{\Phi}_{M_\star,2}^{\star}$} \\
&&&&&& \\
\hline

Star-forming & $-1.34^{+0.01}_{-0.01}$ & $0.00_{--}^{--}$ & $10.92_{-0.05}^{+0.05}$ & $-0.63_{-0.13}^{+0.13}$ & $-3.11_{-0.04}^{+0.04}$ & $\ \ 0.42_{-0.09}^{+0.09}$ \\
Quiescent & $-0.80_{-0.03}^{+0.03}$ & $1.23_{-0.22}^{+0.23}$ & $10.94_{-0.03}^{+0.03}$ & $-0.43_{-0.12}^{+0.13}$ & $-2.67_{-0.03}^{+0.03}$ & $-0.37_{-0.07}^{+0.07}$ \\

\hline
\multirow{2}{*}{Spectral-type} & \multirow{2}{*}{$\beta_1$} & \multirow{2}{*}{$\beta_2$} & \multirow{2}{*}{$\bar{\mathcal{M}}_{B,1}$} & \multirow{2}{*}{$\bar{\mathcal{M}}_{B,2}$} & \multirow{2}{*}{$\bar{\Phi}_{L_B,1}^{\star}$} & \multirow{2}{*}{$\bar{\Phi}_{L_B,2}^{\star}$} \\
&&&&&& \\
\hline

Star-forming & $-1.36^{+0.01}_{-0.01}$ & $0.00_{--}^{--}$ & $-20.99_{-0.08}^{+0.08}$ & $-1.11_{-0.21}^{+0.21}$ & $-2.69_{-0.04}^{+0.04}$ & $-0.16_{-0.10}^{+0.10}$ \\
Quiescent & $-0.68_{-0.02}^{+0.02}$ & $0.00_{--}^{--}$ & $-20.56_{-0.08}^{+0.08}$ & $-1.77_{-0.23}^{+0.23}$ & $-2.68_{-0.03}^{+0.03}$ & $-0.64_{-0.08}^{+0.09}$ \\

\hline
\end{tabular}
\end{table*}

\subsection{Evolution of the stellar mass and luminosity densities since $z=0.7$}\label{sec:results:density}

One of the advantages of using parametric LMFs is that we are able to constrain the cosmic evolution of the stellar mass and luminosity densities (\jm\ and \jb, respectively) through their Schechter parameters. Moreover, as our LMF parameters are also analytic functions of redshift, this allows us to analytically estimate \jm\ and \jb\ at any redshift since $z=0.7$. The expression for the stellar mass and luminosity densities can be written as
\begin{equation}\label{eq:densityb}
\begin{split}
j_B(z) & = \int L_B~\Phi(L_B,z)~\mathrm{d}L_B \\
& = \Phi_{L_B}^{\star}(z)~10^{0.4~(M_{B,\odot}-\mathcal{M}_B(z))}~\Gamma\left( \beta(z)+2\right)\ ,
\end{split}
\end{equation}
\begin{equation}\label{eq:densitym}
\begin{split}
j_M(z) & = \int M_\star~\Phi(M_\star,z)~\mathrm{d}M_\star \\
& = \Phi_{M_\star}^{\star}(z)~\mathcal{M}_\star(z)~\Gamma\left( \alpha(z)+2\right)\ ,
\end{split}
\end{equation}
where $\Gamma$ is the Euler gamma function and $M_{B,\odot}$ is the $B$-band absolute magnitude of the Sun, for which we adopt a value of $M_{B,\odot}=5.38$ \citep{Binney1998}.

\begin{figure*}
\centering
\includegraphics[width=0.45\hsize]{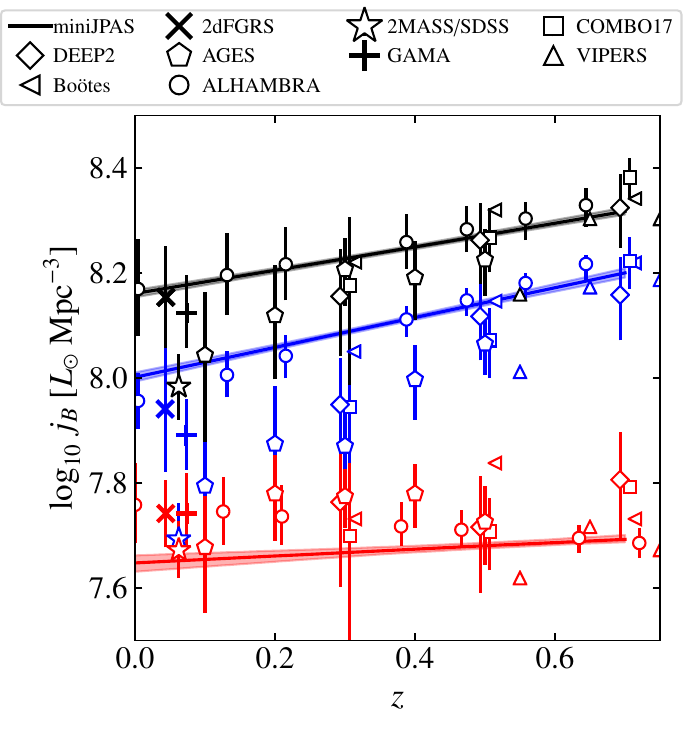}
\includegraphics[width=0.45\hsize]{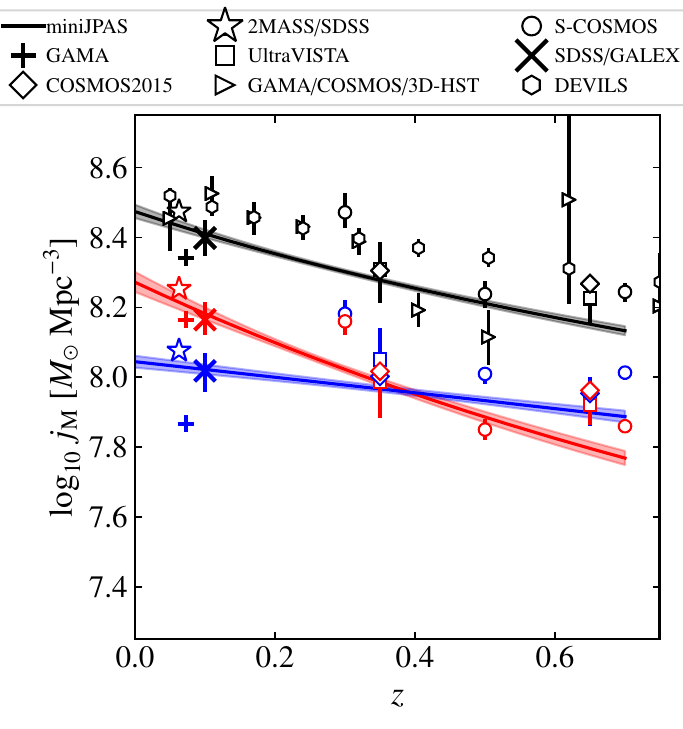}
\caption{Cosmic evolution of the $B$-band luminosity and stellar mass densities (\textit{left and right panels}, respectively) for quiescent, star-forming, and full miniJPAS samples (red, blue, and black lines, respectively). The shaded areas illustrate the uncertainties of these densities. \textit{Left panel}: $B$-band absolute magnitude densities from 2dFGRS \citep[cross markers,][]{Madgwick2002}, 2MASS/SDSS \citep[star markers,][]{Bell2003}, COMBO-17/DEEP2 \citep[squares and diamonds, respectively;][]{Faber2007}, AGES \citep[pentagons,][]{Cool2012}, GAMA \citep[plus marker,][]{Loveday2012}, VIPERS \citep[triangle up,][]{Fritz2014}, Boötes field \citep[triangle left,][]{Beare2015}, and ALHAMBRA \citep[circle,][]{LopezSanjuan2017} surveys are included for comparison. \textit{Right panel}: stellar mass densities from 2MASS/SDSS \citep[star markers,][]{Bell2003},  S-COSMOS \citep[circles,][]{Ilbert2010}, GAMA \citep[plus markers,][]{Baldry2012}, UltraVISTA \citep[squares,][]{Ilbert2013}, SDSS/GALEX \citep[crosses,][]{Moustakas2013}, COSMOS2015 \citep[diamonds,][]{Davidzon2017}, GAMA/G10-COSMOS/3D-HST \citep[triangle right,][]{Wright2018}, and DEVILS \citep[hexagons,][]{Thorne2021}.} 
\label{fig:jBM}
\end{figure*}

As a result, the cosmic evolution of all the miniJPAS galaxies indicates that the global evolution of the $B$-band density decreases at decreasing redshift (see black line in left panel of Fig.~\ref{fig:jBM}). In this sense, the global evolution of the total sample goes from a value of $8.4$ to $8.2$~dex. Therefore the luminosity of galaxies per unit of volume is greater at higher redshift. However, this trend is different when we take the spectral-type of galaxies into account. For quiescent galaxies, the $B$-band luminosity density remains roughly constant with a value of $7.7$~dex (see red line in left panel of Fig.~\ref{fig:jBM}). Consequently, the decrease of the luminosity density is mainly lead by star-forming galaxies. For this kind of galaxies, the luminosity density decreases from $8.2$ to $8.0$~dex (see blue line in left panel of Fig.~\ref{fig:jBM}). Consequently, we find that, except for quiescent galaxies, there is a global decrease of $0.2$~dex.

The cosmic evolution of the stellar mass density exhibits larger differences with respect to the luminosity case. The stellar mass density of the global sample of miniJPAS galaxies is compatible with a mild evolution of $\sim 0.3$~dex from $z=0.7$ to $z=0$ (see black line in right panel of Fig.~\ref{fig:jBM}). The global value at $z=0.7$ is $\sim 8.2$~dex and $8.5$~dex at the very nearby Universe. Unlike the $B$-band luminosity density, the evolution of the stellar mass density is mainly driven by quiescent galaxies. There is a strong evolution in stellar mass density of quiescent galaxies since $z=0.7$ with a value of $7.8$~dex that evolves up to $8.3$~dex at $z=0$ (see red line in right panel of Fig.~\ref{fig:jBM}). On the other hand, the stellar mass density related to star-forming galaxies slightly increases at decreasing redshift (see blue line in right panel of Fig.~\ref{fig:jBM}), more precisely, from $7.9$ to $8.0$~dex. Interestingly, the stellar mass density of both spectral-types present a similar value of $8.0$~dex at $z=0.4$. Therefore, there is a larger fraction of stellar mass in quiescent galaxies at $z=0$, whereas the major part of stars reside in star-forming galaxies at the highest redshift explored in this work. This is mainly related to the fact that the high-mass end of the stellar mass function of quiescent galaxies is shifted to higher values than for star-forming galaxies and, as a consequence, the most massive galaxies at the nearby Universe mostly belong to the quiescent galaxy population. 


\subsection{Fraction of quiescent galaxies since $z=0.7$}\label{sec:results:fraction}

The fraction of quiescent galaxies, $f_\mathrm{r}$, as a function of the $B$-band absolute magnitude and stellar mass at different redshift can be directly inferred from the LMFs by
\begin{equation}\label{eq:fr_mabs}
f_\mathrm{r}(M_B,z) = \frac{\Phi_\mathrm{Q}(M_B,z)}{\Phi_\mathrm{Q}(M_B,z)+\Phi_\mathrm{SF}(M_B,z)}\ ,
\end{equation}
\begin{equation}\label{eq:fr_mass}
f_\mathrm{r}(M_\star,z) = \frac{\Phi_\mathrm{Q}(M_\star,z)}{\Phi_\mathrm{Q}(M_\star,z)+\Phi_\mathrm{SF}(M_\star,z)}\ ,
\end{equation}
where $\Phi_\mathrm{Q}$ and $\Phi_\mathrm{SF}$ are the LMFs of quiescent and star-forming galaxies, respectively.  

\begin{figure*}
\centering
\includegraphics[width=0.45\hsize]{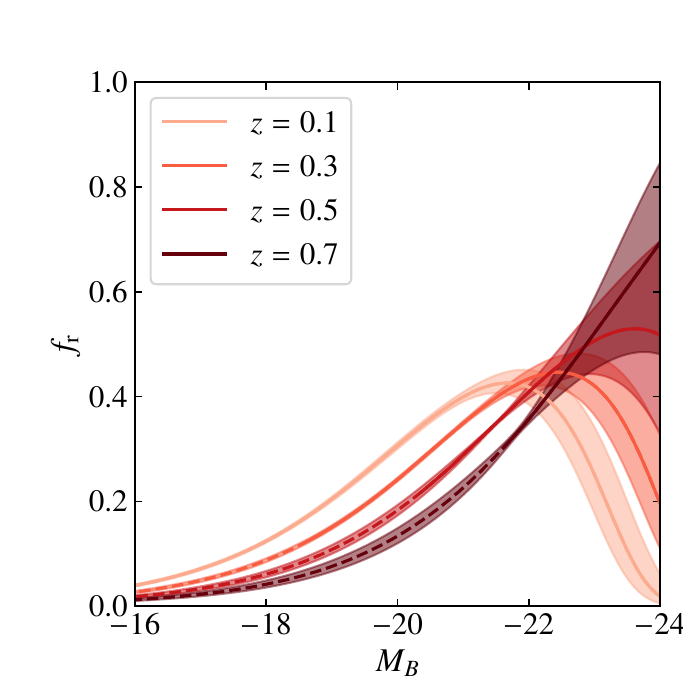}
\includegraphics[width=0.45\hsize]{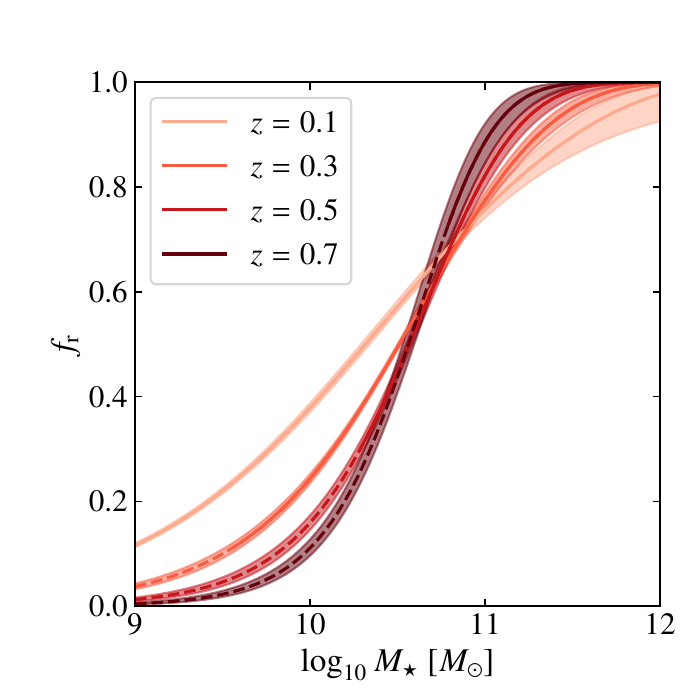}
\caption{Fraction of quiescent galaxies, $f_\mathrm{r}$, as a function of the $B$-band absolute magnitude (\textit{left panel}) and stellar mass (\textit{right panel}) at different redshift (see inset). The dashed lines illustrate the absolute magnitude and stellar mass range for which our quiescent sample is under a completeness level of $\mathcal{C}=0.25$ and $0.05$, respectively, meaning that the fraction of red galaxies are obtained through a extrapolation of our results.} 
\label{fig:fr}
\end{figure*}

Interestingly, the brightest galaxies ($M_B < -22$) are mainly classified as star-forming galaxies at $z < 0.7$ (see left panel in Fig.~\ref{fig:fr}). However, for $B$-band absolute magnitudes fainter than $M_B \sim -23.5$, the fraction of quiescent galaxies starts to increase until reaching a maximum value of $f_\mathrm{r} \sim 0.5$, where the position of this maximum fraction depends on redshift. Hence the $f_\mathrm{r}$ peak moves towards fainter magnitudes at decreasing redshift, more precisely, from $M_B \sim -23$ to $-21$ in the redshift range explored in this work. For fainter magnitudes than the peak previously mentioned, the fraction of quiescent galaxies abruptly decreases reaching values of $f_\mathrm{r} < 0.2$. It is worth mentioning that for the faint regime ($M_B \gtrsim -19$), which in turn exhibits $f_\mathrm{r}$ values below $\sim 0.2$, the fraction of red galaxies are in fact an extrapolation of the $B$-band luminosity functions owing to the completeness lower limit imposed for the maximisation of the luminosity function likelihoods (i.e.~$\mathcal{C}=0.25$). In fact, this magnitude regime is usually reproduced by the second component of a double Schechter function. 

The fraction of red galaxies according to the stellar mass of galaxies also reveals interesting results (see right panel in Fig.~\ref{fig:fr}). On the one hand, the majority of galaxies with stellar mass above $\log_{10}M_\star \gtrsim 10.7$ are actually quiescent galaxies at $z < 0.7$, where the fraction of red galaxies is above $60$~\% in all cases. Moreover, the fraction of quiescent galaxies with stellar masses in the range $\log_{10}M_\star < 10.7$ becomes larger at decreasing redshift, meaning that the number of quiescent galaxies is increasing in importance in more recent epochs. Even though this increase in number, the galaxy sample is largely dominated in number by star-forming galaxies for $\log_{10}M_\star < 10.5$. It is also of note that the values of $f_\mathrm{r}$ for $\log_{10}M_\star \lesssim 10$ should be treated as $f_\mathrm{r}$ lower limits, since we are currently using single Schechter functions for constraining the stellar mass functions of quiescent galaxies owing to the limitations of our sample.

\section{Discussion}\label{sec:discussion}


\subsection{Comparison with previous work}\label{sec:discussion:previous}

During the last decades, the LMFs of galaxies have been widely studied making use of spectroscopic and/or photometric data including galaxies from very early cosmological epochs ($z\sim 5$) to the nearby Universe. This allows us an easy comparison of our results with the parametric LMFs (single or double Schechter functions) obtained in previous works within our redshift range ($0.05 \le z \le 0.7$).

\begin{figure*}
\centering
\includegraphics[width=\hsize,trim={0 1.7cm 0 0},clip=True]{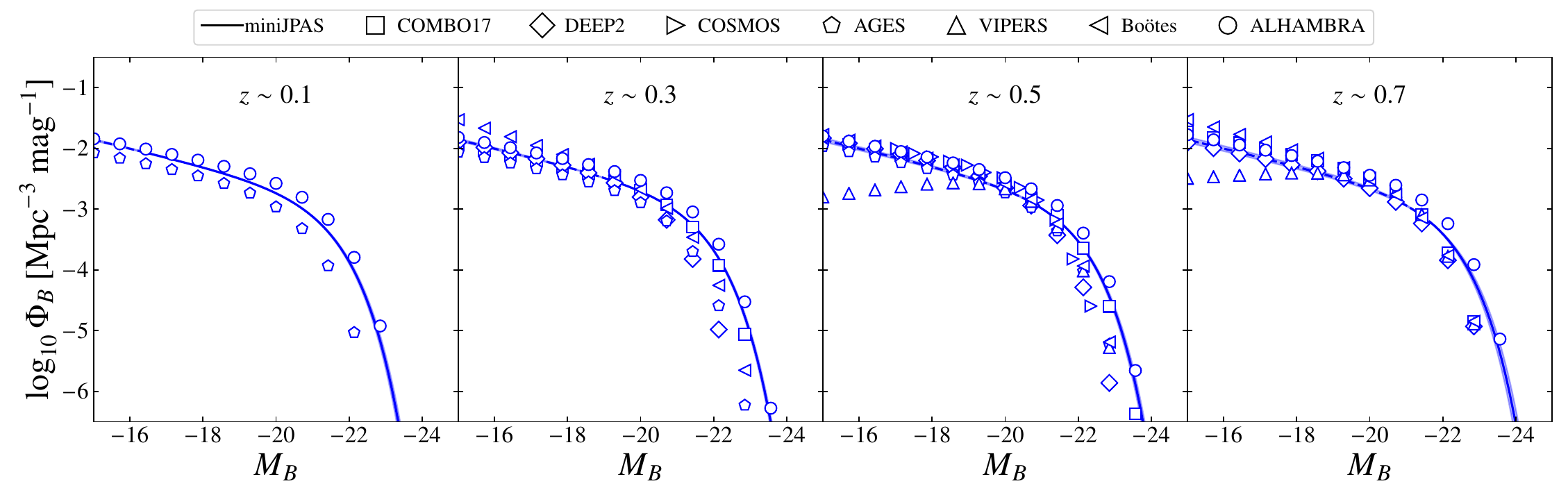}\\
\includegraphics[width=\hsize,trim={0 1.7cm 0 1.35cm},clip=True]{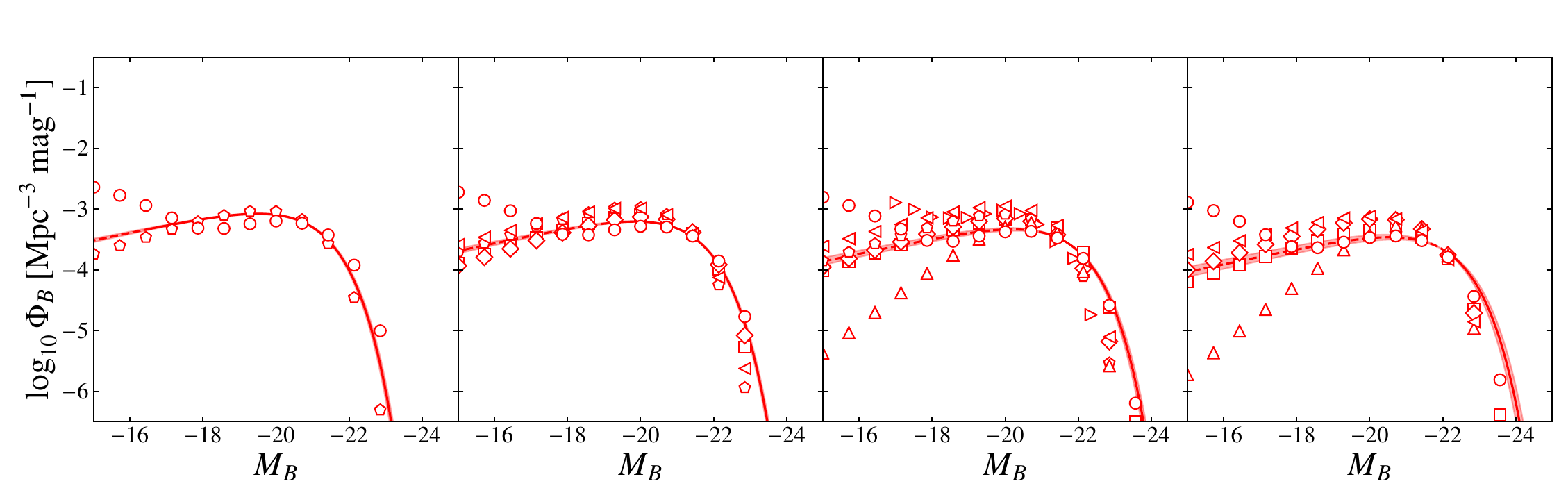}\\
\includegraphics[width=\hsize,trim={0 0 0 1.35cm},clip=True]{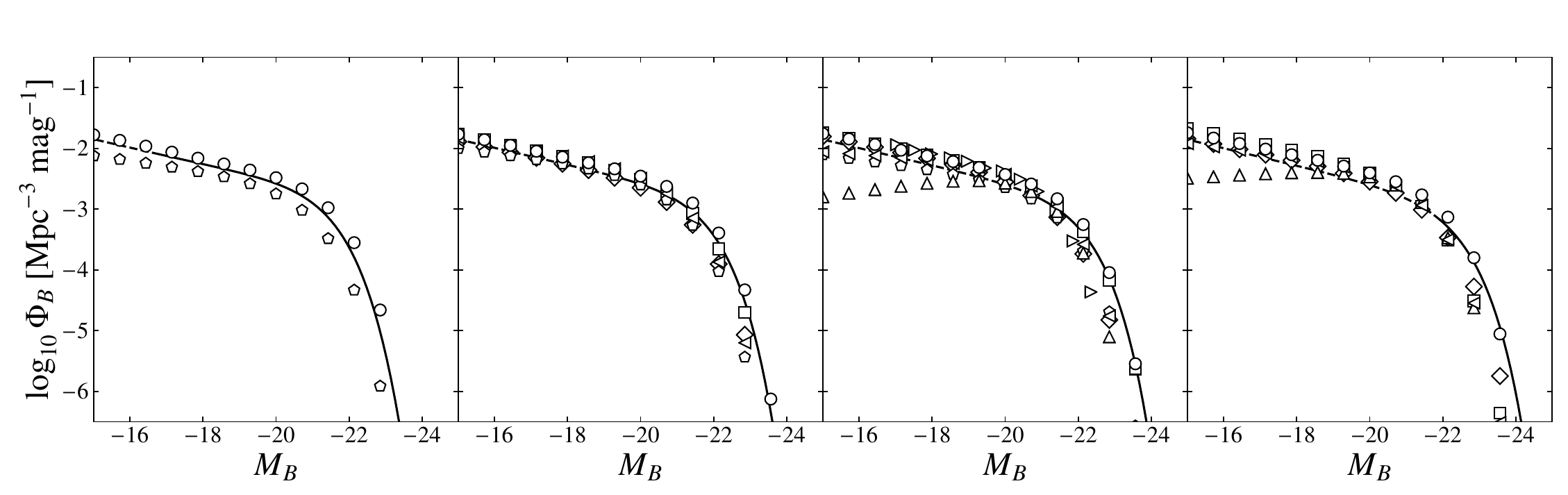}
\caption{Comparison of the $B$-band luminosity function of miniJPAS galaxies (colour lines, \textit{from top to bottom}; star-forming, quiescent, and full galaxy sample) with those from previous work at different redshift. The shaded area illustrates uncertainties in our results. The dashed lines show the $B$-band luminosity functions for magnitudes fainter than the completeness lower limit of the sample employed for their determination (i.e.~$M_B > M_B^{\mathcal{C}=0.25}(z)$). $B$-band luminosity functions from COMBO-17/DEEP2 \citep[squares and diamonds, respectively;][]{Faber2007}, COSMOS \citep[triangle right,][]{Drory2009}, AGES \citep[pentagons,][]{Cool2012}, VIPERS \citep[triangle up,][]{Fritz2014}, Boötes field \citep[triangle left,][]{Beare2015}, and ALHAMBRA \citep[circles,][]{LopezSanjuan2017} surveys are included for comparison.}
\label{fig:LF_comparison}
\end{figure*}

The $B$-band luminosity functions of the miniJPAS galaxies show number densities similar to those obtained in previous researches (see Fig.~\ref{fig:LF_comparison}). In general, we do not find large discrepancies among luminosity functions and both the faint and bright ends of these functions present similar values than in previous work. This conclusion can be extended for quiescent and star-forming galaxies, and hence for the full sample (see top, middle, and bottom panels in Fig.~\ref{fig:LF_comparison}, respectively). Interestingly, the miniJPAS $B$-band luminosity functions are especially similar to those obtained for the ALHAMBRA survey \citep[see circles in Fig.~\ref{fig:LF_comparison} and][]{LopezSanjuan2017}, which were obtained by a statistical Bayesian method with similarities to ours. Except for the ALHAMBRA case, the number density of quiescent galaxies in miniJPAS with absolute magnitudes fainter than $M_B \sim -21$ is $\sim 0.5$~dex lower than for the rest of surveys at $z \gtrsim 0.5$ \citep[see middle panel in Fig.~\ref{fig:LF_comparison}][]{Faber2007,Drory2009,Cool2012,Beare2015}. However, this discrepancy can be explained as a consequence of the use of the MCDE diagram for the galaxy classification. This is because this diagram accounts for extinction effects and commonly yields a lower number of low mass quiescent galaxies in comparison to other diagrams since it removes dusty star-forming galaxies from the quiescent sample \citep[for a further analysis of this effect, see][]{DiazGarcia2019a}. We note that for the ALHAMBRA luminosity functions, as well as its galaxy classification, the authors accounted for extinction and maybe for this reason these luminosity functions are particularly similar to the miniJPAS ones. Finally, we find that the parametric luminosity functions from the VIPERS survey \citep[][]{Fritz2014} are the most different functions with respect to the miniJPAS ones, but also to the rest of cases included here.

\begin{figure*}
\centering
\includegraphics[width=\hsize,trim={0 1.7cm 0 0},clip=True]{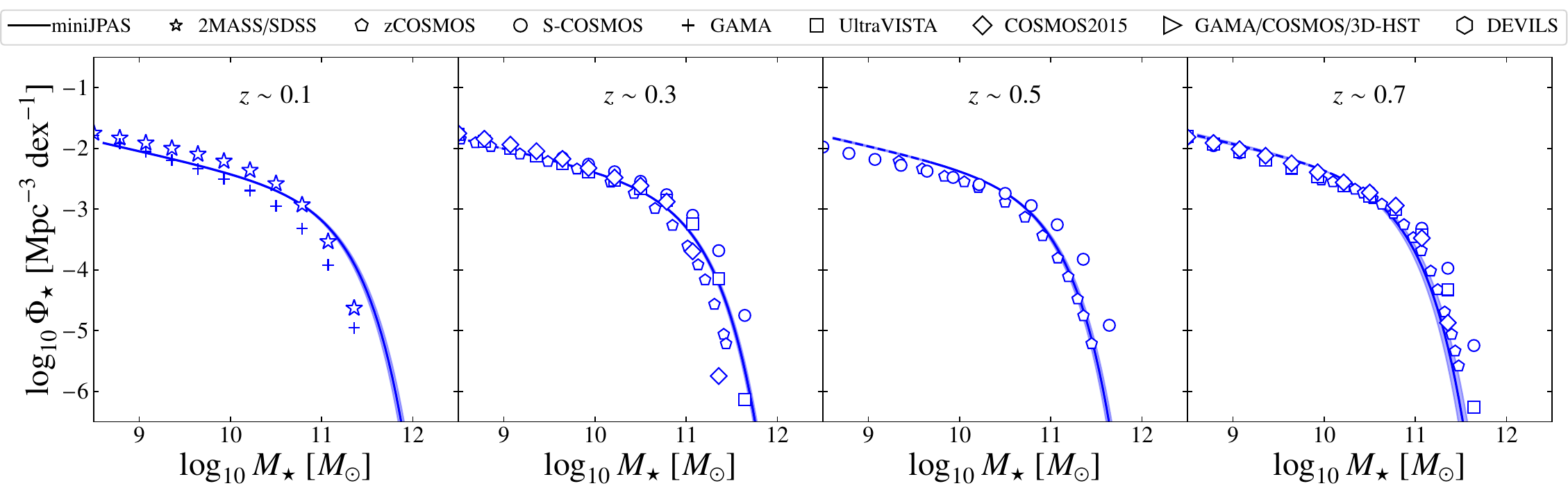}\\
\includegraphics[width=\hsize,trim={0 1.7cm 0 1.35cm},clip=True]{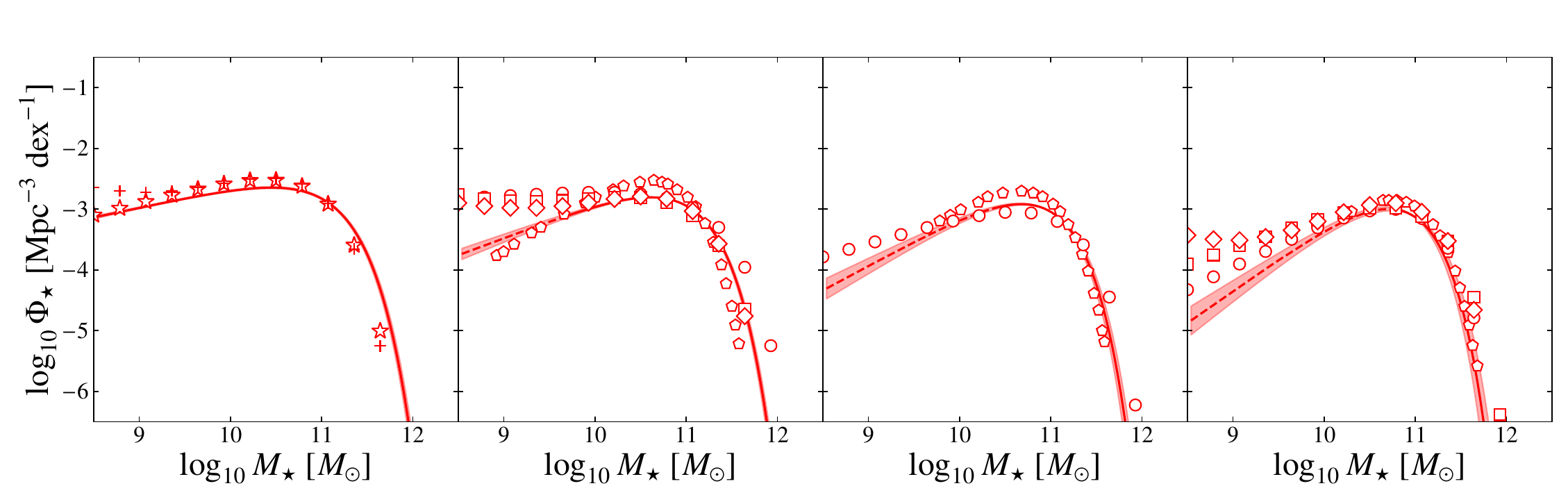}\\
\includegraphics[width=\hsize,trim={0 0 0 1.35cm},clip=True]{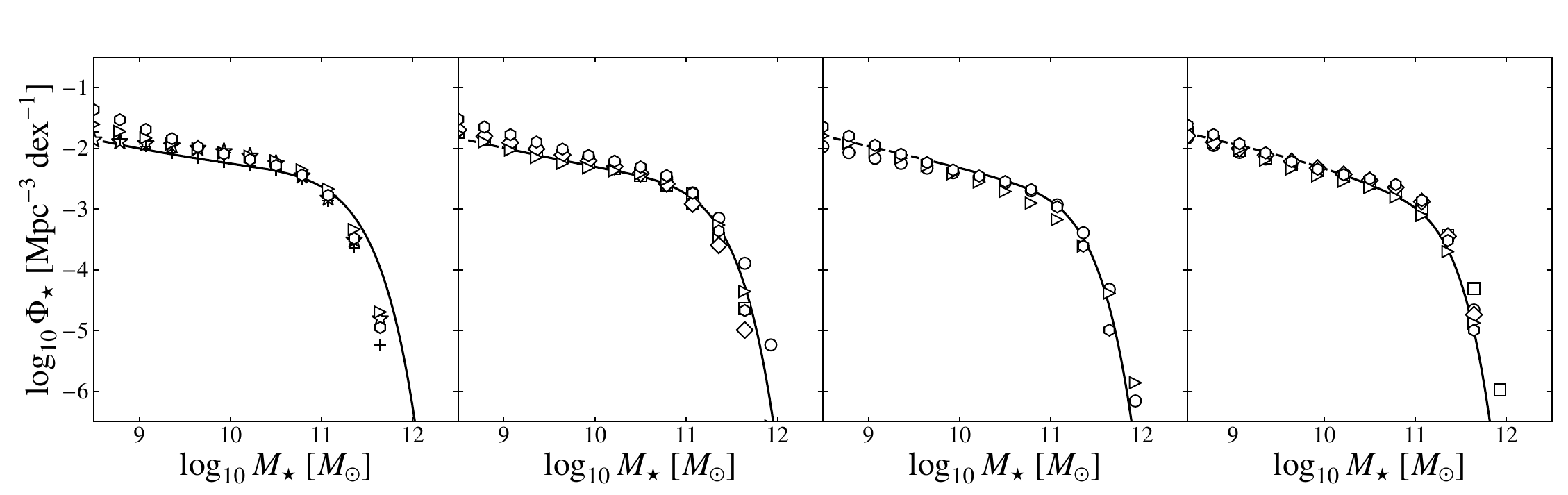}
\caption{As Fig.~\ref{fig:LF_comparison}, but for stellar mass functions. Stellar mass functions from 2MASS/SDSS \citep[star markers,][]{Bell2003}, zCOSMOS \citep[pentagons,][]{Pozzetti2010}, S-COSMOS \citep[circles,][]{Ilbert2010}, GAMA \citep[plus markers,][]{Baldry2012}, UltraVISTA \citep[squares,][]{Ilbert2013}, COSMOS2015 \citep[diamonds,][]{Davidzon2017}, GAMA/COSMOS/3D-HST \citep[triangle right,][]{Wright2018}, and DEVILS \citep[hexagons,][]{Thorne2021} surveys are included for comparison.}
\label{fig:MF_comparison}
\end{figure*}

As revealed by many previous work \citep[see e.g.][]{Bell2003, Bruzual2003, Vazdekis2012, FerreMateu2013, DiazGarcia2019a, DiazGarcia2019b}, the SSP model set along with the IMF adopted for building stellar population synthesis models can introduce a bias or shift in the stellar mass measurements obtained by a stellar population analysis (e.g.~SED-fitting). For instance, we checked that quiescent and star-forming galaxies from miniJPAS are more massive by $0.04$ and $0.11$~dex, respectively, when \citet{Bruzual2003} SSP models (hereafter, BC03) are used instead of the CB17 ones for the MUFFIT SED-fitting analysis, both with the same \citet{Chabrier2003} IMF. \citet{DiazGarcia2019a} also found discrepancies between EMILES SSP models \citep[Extended Medium-resolution Isaac Newton Telescope of Empirical Spectra][]{Vazdekis2016} with a universal \citet{Kroupa2001} IMF for both Padova00 and BaSTI isochrones \citep[][respectively]{Girardi2000,Pietrinferni2004} and the BC03 models with a \citet{Chabrier2003} IMF, where EMILES stellar masses are in average higher by about $0.11$ and $0.15$~dex, respectively. \citet{Bell2003} found that the \citet{Salpeter1955} IMF assumption systematically yields higher stellar masses by about $0.4$ and $0.15$~dex than Kroupa-like and its `diet' \citet{Salpeter1955} IMFs, respectively. In this regard and independently of the galaxy spectral-type, the stellar masses of miniJPAS galaxies determined using MUFFIT with BC03 SSP models and a \citet{Salpeter1955} IMF are systematically $0.24$~dex more massive than using the same model set with a \citet{Chabrier2003} IMF instead. Whenever possible and with the only goal of palliating discrepancies amongst stellar masses owing to model assumptions, and hence performing a proper comparison of the stellar mass function resulting from previous studies, we present all the stellar mass functions to a common stellar mass framework that is scaled according to the \citet{Bruzual2003} model set and a \citet{Chabrier2003} IMF by adding the aforementioned offsets to the characteristic stellar mass (see Fig.~\ref{fig:MF_comparison}). Similarly to the luminosity function case, the stellar mass functions of miniJPAS galaxies exhibit similar trends and values to those obtained in previous studies down to $z=0.7$. However, there is evidence of an excess of massive star-forming galaxies in the miniJPAS at low redshift ($z\sim 0.1$) with respect to results from 2MASS/SDSS and GAMA suveys \citep[see top-left panel in Fig.~\ref{fig:MF_comparison}][respectively]{Bell2003,Baldry2012}. Similarly, the number density of massive quiescent galaxies at $z\sim 0.1$ is also higher in miniJPAS than in previous studies, but less remarkable than for the star-forming case (see middle-left panel in Fig.~\ref{fig:MF_comparison}). Consequently, this excess in number of nearby massive galaxies is also present for the stellar mass function of the full sample and persists when comparing with results based on COSMOS, 3D-HST, and DEVILS results \citep[see bottom-left panel in Fig.~\ref{fig:MF_comparison}][]{Wright2018, Thorne2021}. Owing to the characteristics of the miniJPAS survey and the stellar mass range, it is likely that the miniJPAS survey may be affected of cosmic variance, which would explain this excess in the number density of galaxies at $z\sim 0.1$ (see also Sect.~\ref{sec:discussion:variance}). At $z > 0.2$, discrepancies between the stellar mass functions of both miniJPAS star-forming and the full sample of galaxies and those from other surveys \citep[see top and bottom panels in Fig.~\ref{fig:MF_comparison} and also][]{Pozzetti2010, Ilbert2010,Ilbert2013,Davidzon2017} are small and may partly result of the use of different stellar population synthesis models and techniques. Overall, we find that the stellar mass functions of quiescent galaxies are in good agreement with results from other surveys (see middle panels in Fig.~\ref{fig:MF_comparison}). Nevertheless, there are slight discrepancies at the low mass regime ($\log_{10}M_\star \lesssim 10$~dex at $z > 0.3$), meaning that our parametrised stellar mass functions at the stellar mass range for which our sample is not complete ($\mathcal{C} < 0.05$) show lower number densities than in other works. Part of these discrepancies may be explained by the use of the MCDE for the galaxy classification in miniJPAS (see Sect.~\ref{sec:qprob}), because other diagrams based on rest-frame colours (e.g.~the $UVJ$ diagram) and used for the same aim are commonly contaminated by a $20$~\% fraction of dusty star-forming galaxies \citep[][]{DiazGarcia2019a}, especially at low mass. 

The cosmic evolution of the $B$-band luminosity and stellar mass densities go hand in hand with the evolution of the LMFs. Thus, the good agreement between the miniJPAS LMFs and those from previous studies is also reflected in the $B$-band luminosity and stellar mass densities (see left and right panels in Fig.~\ref{fig:jBM}, respectively). For both densities, there is good quantitative and qualitative agreement with previous studies at $0.05 < z < 0.7$, especially when uncertainties and abundance variance are taken into account (see also Sect.~\ref{sec:discussion:variance}). In this regard, the $B$-band luminosity densities obtained with galaxies from the ALHAMBRA survey \citep[][]{LopezSanjuan2017} are particularly similar to ours, especially the densities of the full sample. For the rest of $B$-band luminosity densities included here for comparison \citep[including some of the most extended surveys such as 2dFGRS, COMBO-17, DEEP2, GAMA, and VIPERS][]{Madgwick2002,Faber2007,Loveday2012,Fritz2014,Beare2015}, we find little numerical differences that in average do not exceed $0.1$~dex. At $z<0.2$, results involving data from 2MASS, SDSS, and AGES \citep[see][]{Bell2003,Cool2012} show larger differences, $>0.2$~dex, which is more remarkable for star-forming galaxies but also in disagreement with the trends of other studies (see left panel in Fig.~\ref{fig:jBM}). Regarding the stellar mass densities, as expected, the variance or range of values obtained for this parameter in previous researches is wider than for the $B$-band luminosity density (see right panel in Fig.~\ref{fig:jBM}). In any case, we find that estimations from previous work in the literature are below $0.2$~dex in the redshift range explored here. In fact, the stellar mass densities for quiescent and star-forming miniJPAS galaxies properly reproduce the evolutionary trends previously constrained in studies involving a galaxy spectral-type classification \citep[including galaxies and data from 2MASS, SDSS, COSMOS, GAMA, UltraVISTA, and GALEX][]{Bell2003,Ilbert2010,Baldry2012,Ilbert2013,Moustakas2013,Davidzon2017}. For the total stellar mass density, there is also evidence of good agreement with the aforementioned and recent works \citep[e.g.][]{Davidzon2017,Wright2018}. However, we note that the stellar mass densities obtained using data from the DEVILS survey \citep[][]{Thorne2021} are systematically higher at $z<0.7$ than our values (see right panel in Fig.~\ref{fig:jBM}), although we find a similar evolution of $0.4$~dex from $z=0.7$ to $0$ as in \citet{Thorne2021}.


\subsection{Cosmic variance of the miniJPAS sample}\label{sec:discussion:variance}

As a consequence of the non-homogeneity of the Universe at small scales \citep[$\lesssim 1$~Gpc, e.g.][]{Davis1985}, the number density of galaxies in small volumes may present variations amongst different pointings much larger than those predicted from Poisson statistics. These variations in the apparent abundance of galaxies, a.k.a.~cosmic variance, may be particularly significant for the so-called `pencil beam' surveys ($<1$~deg$^2$) to the point that can be one of the largest sources of uncertainty for intermediate and high-redshift surveys \citep[see e.g.][]{Newman2002,Driver2010,Moster2011}. Cosmic variance relies on multiple factors such as the aspect ratio of the survey footprint, the imaged cosmological volume (i.e.~area and redshift range), the contiguity of the area surveyed (correlated or independent fields), and the galaxy stellar mass or luminosity \citep[see e.g.][]{Driver2010,Moster2011}.

With this in mind and taking into account that the LMFs themselves are a direct measurement of the abundance of galaxies with different properties, we infer that cosmic variance may play a role in our LMF results and must be quantified according to the miniJPAS characteristics. As a consequence, we expect that the uncertainties of the Schechter parameters describing the miniJPAS LMFs (reported in Sect.~\ref{sec:results:lmf}) only account for counting errors and may be underestimated. In order to address this, we perform a rough estimation of the cosmic variance affecting our results, meaning that we roughly approach the uncertainties on the number densities of galaxies at different stellar mass, luminosity, and redshift. This in turn can explain part of the discrepancies and differences found when comparing with previous studies (see Sect.~\ref{sec:discussion:previous}). For this aim, we adopt the equations proposed by \citet{Moster2011} for setting constraints on the cosmic variance of galaxies by using predictions from $\Lambda$CMD theory and the galaxy bias obtained via a halo occupation model to predict galaxy clustering. In particular, we choose all the parameters established for the EGS field owing to the geometry and size of this field is close to the miniJPAS one. Regarding the cosmic variance of galaxies for the $B$-band luminosity, we take advantage of the well-known stellar mass versus mass-to-light ratio correlation \citep[see e.g.][]{Bell2003}. Making use of the SED-fitting results obtained with MUFFIT and miniJPAS data, we construct the distribution of $B$-band absolute magnitudes for each of the stellar mass bins included in the cosmic variance recipe proposed by \cite{Moster2011}, all this at different redshift and for a constant width of $\Delta z =0.2$. Hereafter, we assume that the cosmic variance at a luminosity equal to the mode of each of the $B$-band absolute magnitude distributions is roughly the same than the cosmic variance of the stellar mass bin employed for constructing each of these distributions.

\begin{figure}
\centering
\includegraphics[width=\hsize]{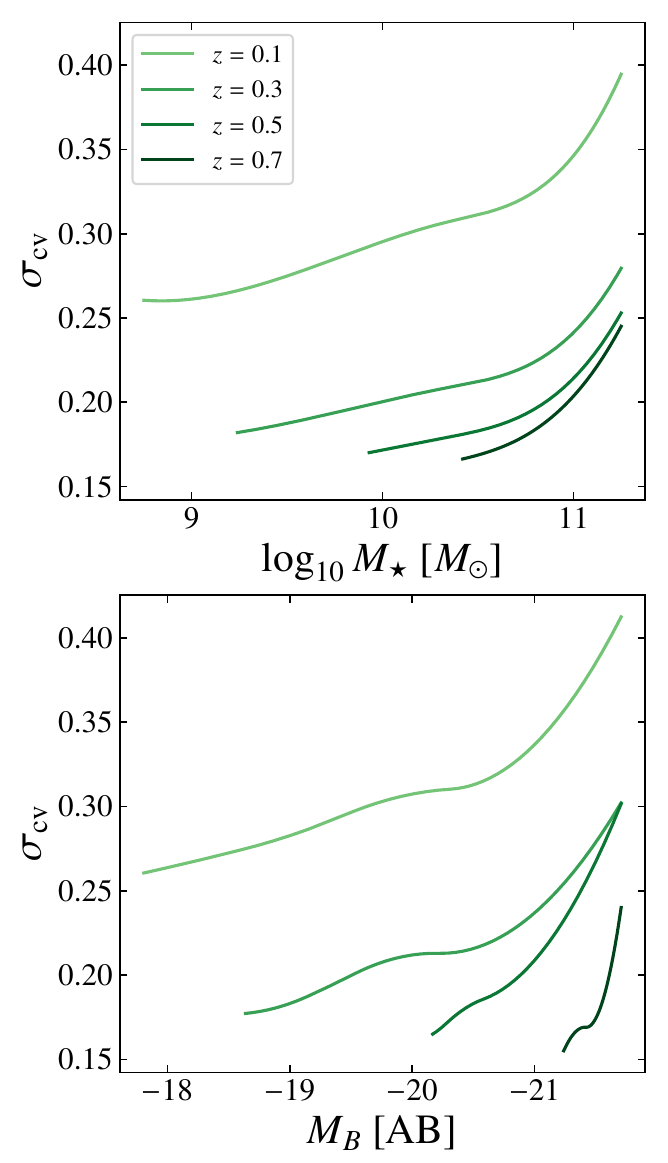}
\caption{Fractional cosmic variance of our sample ($Y$-axis) at different redshift (see inset), stellar mass ($X$-axis, \textit{top panel}), and $B$-band absolute magnitude ($X$-axis, \textit{bottom panel}) for redshift and stellar mass bins of width $\Delta z =0.2$ and $\Delta \log_{10}M_\star = 0.5$, respectively.}
\label{fig:cv}
\end{figure}

Overall, the cosmic variance of the miniJPAS survey present theoretical values ranging from $15$ to $40$~\% (see Fig.~\ref{fig:cv}), as expected for an EGS-like survey \citep[details in][]{Moster2011} of similar area. Despite these fractional values are only indicative, they are very useful for understanding how cosmic variance may affect our results. Firstly, for a constant redshift bin width, $\Delta z$, the miniJPAS cosmic variance decreases at increasing redshift as a consequence of a larger surveyed comoving volume (see light and dark green lines in Fig.~\ref{fig:cv}). On the contrary, for a constant volume case, the cosmic variance increases at increasing redshift since $\Delta z$ is smaller \citep[see][]{Moster2011}. As a consequence, our sample at the nearby Universe is subject to a greater cosmic variance than at high redshift. Secondly, the cosmic variance of both bright and massive galaxies is greater than for their faint and less massive counterparts at a given redshift and volume. This implies that at a fixed redshift, the bright or massive ends of LMFs are going to be worse determined than the less massive parts. This fact also reflects that the brightest and most massive galaxies are less frequent in cosmological surveys, and therefore, these are subject to greater cosmic variance and Poisson variations. 

Cosmic variance is a general uncertainty that, a priori, may also affect the values of the Schechter parameters of the LMFs obtained during our likelihood maximisation. A precise determination of how cosmic variance is affecting our results is beyond the scope of this work because this would imply a more sophisticated analysis, which should include all the aspects embedded in the novel methodology developed in this work. However, previous works based on semi-analytic models and real observations showed that different luminosity bins are highly correlated, especially for values below the characteristic luminosity \citep[][]{Smith2012,LopezSanjuan2017}. This bin-to-bin covariance is mainly due to sample variance arising from the presence of large-scale structures and valid for both volume- and flux-limited samples \citep[][]{Smith2012}. This means that if there is an upward/downward fluctuation in the average number of galaxies in one LMF bin, the other bins will share an upward/downward fluctuation, meaning that cosmic variance is not a random process along the stellar mass or luminosity. Consequently, we expect that the Schechter parameter most affected by cosmic variance is the LMF normalization \citep[see also][]{Trenti2008,Smith2012}, whereas the LMF shape may be slightly affected by it (see also Sect.~\ref{sec:discussion:lessons}). For the incoming J-PAS survey ($\sim8000$~deg$^2$), we expect that these LMFs are going to be hardly affected by cosmic variance, which in turn will be useful for determining how cosmic variance affects our methods depending on the geometry and volume used for constructing galaxy subsamples \citep[see][]{LopezSanjuan2017}.

\subsection{Lessons learned for J-PAS and future prospects}\label{sec:discussion:lessons}

The analysis of the miniJPAS sample does not show striking or new results about the LMFs of galaxies down to $z=0.7$, since these functions have been largely studied in the last decades. In general, the values of the characteristic parameters, the low-mass and faint-end slopes, and normalisations of the LMFs present values that are in the range of values obtained in recent and previous studies (see Table~\ref{tab:lmf_results} and Sect.~\ref{sec:discussion:previous}). 

In fact, the good agreement between the $B$-band luminosity functions and those from other surveys reflects that current star and galaxy classification, completeness limits, photo-$z$ constraints, and aperture photometry are yielding results that are self-consistent and in good agreement with previous work. Moreover, the great degree of similarity with the ALHAMBRA luminosity functions contributes to think that the spectral classification of galaxies along with the completeness of the miniJPAS sample are properly determined for miniJPAS and J-PAS owing to the concordance with the results from this much deeper survey. This also supports the idea that the statistical treatment of the data via probability distribution functions, as the obtained from MUFFIT and the other codes in the J-PAS collaboration \citep[see][]{GonzalezDelgado2021}, is highly recommended for shedding light on the determination of some of the stellar population properties of galaxies. In addition, the inclusion of the PDZs in the SED-fitting analysis is necessary for a proper estimation of the uncertainties affecting some of the stellar population parameters such as stellar mass and luminosity, especially at fainter magnitudes \citep[see also Fig.~\ref{fig:errors} in this work and Fig.~12 in][]{GonzalezDelgado2021}. In particular, J-PAS will be a survey that can accurately constrain the stellar mass of luminous red galaxies (LRGs) with an apparent magnitude of $r_\mathrm{SDSS} = 22$ with an average precision of $\sim0.2$~dex.

Based on the results obtained in Sect.~\ref{sec:results:fraction} (see also Fig.~\ref{fig:fr}), the LMFs obtained in this work can be used to provide a more robust spectral-type classification of galaxies in future studies. For instance, our results illustrate that, after the Eddington bias correction, the most massive galaxies at low redshift are almost entirely quiescent galaxies. This means that even when a very massive galaxy in the nearby Universe exhibits a value of $P_\mathrm{Q} \sim 0.5$, this is actually a quiescent galaxy that owing to errors is scattered close to the colour limits used to separate both types (formally expressed by Eq.~\ref{eq:mcde} and Table~\ref{tab:mcde}). Therefore, the LMFs describe the probability of observing a galaxy of each type within the MCDE diagram that in turn can be used as a prior to perform a more robust classification by a Bayesian PDF analysis \citep[see e.g.][]{LopezSanjuan2019}. Nevertheless, we note that this refinement should not be addressed before an LMF analysis, since this might introduce a bias in the results. We plan to tackle this more refined galaxy classification in a future work.

On the other hand, we find necessary a revision of the LMFs at the very nearby Universe with J-PAS. The access to many more nearby galaxies ($z < 0.2$) will allow us to perform a better determination of the completeness limits at the very nearby Universe, as well as to open the possibility of using double Schechter functions for the parametrisation of the LMFs of quiescent galaxies. In addition, a large volume of galaxies will allow us to discard cosmic variance effects, thus reducing the uncertainties in the determination of these functions at any redshift. These facts will facilitate the proper determination of the LMFs at the low mass and faint regimes, which are underestimated in comparison with previous results, hence discarding whether this results of the use of the MCDE diagram for the spectral-type classification of galaxies. In turn, this will permit to discern whether the relative excess in number of massive galaxies spotted at $z \sim 0.1$ (see Sect.~\ref{sec:discussion:previous} and top left panel in Fig.~\ref{fig:MF_comparison}) results of the small area observed in miniJPAS and/or it is a discrepancy with respect to previous works that is ultimately linked to the methodologies or models employed for the stellar mass determination.

In light of our results and based on our experience for the determination of the miniJPAS luminosity functions, we recommend the inclusion of the \textit{odds} parameter in future studies involving J-PAS photo-$z$ and/or PDZs, or at least considering whether there may be an impact and/or bias on the results when accounting for \textit{odds}. This arises from the fact that faint miniJPAS galaxies ($r_\mathrm{SDSS}\sim 22$) usually present lower \textit{odds} values that in turn have photo-$z$ biased to higher values after comparing with spectroscopic redshifts, which is also supported by the luminosity functions constrained in this work. Otherwise, the $\beta$ parameter in the Schechter function for luminosity (see Eq.~\ref{eq:schechter_lum}) would be overestimated and more rapidly evolved after comparing with previous results from the literature. It is also of note that a definition of the sample based on the \textit{odds} parameter (e.g.~by choosing only sources with \textit{odds} higher than a certain limiting value) can also result in a selection bias of the data set.

\begin{figure}
\centering
\includegraphics[trim={0 1.73cm 0 0},clip=True,width=\hsize]{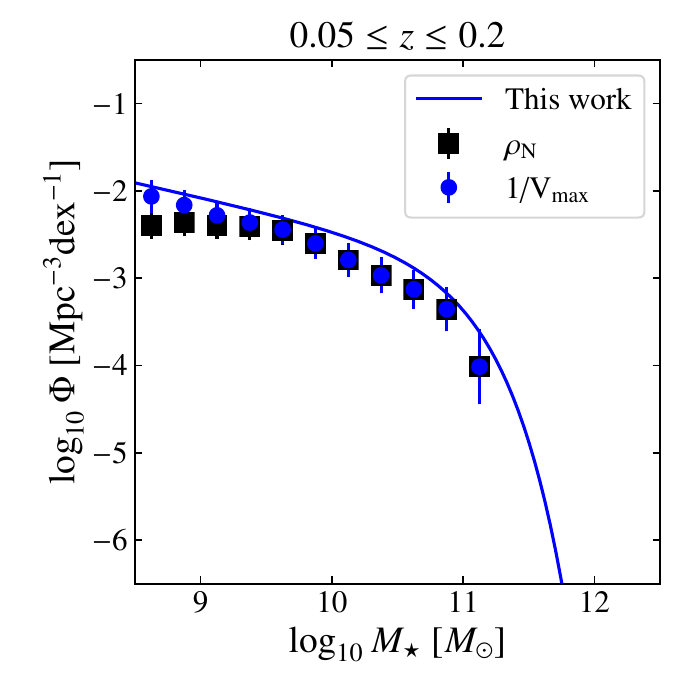}\\
\includegraphics[trim={0 0 0 0.99cm},clip=True,width=\hsize]{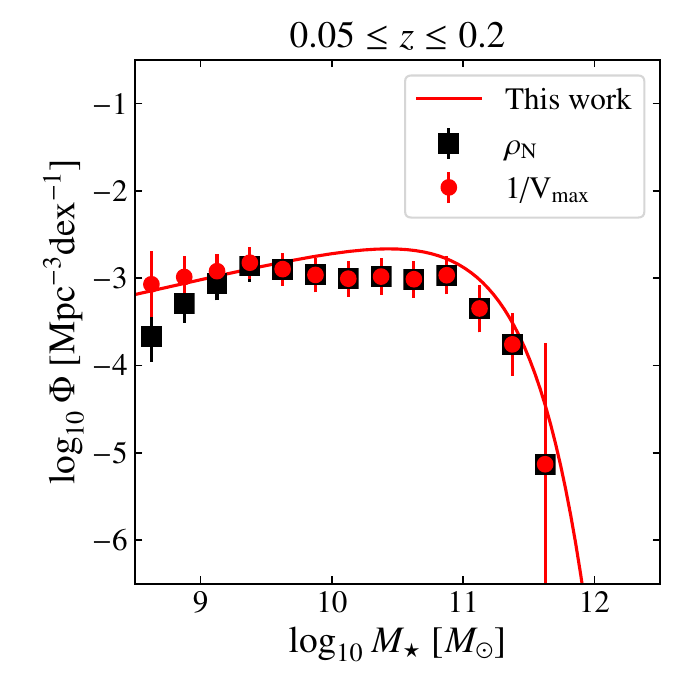}
\caption{Non-parametric stellar mass functions ($1/V_\mathrm{max}$ estimator, dot markers) and comoving number densities ($\rho_\mathrm{N}$, square markers) of star-forming and quiescent galaxies from miniJPAS at $0.5 \le z \le 0.2$ (top and bottom panels, respectively) versus the parametric stellar mass functions determined in this work with our novel methodology (solid lines) at the mean redshift of the bin ($z = 0.12$). The vertical bars illustrate uncertainties in the measurements, which account for Poisson errors and cosmic variance.}
\label{fig:vmax}
\end{figure}

Finally, we find that there is a deficit of galaxies at the nearby Universe in the miniJPAS survey, which in turn supports the discussion carried out by \citet{HernanCaballero2023} about the miniJPAS photo-$z$ distribution. More precisely, the authors found that there was an excess of galaxies at $z < 0.2$ in the Javalambre North Ecliptic Pole survey \citep[J-NEP, a survey imaged by $0.5$--$1.0$ magnitudes deeper and with the same configuration than miniJPAS, further details in][]{HernanCaballero2023} after comparing with the miniJPAS photo-$z$ distribution. All this by using the JPHOTOZ package with exactly the same configuration than for miniJPAS. These `missed' galaxies can be also found after comparing the results from our parametrised LMFs, whose parameters are redshift dependent functions, and those from non-parametric estimators such as the $1/V_\mathrm{max}$ method \citep[e.g.~][]{Schmidt1968,Marshall1985,Ilbert2005}. As shown in Fig.~\ref{fig:vmax}, the $1/V_\mathrm{max}$ method estimator systematically yields lower densities for both star-forming and quiescent galaxies at $z < 0.2$ than those obtained with our novel and more sophisticated methodology, hence supporting the J-NEP results. This lack of galaxies is generalised and independent of the stellar mass range, although it is more severe or remarkable in the stellar mass range of $10 < \log_{10}M_\star < 11$. According to Poisson statistics and the miniJPAS cosmic variance (see vertical bars in Fig.~\ref{fig:vmax} and Sect.~\ref{sec:discussion:variance}), these generalised discrepancies may be largely explained as a result of the cosmic variance. After subtracting both results, we conclude that there is evidence of a lack of galaxies, amounting to an average $\sim30$~\% fraction of galaxies with stellar mass above $9.5$~dex (see Fig.~\ref{fig:vmax}), which roughly involve to around $200$ galaxies. It is worth mentioning that the effects of this lack of galaxies were mitigated thanks to the inclusion of a linear function for the LMF normalisations (i.e.~$\log_{10}\Phi^\star (z) = \bar{\Phi}^\star_1 + \bar{\Phi}^\star_2 \times z$), which supports the idea that our methodology may be less sensitive to cosmic variance effects. It is also of note that the comoving number density of galaxies ($\rho_\mathrm{N}$) points to the same conclusion than the $1/V_\mathrm{max}$ estimator (see Fig.~\ref{fig:vmax}). However, the $\rho_\mathrm{N}$ values are affected by incompleteness at lower stellar masses, and in a more remarkable way, than the $1/V_\mathrm{max}$ case. As a consequence, when $\rho_\mathrm{N}$ is lower than the $1/V_\mathrm{max}$ values, this is an indicative that the galaxy sample is not complete ($\mathcal{C} < 0.95$). For the miniJPAS case at $0.05 < z < 0.2$, this happens for galaxies less massive than $\sim9.5$~dex, which matches with our completeness predictions (see Sect.~\ref{sec:completeness} and Fig.~\ref{fig:completeness}).

\section{Summary and conclusions}\label{sec:summary}


This work is mainly focused on the development of a robust methodology for the determination of the luminosity and stellar mass functions (LMFs) of galaxies by solely using data from large scale multi-filter surveys. Even though this methodology can be easily extended to other kind of surveys, we are especially interested in the application of our methods in the incoming J-PAS survey, putting emphasis in the proper determination of the LMFs of galaxies up to $z \sim 0.7$. In addition, we are especially interested in a method involving the value-added catalogues and the typical outputs from the SED-fitting codes used in the J-PAS collaboration in order to perform a solid and statistical analysis accounting for the uncertainties, degeneracies, and correlations amongst all the involved parameters.

As J-PAS is still an ongoing survey, we use a previous data set referred as miniJPAS \citep[a stripe of $\sim 1$~deg$^2$ centred at the EGS field, see][]{Bonoli2021} that was dictated according to the J-PAS strategy and specifically conceived for testing the performance and potential of the future J-PAS. Part of the ingredients needed for the determination of the LMFs were previously calculated by the J-PAS collaboration and included in miniJPAS value-added catalogues. More precisely, we make use of the `default' photometric redshift constraints \citep[photo-$z$,][]{HernanCaballero2021}, the star and galaxy classification of sources \citep[][]{LopezSanjuan2019}, along with the forced-aperture photometry included in the general miniJPAS catalogues \citep[][]{Bonoli2021}. However, the determination of LMFs requires additional inputs (e.g.~stellar mass and luminosity constraints for each of the miniJPAS galaxies, spectral-type classification of galaxies, and completeness of the flux-limited sample), that are constrained by a set of techniques specifically designed and built from scratch for this study.

Firstly, we determine the probability distribution functions of the $B$-band luminosity and stellar mass via SED-fitting techniques. For this purpose, we use an updated version of the SED-fitting code dubbed MUFFIT \citep[MUlti-Filter FITting for stellar population diagnostics,][]{DiazGarcia2015}, which is carefully developed and optimised to deal with multi-band photometric data. In brief, the SED-fitting analysis performed by MUFFIT is solely based on the stellar continuum of galaxies or colours (i.e.~nebular and AGN emission lines are removed during the analysis), using composite models of stellar populations (mixtures of two SSPs, i.e.~non-parametric star formation history) for a recent version of the \citet{Bruzual2003} SSP models (CB17), the attenuation law of \citet{Calzetti2000}, the miniJPAS fluxes obtained for \texttt{AUTO} aperture photometry, and the photo-$z$ probability distribution functions provided for each of the miniJPAS sources. Secondly, we adopt the stellar mass versus rest-frame colour diagram corrected for extinction \citep[MCDE,][]{DiazGarcia2019a} for performing a spectral-type classification of galaxies (quiescent and star-forming), hence preventing the inclusion of dusty star-forming galaxies in the quiescent sample. As a result, we assign probabilities of being quiescent galaxies according to the position of sources within this diagram, uncertainties, and degeneracies in the involved parameters. Thirdly, we build parametric functions for the stellar mass and $B$-band luminosity completeness of our sample as a function of redshift and for different magnitude limits of the $r_\mathrm{SDSS}$ detection band.

The LMFs of the miniJPAS galaxies are formally characterised by single-Schechter functions and according to the spectral-type classification of galaxies, while the LMFs of the full sample result of the sum of the quiescent and star-forming ones. The determination of the parameters defining the LMFs are constrained by a novel maximum likelihood method, which makes the most of our sample and all the statistical results obtained within the J-PAS collaboration. Thanks to the inclusion of weights, we are able to include the probability distribution functions and correlations of the parameters involved (photo-$z$, stellar mass, and $B$-band luminosity), the sample completeness, various probabilities (e.g.~the star-galaxy classification), and priors (e.g.~the range of values for the Schechter parameters) for each of the galaxy spectral-types explored in this work (i.e.~quiescent and star-forming) in a robust statistical way. Moreover, the miniJPAS LMFs are subsequently corrected for the so-called Eddington bias effect.

Broadly, the LMFs obtained by miniJPAS galaxies point to smooth evolution with redshift down to $z=0.7$, where their shapes or Schechter parameters mainly rely on the galaxy spectral-type. The LMF variations with redshift for star-forming galaxies mainly affects the bright and high-mass ends, while their faint and low-mass ends remain roughly constant. In this sense, we find that there is a decreasing number of bright star-forming galaxies ($M_B < -23$) at decreasing redshift, while our results are compatible with a subtle increase of massive star-forming galaxies ($\log_{10}M_\star > 11$~dex). Regarding quiescent galaxies, we reach similar conclusions for the bright and high-mass ends of the LMFs than for the star-forming case. However, there is evidence of a greater evolution in the number density of low/intermediate-mass and faint quiescent galaxies ($\log_{10}M_\star < 10$ and $M_B > -20$, respectively). In addition, the cosmic evolution of the global $B$-band luminosity density is undergoing a slight decrease of $\sim0.1$~dex from $z=0.7$ to $z=0$. There is evidence that this density decrement is mainly driven by star-forming galaxies, since the $B$-band luminosity density of quiescent galaxies is compatible with a non-evolution. On the other hand, the global stellar mass density increases $\sim0.3$~dex at the same redshift range, which is mainly driven by the quiescent galaxy population. At $z < 0.7$, the $B$-band luminosity functions point out that the brightest galaxies are mainly star-forming galaxies, whereas more than a $60$~\% fraction of massive galaxies ($\log_{10} M_\star > 10.7$~dex) are quiescent galaxies. The fraction of red galaxies for the faint and low mass regimes ($M_B \gtrsim -22$ and $\log_{10}M_\star < 10.5$~dex, respectively) is below $20$~\%, meaning that star-forming galaxies dominate in number and are much more frequent than quiescent galaxies in these regimes. Nevertheless, the fraction of red galaxies with $M_B \gtrsim -22$ and/or $\log_{10}M_\star < 10.5$~dex is gaining prominence at decreasing redshift.

After comparing with LMFs from previous work involving photometric and/or spectroscopic data at redshift $0.05 \le z \le 0.7$ (e.g.~SDSS, DEEP2, zCOSMOS, UltraVISTA, 3D-HST, etc.), we find a reasonable good agreement with the miniJPAS LMFs for both quiescent and star-forming spectral-types. Likewise, the cosmic evolution of the $B$-band luminosity and stellar mass densities show similar values and trends than those obtained in similar researches. This highlights that the miniJPAS photo-$z$ constraints, star and galaxy classification, spectral-type, stellar mass, and rest-frame luminosities computed within the J-PAS collaboration, along with the methodology introduced in this manuscript, are self-consistent as to determine the LMFs of galaxies down to $z=0.7$ in good agreement with previous work.


\begin{acknowledgements}

L.A.D.G., R.M.G.D., R.G.B., G.M.S., J.E.R.M., I.M., and J.M.V.~acknowledge financial support from the State Agency for Research of the Spanish MCIU through `Center of Excellence Severo Ochoa' award to the Instituto de Astrof\'isica de Andaluc\'ia (SEV-2017-0709) and CEX2021-001131-S funded by MCIN/AEI/10.13039/501100011033. L.A.D.G., R.M.G.D., R.G.B., G.M.S., and J.E.R.M.~are also grateful for financial support from the project PID-2019-109067-GB100. A.H.C.~receives financial support from the Spanish Ministry of Science and Innovation (MCIN/AEI/10.13039/501100011033) and `ERDF A way of making Europe' through the grant PID2021-124918NB-C44. In addition, I.M.~thanks for financial support from the project PID2019-106027GB-C41 and J.M.V.~from project PID2019-107408GB-C44. 

Based on observations made with the JST/T250 telescope and PathFinder camera for the miniJPAS project at the Observatorio Astrof\'{\i}sico de Javalambre (OAJ), in Teruel, owned, managed, and operated by the Centro de Estudios de F\'{\i}sica del Cosmos de Arag\'on (CEFCA). We acknowledge the OAJ Data Processing and Archiving Unit (UPAD) for reducing and calibrating the OAJ data used in this work. Funding for OAJ, UPAD, and CEFCA has been provided by the Governments of Spain and Arag\'on through the Fondo de Inversiones de Teruel; the Arag\'on Government through the Research Groups E96, E103, and E16\_17R; the Spanish Ministry of Science, Innovation and Universities (MCIU/AEI/FEDER, UE) with grant PGC2018-097585-B-C21; the Spanish Ministry of Economy and Competitiveness (MINECO/FEDER, UE) under AYA2015-66211-C2-1-P, AYA2015-66211-C2-2, AYA2012-30789, and ICTS-2009-14; and European FEDER funding (FCDD10-4E-867, FCDD13-4E-2685).

\end{acknowledgements}


%
%

\bibliographystyle{aa}
\bibliography{LMF_v1.1}


\begin{appendix}


\section{Stellar mass and $B$-band luminosity completeness for samples with different magnitude cuts}\label{sec:appendix:completeness}

In this appendix, we include the coefficients defining the stellar mass and $B$-band luminosity completeness of the miniJPAS galaxies for samples with different magnitude cuts following the methodology detailed in Sect.~\ref{sec:completeness}. Even though these coefficients were not explicitly used in this research, they can be very useful to define the stellar mass or luminosity completeness of samples in future works involving miniJPAS data. In particular, we present the coefficients for magnitude cuts of $r_\mathrm{SDSS} \le 21.5$, $21.75$, $22.25$, and $22.5$ in the detection band (Tables~\ref{tab:completeness_r215}--\ref{tab:completeness_r225}, respectively). 

\begin{table*}
\caption{Coefficients determining the stellar mass and $B$-band luminosity completeness (\textit{top} and \textit{bottom panels}, respectively) of the star-forming and quiescent galaxies from miniJPAS for our flux-limited sample at $r_\mathrm{SDSS} \le 21.5$.}
\label{tab:completeness_r215}
\centering
\begin{tabular}{c c c c c c c}
\hline\hline
& \multicolumn{3}{c}{Star-forming} & \multicolumn{3}{c}{Quiescent}\\
\multirow{2}{*}{$\mathcal{C}$} & \multirow{2}{*}{$\mu$} & \multirow{2}{*}{$\nu$} & \multirow{2}{*}{$\gamma$} & \multirow{2}{*}{$\mu$} & \multirow{2}{*}{$\nu$} & \multirow{2}{*}{$\gamma$}\\
 & & & & & & \\
\hline
$0.05$ & $10.607 \pm 0.117$ & $0.128 \pm 0.031$ & $0.000 \pm 0.066$ & $11.595 \pm 0.102$ & $0.141 \pm 0.026$ & $0.016 \pm 0.044$\\
$0.50$ & $11.143 \pm 0.064$ & $0.129 \pm 0.016$ & $0.000 \pm 0.032$ & $11.776 \pm 0.071$ & $0.134 \pm 0.018$ & $0.015 \pm 0.033$\\
$0.95$ & $11.700 \pm 0.080$ & $0.129 \pm 0.020$ & $0.000 \pm 0.043$ & $11.979 \pm 0.128$ & $0.131 \pm 0.033$ & $0.020 \pm 0.064$\\
\hline
\multirow{2}{*}{$\mathcal{C}$} & \multirow{2}{*}{$\delta$} & \multirow{2}{*}{$\epsilon$} & \multirow{2}{*}{$\zeta$} & \multirow{2}{*}{$\delta$} & \multirow{2}{*}{$\epsilon$} & \multirow{2}{*}{$\zeta$}\\
 & & & & & & \\
\hline
$0.05$ & $-22.627 \pm 0.054$ & $0.150 \pm 0.009$ & $0.000 \pm 0.019$ & $-23.205 \pm 0.060$ & $0.167 \pm 0.008$ & $0.000 \pm 0.015$\\
$0.50$ & $-22.785 \pm 0.109$ & $0.177 \pm 0.017$ & $0.086 \pm 0.043$ & $-23.335 \pm 0.061$ & $0.198 \pm 0.009$ & $0.084 \pm 0.020$\\
$0.95$ & $-21.702 \pm 0.699$ & $0.277 \pm 0.042$ & $0.389 \pm 0.118$ & $-22.715 \pm 0.414$ & $0.270 \pm 0.029$ & $0.284 \pm 0.072$\\
\hline
\end{tabular}
\end{table*}

\begin{table*}
\caption{As Table~\ref{tab:completeness_r215}, but for a flux-limited sample of $r_\mathrm{SDSS} \le 21.75$.}
\label{tab:completeness_r2175}
\centering
\begin{tabular}{c c c c c c c}
\hline\hline
& \multicolumn{3}{c}{Star-forming} & \multicolumn{3}{c}{Quiescent}\\
\multirow{2}{*}{$\mathcal{C}$} & \multirow{2}{*}{$\mu$} & \multirow{2}{*}{$\nu$} & \multirow{2}{*}{$\gamma$} & \multirow{2}{*}{$\mu$} & \multirow{2}{*}{$\nu$} & \multirow{2}{*}{$\gamma$}\\
 & & & & & & \\
\hline
$0.05$ & $10.459 \pm 0.082$ & $0.127 \pm 0.028$ & $0.023 \pm 0.070$ & $11.560 \pm 0.128$ & $0.136 \pm 0.031$ & $0.001 \pm 0.048$\\
$0.50$ & $11.049 \pm 0.083$ & $0.128 \pm 0.023$ & $0.000 \pm 0.051$ & $11.620 \pm 0.058$ & $0.126 \pm 0.015$ & $0.002 \pm 0.028$\\
$0.95$ & $11.658 \pm 0.153$ & $0.136 \pm 0.038$ & $0.000 \pm 0.079$ & $11.687 \pm 0.130$ & $0.113 \pm 0.031$ & $0.000 \pm 0.060$\\
\hline
\multirow{2}{*}{$\mathcal{C}$} & \multirow{2}{*}{$\delta$} & \multirow{2}{*}{$\epsilon$} & \multirow{2}{*}{$\zeta$} & \multirow{2}{*}{$\delta$} & \multirow{2}{*}{$\epsilon$} & \multirow{2}{*}{$\zeta$}\\
 & & & & & & \\
\hline
$0.05$ & $-22.303 \pm 0.069$ & $0.164 \pm 0.013$ & $0.040 \pm 0.031$ & $-23.016 \pm 0.072$ & $0.177 \pm 0.013$ & $0.010 \pm 0.024$\\
$0.50$ & $-22.345 \pm 0.144$ & $0.198 \pm 0.016$ & $0.143 \pm 0.042$ & $-23.043 \pm 0.110$ & $0.209 \pm 0.014$ & $0.101 \pm 0.031$\\
$0.95$ & $-21.761 \pm 0.347$ & $0.260 \pm 0.022$ & $0.335 \pm 0.063$ & $-22.433 \pm 0.471$ & $0.268 \pm 0.032$ & $0.281 \pm 0.082$\\
\hline
\end{tabular}
\end{table*}

\begin{table*}
\caption{As Table~\ref{tab:completeness_r215}, but for a flux-limited sample of $r_\mathrm{SDSS} \le 22.25$.}
\label{tab:completeness_r2225}
\centering
\begin{tabular}{c c c c c c c}
\hline\hline
& \multicolumn{3}{c}{Star-forming} & \multicolumn{3}{c}{Quiescent}\\
\multirow{2}{*}{$\mathcal{C}$} & \multirow{2}{*}{$\mu$} & \multirow{2}{*}{$\nu$} & \multirow{2}{*}{$\gamma$} & \multirow{2}{*}{$\mu$} & \multirow{2}{*}{$\nu$} & \multirow{2}{*}{$\gamma$}\\
 & & & & & & \\
\hline
$0.05$ & $10.600 \pm 0.134$ & $0.134 \pm 0.046$ & $0.000 \pm 0.114$ & $11.248 \pm 0.120$ & $0.140 \pm 0.041$ & $0.010 \pm 0.072$\\
$0.50$ & $10.953 \pm 0.120$ & $0.127 \pm 0.037$ & $0.000 \pm 0.095$ & $11.419 \pm 0.054$ & $0.131 \pm 0.018$ & $0.006 \pm 0.033$\\
$0.95$ & $11.345 \pm 0.127$ & $0.122 \pm 0.038$ & $0.000 \pm 0.104$ & $11.562 \pm 0.104$ & $0.117 \pm 0.031$ & $0.000 \pm 0.059$\\
\hline
\multirow{2}{*}{$\mathcal{C}$} & \multirow{2}{*}{$\delta$} & \multirow{2}{*}{$\epsilon$} & \multirow{2}{*}{$\zeta$} & \multirow{2}{*}{$\delta$} & \multirow{2}{*}{$\epsilon$} & \multirow{2}{*}{$\zeta$}\\
 & & & & & & \\
\hline
$0.05$ & $-21.985 \pm 0.133$ & $0.182 \pm 0.018$ & $0.056 \pm 0.047$ & $-22.548 \pm 0.094$ & $0.173 \pm 0.017$ & $0.000 \pm 0.032$\\
$0.50$ & $-21.492 \pm 0.448$ & $0.241 \pm 0.032$ & $0.242 \pm 0.091$ & $-22.622 \pm 0.122$ & $0.216 \pm 0.014$ & $0.097 \pm 0.032$\\
$0.95$ & $-18.309 \pm 1.638$ & $0.404 \pm 0.074$ & $0.761 \pm 0.225$ & $-21.452 \pm 2.461$ & $0.323 \pm 0.146$ & $0.357 \pm 0.354$\\
\hline
\end{tabular}
\end{table*}

\begin{table*}
\caption{As Table~\ref{tab:completeness_r215}, but for a flux-limited sample of $r_\mathrm{SDSS} \le 22.5$.}
\label{tab:completeness_r225}
\centering
\begin{tabular}{c c c c c c c}
\hline\hline
& \multicolumn{3}{c}{Star-forming} & \multicolumn{3}{c}{Quiescent}\\
\multirow{2}{*}{$\mathcal{C}$} & \multirow{2}{*}{$\mu$} & \multirow{2}{*}{$\nu$} & \multirow{2}{*}{$\gamma$} & \multirow{2}{*}{$\mu$} & \multirow{2}{*}{$\nu$} & \multirow{2}{*}{$\gamma$}\\
 & & & & & & \\
\hline
$0.05$ & $10.709 \pm 0.163$ & $0.138 \pm 0.060$ & $0.000 \pm 0.139$ & $11.163 \pm 0.105$ & $0.140 \pm 0.038$ & $0.007 \pm 0.067$\\
$0.50$ & $10.885 \pm 0.117$ & $0.120 \pm 0.037$ & $0.000 \pm 0.103$ & $11.324 \pm 0.051$ & $0.131 \pm 0.017$ & $0.003 \pm 0.032$\\
$0.95$ & $11.143 \pm 0.137$ & $0.116 \pm 0.046$ & $0.000 \pm 0.132$ & $11.486 \pm 0.094$ & $0.119 \pm 0.028$ & $0.000 \pm 0.055$\\
\hline
\multirow{2}{*}{$\mathcal{C}$} & \multirow{2}{*}{$\delta$} & \multirow{2}{*}{$\epsilon$} & \multirow{2}{*}{$\zeta$} & \multirow{2}{*}{$\delta$} & \multirow{2}{*}{$\epsilon$} & \multirow{2}{*}{$\zeta$}\\
 & & & & & & \\
\hline
$0.05$ & $-21.627 \pm 0.401$ & $0.205 \pm 0.041$ & $0.110 \pm 0.110$ & $-22.346 \pm 0.063$ & $0.175 \pm 0.012$ & $0.000 \pm 0.023$\\
$0.50$ & $-21.290 \pm 0.424$ & $0.245 \pm 0.030$ & $0.240 \pm 0.085$ & $-22.410 \pm 0.125$ & $0.219 \pm 0.014$ & $0.095 \pm 0.032$\\
$0.95$ & $-20.478 \pm 1.526$ & $0.298 \pm 0.083$ & $0.422 \pm 0.248$ & $-20.746 \pm 2.754$ & $0.352 \pm 0.153$ & $0.417 \pm 0.371$\\
\hline
\end{tabular}
\end{table*}


\section{Constraints on the parameters defining the miniJPAS LMFs}\label{sec:appendix:LMF}

\begin{figure*}
\centering
\includegraphics[width=0.6\hsize]{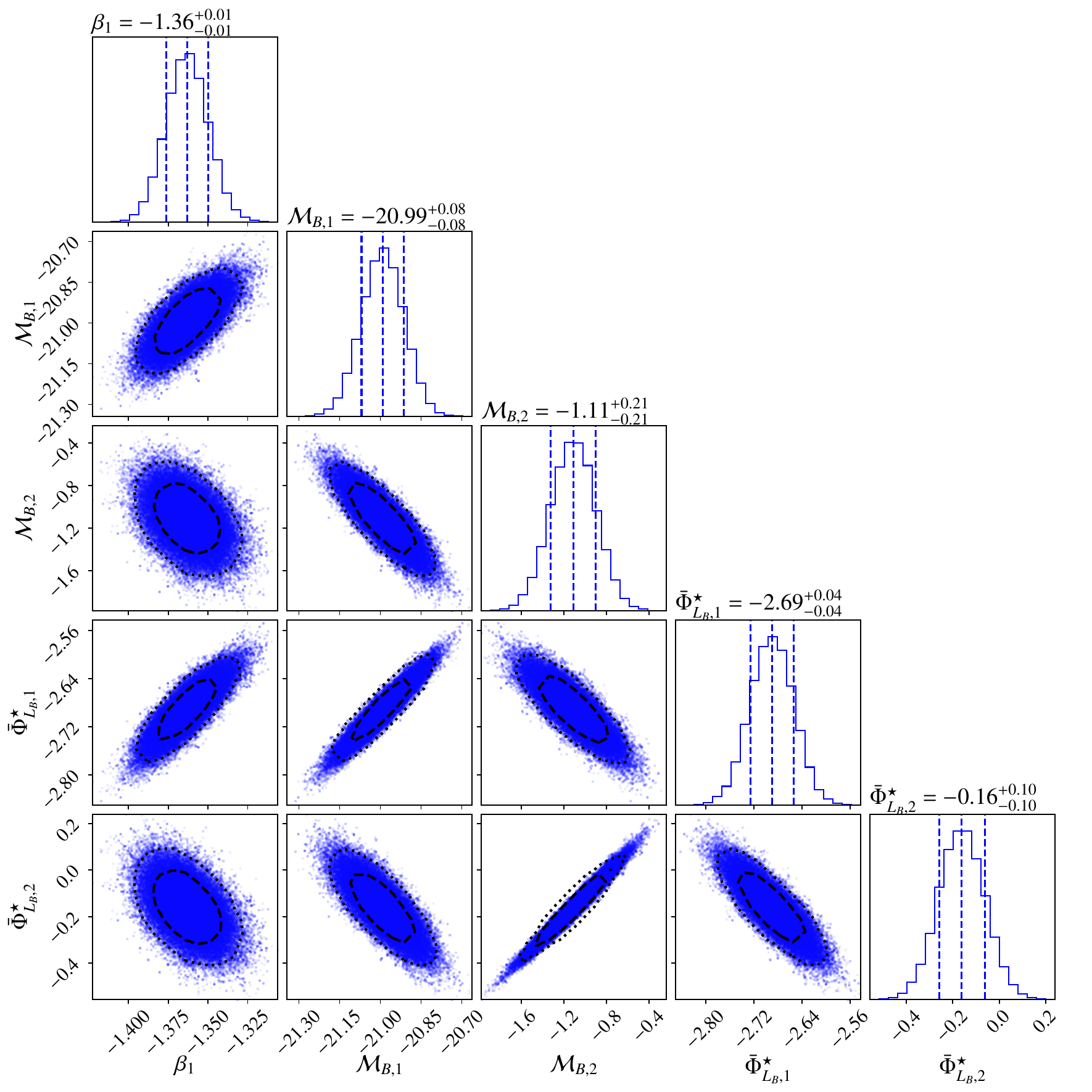}
\includegraphics[width=0.6\hsize]{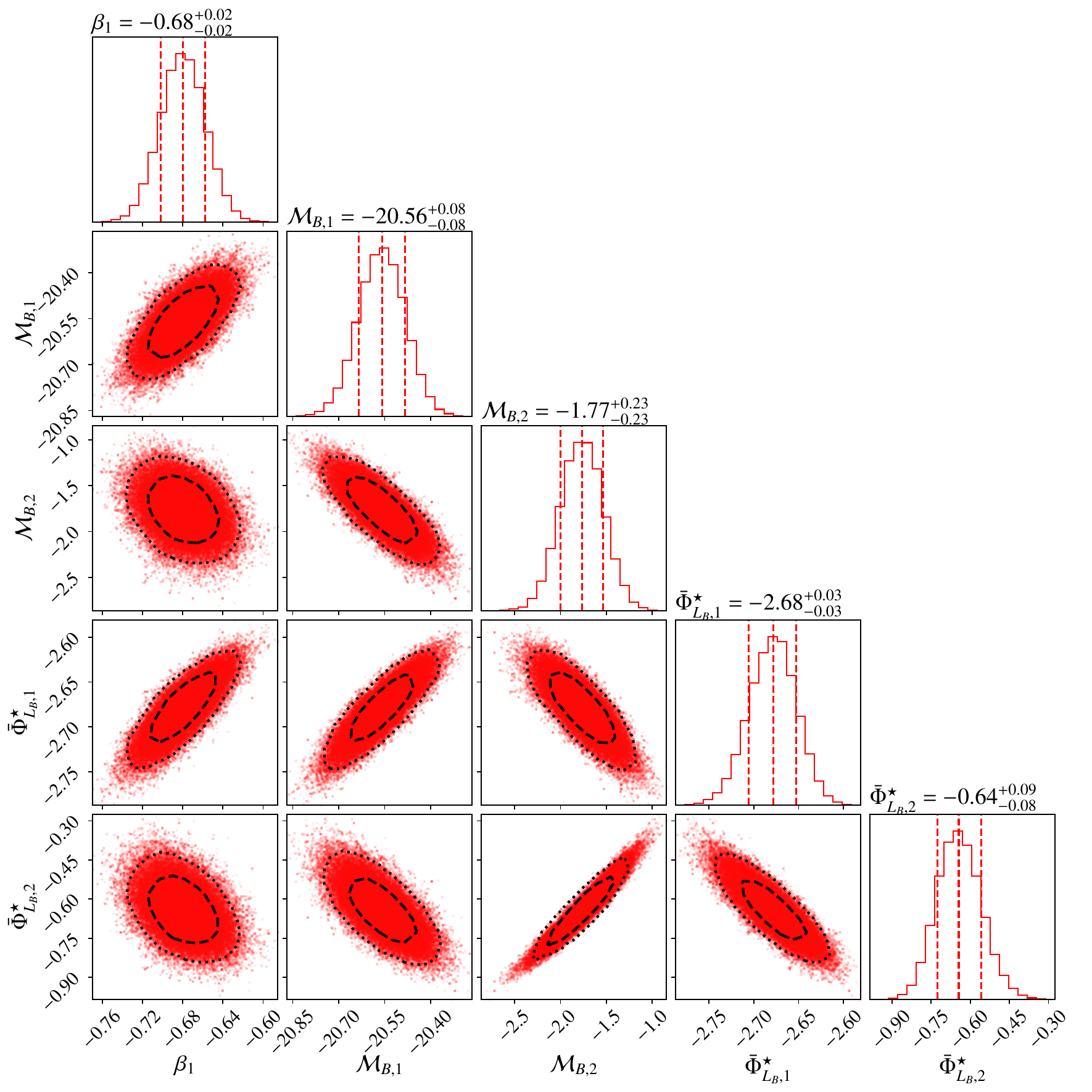}
\caption{Constraints on the parameters defining the $B$-band luminosity functions of miniJPAS star-forming and quiescent galaxies (\textit{top and bottom corner panels}) according to the methodology developed in Sect.~\ref{sec:zsty}. Each dot correspond to one of the solutions contained in the MCMC chains. The dashed and dotted black lines illustrate the confidence regions containing the $68$ and $90$~\% of probability, respectively. The dashed vertical lines establish the 1$\sigma$ confidence interval and median values of the parameter histograms.}
\label{fig:corner_LF}
\end{figure*}

\begin{figure*}
\centering
\includegraphics[width=0.6\hsize]{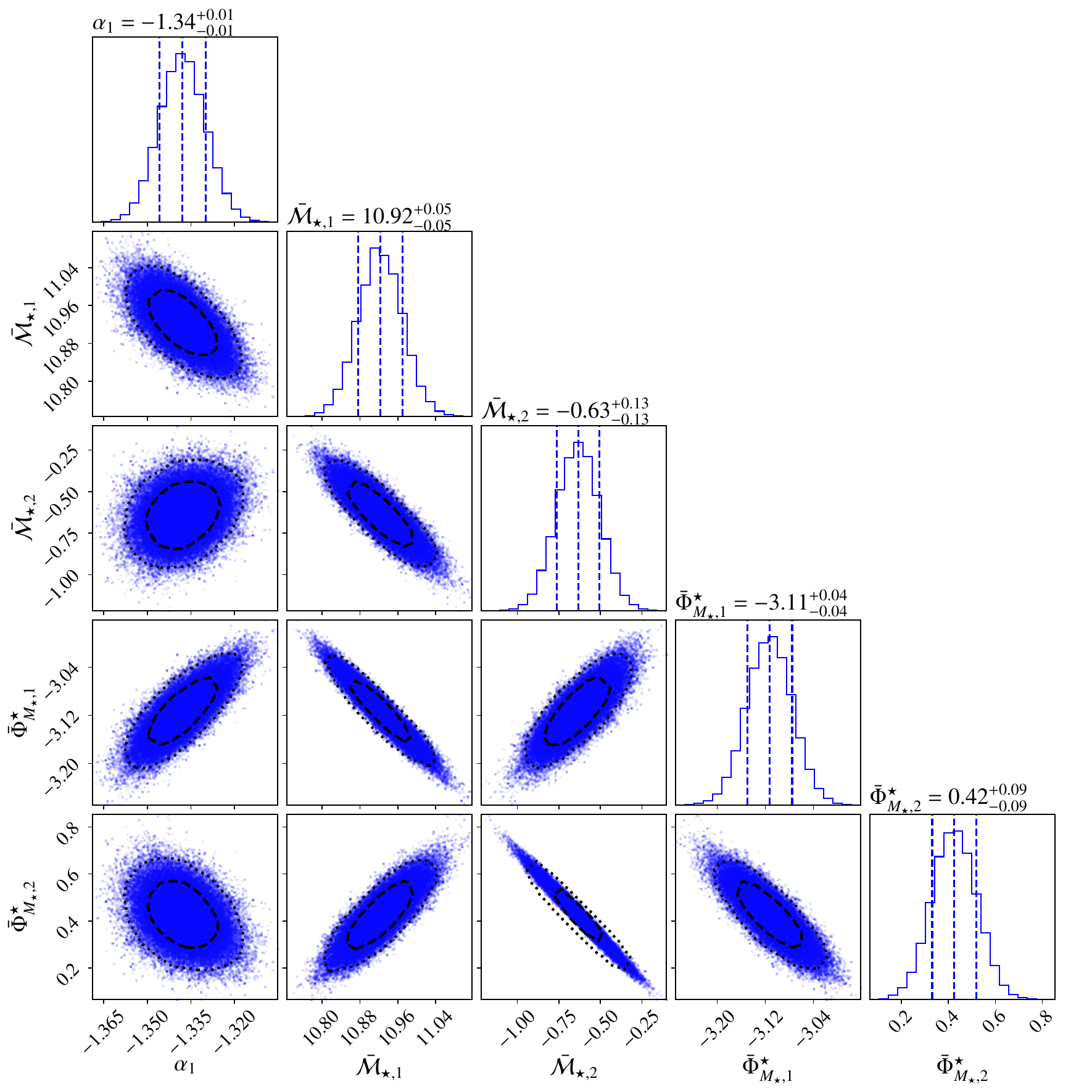}
\includegraphics[width=0.7\hsize]{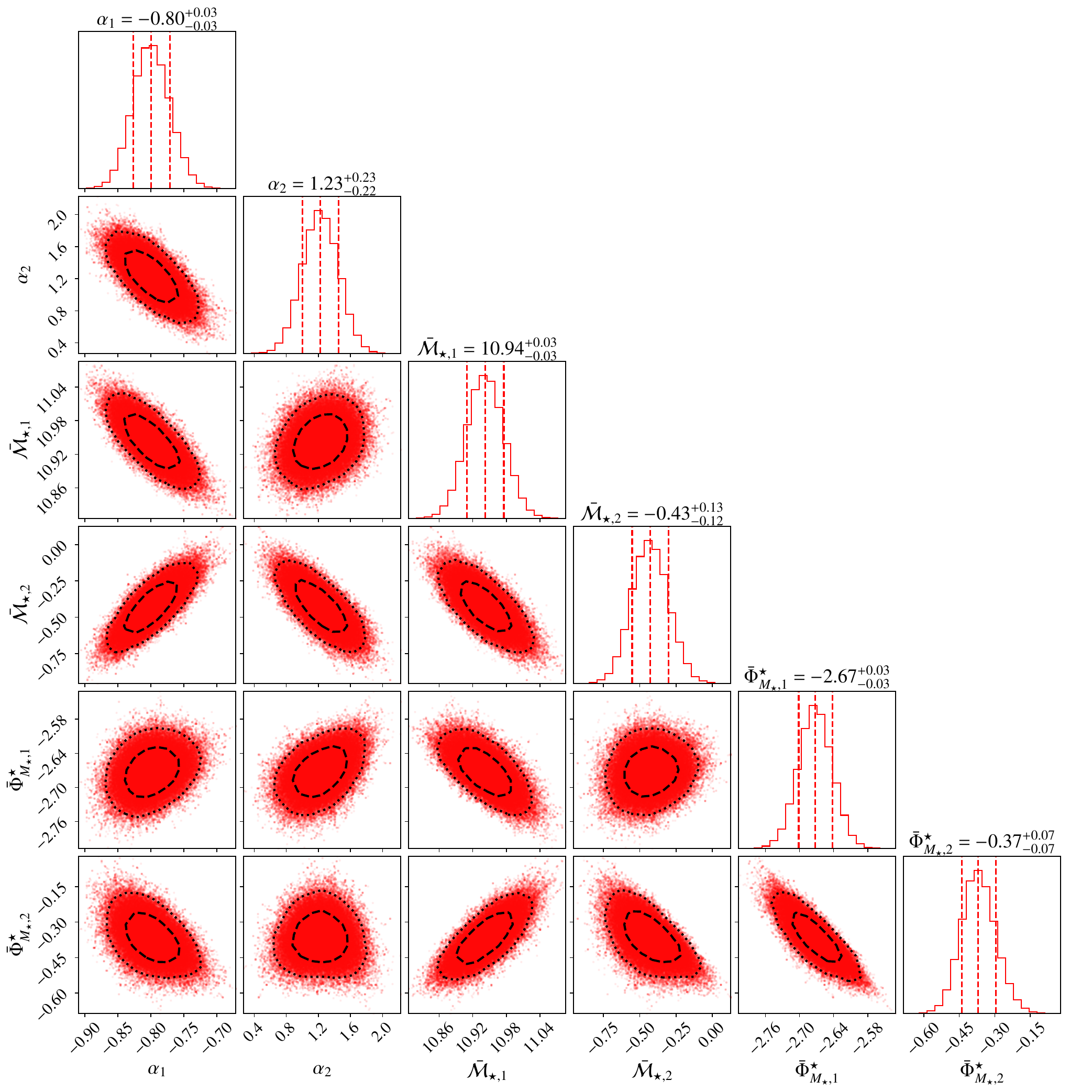}
\caption{As Fig.~\ref{fig:corner_LF}, but for stellar mass functions.}
\label{fig:corner_MF}
\end{figure*}

In this section, we include the sampling of the posterior distributions of the coefficients defining the parametric $B$-band luminosity and stellar mass functions (Figs.~\ref{fig:corner_LF} and \ref{fig:corner_MF}, respectively) that resulted from the likelihood maximisation of Eq.~\ref{eq:likelihood_z} (largely detailed in Sects.~\ref{sec:zsty:par}--\ref{sec:zsty:norm}) and the subsequent correction for Eddington bias effects (see Sect.~\ref{sec:zsty:eddington}). These results are obtained according to the galaxy spectral-type and constrain the diverse evolutionary tracks followed by the two kind of galaxies explored in this work (i.e.~quiescent and star-forming galaxies), as well as the degeneracies and/or correlations amongst the parameters involved in the single Schechter functions (see Sect.~\ref{sec:results:lmf} for a further description).

\end{appendix}

\end{document}